\documentclass[12pt]{article}

\usepackage{epsf,epsfig,epstopdf}
\usepackage{graphics}

\usepackage{url}
\usepackage{amsmath,amssymb}
\usepackage{cite}
\usepackage{a4wide}
\numberwithin{equation}{section}

\newcommand{\bff}[1]{\mbox{\boldmath ${#1}$}}

\begin{document}

\allowdisplaybreaks
\thispagestyle{empty}

\begin{flushright}
{\small
TTK-11-38\\
ITP-UU-11/26\\
SPIN-11/19 \\
FR-PHENO-2011-015\\
SFB/CPP-11-49\\[0.2cm]
September 7, 2011}
\end{flushright}

\vspace{\baselineskip}

\begin{center}
\vspace{0.5\baselineskip}
\textbf{\Large\boldmath
Hadronic top-quark pair production with NNLL\\[0.2cm]  
threshold resummation
}
\\
\vspace{3\baselineskip}
{\sc M.~Beneke$^a$, P.~Falgari$^b$, S.~Klein$^a$, C.~Schwinn$^c$}\\
\vspace{0.7cm}
{\sl ${}^a$Institute f\"ur Theoretische Teilchenphysik und 
Kosmologie,\\
RWTH Aachen University, D--52056 Aachen, Germany\\
\vspace{0.3cm}
${}^b$Institute for Theoretical Physics and Spinoza Institute,\\
Utrecht University, 3508 TD Utrecht, The Netherlands\\
\vspace{0.3cm}
${}^c$ Albert-Ludwigs Universit\"at Freiburg, 
Physikalisches Institut, \\
D-79104 Freiburg, Germany }

\vspace*{1.2cm}
\textbf{Abstract}\\
\vspace{1\baselineskip}
\parbox{0.9\textwidth}{ 
We compute the total top-quark pair production cross 
section at the Tevatron and LHC based on approximate NNLO results, 
and on the summation of threshold logarithms and Coulomb enhancements 
to all orders with next-to-next-to-leading logarithmic (NNLL) 
accuracy, including bound-state effects. We find 
\begin{eqnarray*}
&&\sigma_{t\bar t}(\mbox{Tevatron}) = 
(7.22^{+0.31}_{-0.47}{}^{+0.71}_{-0.55})\,\mbox{pb}
\\[0.2cm]
&&\sigma_{t\bar t}(\mbox{LHC,}\,\sqrt{s}=7\,\mbox{TeV}) = 
(162.6^{+7.4}_{-7.6}{}^{+15.4}_{-14.7})\,\mbox{pb}
\end{eqnarray*}
for $m_t=173.3\,$GeV. The implementation of joint soft and Coulomb 
resummation, its ambiguities, and the present theoretical uncertainty 
are discussed in detail. We further obtain new approximate results at 
N$^{3}$LO. 
} 
\end{center}

\newpage
\setcounter{page}{1}

\section{Introduction}

Fifteen years after the discovery of the top quark, the experimental
uncertainty of the measurement of the top-pair production cross section
at the Tevatron has dropped below ten
percent~\cite{Aaltonen:2010ic,cdf:sigmatt,Abazov:2011mi,Abazov:2011cq}. At
the LHC, the precision of the measurements quickly approaches that
reached at the
Tevatron~\cite{atlas:sigmattlatest,CMS:2011yy}. Therefore precision
studies of the heaviest quark known today can be expected in the near
future.  An interesting use of the total cross section measurement is
the extraction of the top-quark
mass~\cite{Langenfeld:2009wd,Abazov:2011pt,atlas:mtt} in a cleaner (albeit less
precise) way than the direct measurements.  Furthermore, the total
cross section provides constraints on new-physics models that try to
explain the forward-backward asymmetry anomaly observed at
CDF~\cite{Aaltonen:2011kc}. More generally, its experimental
reconstruction involves several aspects, such as missing energy and
$b$-quark tagging, relevant to the search for heavy particles in
extensions of the Standard Model.

The precision of experiments motivates improving the
accuracy of the theoretical calculation of the total top-pair
production cross section beyond that of the next-to-leading
order~(NLO) result~\cite{Nason:1987xz}, and of the summation of 
leading-logarithmic (LL) or next-to-leading logarithmic~(NLL) soft-gluon
corrections~\cite{Laenen:1991af,Catani:1996dj,Berger:1996ad,Kidonakis:1996zd,Bonciani:1998vc,Kidonakis:2001nj}. Work on the complete next-to-next-to-leading order
(NNLO) corrections is in progress and several ingredients are
already known~\cite{Czakon:2008zk,Bonciani:2008az,Bonciani:2009nb,Korner:2008bn,Anastasiou:2008vd,Kniehl:2008fd,Dittmaier:2007wz,Bonciani:2010mn,Czakon:2011ve,Bierenbaum:2011gg}.
Meanwhile, progress in the understanding
of the infrared structure of massive two-loop scattering amplitudes in
QCD~\cite{Kidonakis:2009ev,Mitov:2009sv,Becher:2009kw,Ferroglia:2009ep,Ferroglia:2009ii,Mitov:2010xw}
has provided the prerequisites to compute soft-gluon effects at next-to-next-to-leading logarithmic~(NNLL)
accuracy~\cite{Beneke:2009rj,Czakon:2009zw,Beneke:2009ye,Ahrens:2009uz,Ahrens:2010zv,Beneke:2010da,Beneke:2010fm,Kidonakis:2010dk,Ahrens:2011mw}.

In this paper we are concerned with resumming higher-order corrections
to the partonic cross sections, $\hat\sigma_{pp'}$, for the processes 
\begin{equation}
p(k_1)
p'(k_2)\rightarrow t(p_1)\bar t (p_2)+X \, ,
\label{eq:ttbar-proc}  
\end{equation}
that become large in the partonic threshold limit,
\begin{equation}
\label{eq:def-thresh}
\hat s\equiv (k_1+k_2)^2 \to 4m_t^2 \, ,
\end{equation}
where the top quarks have a small relative velocity 
$\beta=\sqrt{1-\frac{4m_t^2}{\hat s}}$. In this limit corrections due
to the radiation of soft gluons and the exchange of virtual Coulomb
gluons are both
enhanced, resulting in terms of the form $\alpha_s\log^{2,1}\beta$ 
and $\alpha_s/\beta$ at every order of the coupling expansion. Near the 
partonic threshold, $\beta\approx 0$, these corrections become large and 
should be resummed to all orders. Instead of organizing radiative 
corrections by the power of the strong coupling constant it is then 
appropriate to count both  $\alpha_s \ln\beta$ and $\alpha_s/\beta$ as 
quantities of order one. Exploiting the exponentiation of double 
logarithmic terms leads to a parametric representation of the expansion 
of the cross section in the form
\begin{eqnarray}
\label{eq:syst}
\hat{\sigma}_{p p'} &=& \,\hat \sigma^{(0)}_{p p'}\, 
\sum_{k=0} \,\left(\frac{\alpha_s}{\beta}\right)^{\!k} \,
\exp\Big[\underbrace{\ln\beta\,g_0(\alpha_s\ln\beta)}_{\mbox{(LL)}}+ 
\underbrace{g_1(\alpha_s\ln\beta)}_{\mbox{(NLL)}}+
\underbrace{\alpha_s g_2(\alpha_s\ln\beta)}_{\mbox{(NNLL)}}+\ldots\Big]
\nonumber\\[0.2cm]
&& \,\times
\left\{1\,\mbox{(LL,NLL)}; \alpha_s, \beta \,\mbox{(NNLL)}; 
\alpha_s^2,\alpha_s \beta ,\beta^2 \,\mbox{(N$^3$LL)};
\ldots\right\}\, . 
\end{eqnarray}
We remark that the structure of the expansion is significantly more 
complicated than for Drell-Yan or Higgs production. The presence of 
Coulomb enhancements, $(\alpha_s/\beta)^k$, not only calls for a 
joint summation of soft-gluon and Coulomb effects, but also implies 
that one must address $O(\beta)$ corrections from soft emission 
at NNLL~\cite{Beneke:2009ye,Beneke:2010da}, which count as a 
power-suppressed effect in Drell-Yan and Higgs production. In general, 
increasing the order in (\ref{eq:syst}) requires higher-loop corrections to 
anomalous dimensions to sum logarithms, and the inclusion of power 
corrections of increasingly higher order.

In order to obtain the total hadronic cross section, the partonic
cross section is convoluted with the parton distribution 
functions.  Both at Tevatron and LHC, the top anti-top invariant-mass 
distribution peaks at about $400\,$GeV, which corresponds 
(in the absence of radiation) to $\beta \approx 0.5$. The convolution of 
the partonic cross section with the parton luminosity is therefore 
dominated by the region $\beta>0.3$, where the threshold approximation 
is not valid (see e.g~\cite{Beneke:2010fm}). Nevertheless, one often 
finds that the threshold expansion provides a reasonable approximation 
even outside its domain of validity. For instance, in the absence of 
the exact NLO calculation one would do much better including the 
threshold-enhanced NLO terms rather than staying with the LO 
approximation. In the same spirit, we consider NNLL threshold 
resummation, which includes the singular terms at threshold at NNLO 
as a subset, a useful step towards the full NNLO result. Even when 
the latter is finally known, the resummed result contains a set 
of higher-order terms that may shed light on the fixed-order expansion 
of the total cross section.

The resummation of threshold logarithms in top-pair production 
has been conventionally
performed in Mellin-moment space~\cite{Sterman:1986aj,Catani:1989ne},
and numerical results for resummed
corrections~\cite{Bonciani:1998vc,Moch:2008qy,Cacciari:2008zb} at NLL 
accuracy, or
partial higher-order results obtained from expansions of the resummed
cross section~\cite{Kidonakis:2001nj,Kidonakis:2008mu},
have been known for some time.  In these works the Coulomb
corrections were included in fixed order, which corresponds to
setting $k=0$ or $1$ in~\eqref{eq:syst}.  An approximate NNLO correction
based on the expansion of the NNLL result up to order $\alpha_s^2$ has
been obtained in~\cite{Beneke:2009ye}, while the effect of Coulomb
corrections in the threshold region has been considered
in~~\cite{Hagiwara:2008df,Kiyo:2008bv}.  A combined
resummation of soft and Coulomb corrections has not yet been performed, 
and will be done in this paper at NNLL accuracy.  We employ the
momentum-space approach to threshold
resummation~\cite{Becher:2006nr,Becher:2006mr,Becher:2007ty}, and the
next-to-leading order Coulomb Green function obtained in the context
of top-quark production in electron-positron
collisions~\cite{Beneke:1999qg,Hoang:2000yr}. We include all
corrections at NNLL accuracy defined by~\eqref{eq:syst}, apart from a
class of higher-order non-relativistic (as opposed to soft) logarithms, 
whose summation would require an extension of results
of~\cite{Hoang:2001mm,Pineda:2006ri}. The fixed-order logarithmic
correction at $\mathcal{O}(\alpha_s^2)$ was computed
in~\cite{Beneke:2009ye} and is included in our results. 
The numerical impact of the resummation of these logarithms is 
expected to be in the per-mille range and their calculation is left for 
future work.

It should be mentioned that beyond the higher-order QCD effects
discussed in our work, electroweak corrections and finite-width
effects can become relevant at the level of precision aimed at by our
NNLL resummation. The former have been computed
in~\cite{Bernreuther:2006vg,Kuhn:2006vh} and are of the order of
$-2\%$ for the total cross section at the LHC, and smaller at Tevatron,
while the effect on distributions can be larger. Finite-width effects
on the total cross section are expected to be of the order
$\Gamma_t/m_t\lesssim 1\%$, as has been confirmed by recent
calculations of the NLO QCD corrections to the full $b\bar b
W^+W^-$ final state~\cite{Denner:2010jp,Bevilacqua:2010qb}.

Recently higher-order expansions and NNLL resummations in
different kinematical variables, such as the invariant mass
distribution and so called one-particle inclusive
kinematics~\cite{Ahrens:2009uz,Ahrens:2010zv,Kidonakis:2010dk,Ahrens:2011mw}, 
became available. In these calculations different logarithms than the 
ones in the threshold expansion~\eqref{eq:syst} are resummed. With 
respect to the total cross section, this amounts to including 
a particular set of higher-order terms in $\beta$, while excluding 
some of the more singular terms. Since the total cross 
section receives important contributions from outside the threshold 
region, the two resummation procedures are complementary. The numerical
results of these studies appear to differ beyond the quoted uncertainties, 
which calls for particular attention to a realistic estimate
of the remaining uncertainty inherent in the threshold expansion.

The paper is organized as follows. In Section~\ref{sec:basics} we
review the combined soft and Coulomb resummation developed
in~\cite{Beneke:2009rj,Beneke:2010da} and provide the ingredients
required to achieve NNLL resummation for top-quark pair production. 
We expand the NNLL resummed cross section to order $\alpha_s^3$ 
and present a new result of exact terms beyond NNLO 
in Section~\ref{sec:n3lo}. 
In Section~\ref{sec:implementation} we discuss and define our 
implementation of resummation and the method to estimate the 
uncertainties resulting from the threshold approximation. 
Section~\ref{sec:results} contains our final results for the top-quark
pair production cross section at Tevatron and LHC for a range of  
top quark masses.  Details of the 
evaluation of the NLO Coulomb correction, including the bound-state 
contribution, the complete expressions for the expansion up to 
order $\alpha_s^3$, and a discussion of our error estimate are contained 
in the appendices.

\section{Basics of resummation}
\label{sec:basics}

\subsection{Summary of the resummation formula}

Resummation of threshold logarithms is based on the factorization of
the partonic cross section in the threshold region into several 
functions~\cite{Sterman:1986aj,Kidonakis:1997gm} receiving contributions 
from different kinematical regions (soft, hard, collinear...). 
The independence of the cross section on artificial
scales introduced in this factorization implies evolution equations,
whose solutions resum threshold logarithms to all orders. For the case
of top pairs (and other heavy coloured particles) the situation is
more complicated, since in the threshold region soft gluons and the
non-relativistic top quarks have comparable energies of order
$m_t\beta^2$, so that energy is modified by a
relevant amount due to soft
radiation. This is in contrast to the case usually considered in
deriving factorization formulas at partonic thresholds, where all
particles are taken to be highly energetic compared to the scale
of the soft radiation.  As a result, the factorization formula for the
hard-scattering total cross section for the partonic
subprocesses~\eqref{eq:ttbar-proc} near the partonic
threshold~\eqref{eq:def-thresh} factorizes into three
contributions~\cite{Beneke:2010da}, with a potential function $J$ in
addition to the hard and soft functions $H$ and $W$:
\begin{equation}
\label{eq:fact}
  \hat\sigma_{pp'}(\hat s,\mu)
= \sum_{R={\bf 1},{\bf 8}}H^{R}_{pp'}(m_t,\mu)
\;\int d \omega\;
J_{R}(E-\frac{\omega}{2})\,
W^{R}(\omega,\mu)\, .
\end{equation}
Here $E=\sqrt{\hat s}-2 m_t$ is the energy relative to the production
threshold. The sum is over the colour representations of the final
state top-pair system, i.e.  the colour-singlet and octet states.  The
hard functions $H$ can be obtained from the colour-separated
hard-scattering amplitudes for the process $pp'\to t\bar t$ in the
threshold limit, as described in~\cite{Beneke:2010da}.  The soft
function $W^R$ is defined as a vacuum expectation value of soft Wilson
lines, and can be reduced to one for a fictitious $2 \to 1$ scattering
processes with a single final-state particle in the colour
representation $R$, provided an appropriate $s$-channel colour basis
is chosen~\cite{Beneke:2009rj}. Such a basis has been adopted
in~\eqref{eq:fact}. The potential function $J_R$ sums Coulomb gluon exchange
related to the attractive or repulsive Coulomb force in the colour
singlet and octet channels, respectively.

In the momentum-space 
approach~\cite{Becher:2006nr,Becher:2006mr,Becher:2007ty} 
resummation is performed by evaluating the hard function $H(\mu)$ at a
hard matching scale $\mu_h\sim 2m_t$ and the soft function
$W(\omega,\mu)$ at a soft matching scale of the order of $\mu_s\sim
m_t \beta^2$. Both are then evolved to an intermediate factorization scale
$\mu_f$ using evolution equations. For the case of heavy pair
production at threshold the relevant evolution equations and the
solutions have been given in~\cite{Beneke:2009rj,Beneke:2010da}, so we
will only summarize the solutions below.

The convolution of the soft and potential functions in~\eqref{eq:fact} 
accounts for the energy loss of the top pair into soft radiation. 
In our study of NLL resummation for squark anti-squark
production~\cite{Beneke:2010da} we observed a non-negligible
numerical effect from the combined soft and Coulomb resummation, and 
also found a reduced scale dependence. For top anti-top production  
the Coulomb corrections are smaller, so we expect less sizeable
effects for the case of top quarks.

The formula~\eqref{eq:fact} has been derived in an effective field
theory for the production of heavy particles in an $S$-wave
state\footnote{Therefore the formula applies to the $q\bar q$ and
  $gg$ initiated partonic channels but not to contributions from
  $qg,\bar qg$ initial states, that are suppressed by a factor
  $\beta^2$ and therefore formally of N$^3$LL order. 
  Results for the leading threshold-enhanced NNLO
  contributions for those channels have been given
  in~\cite{Langenfeld:2009wd}.}  and is valid for the leading
effective Lagrangian.  Corrections from subleading terms in the
effective Lagrangian, or higher-dimensional production operators,
potentially lead to the corrections ${\cal O}(\beta)$ (NNLL) and
${\cal O}(\alpha_s\beta,\beta^2)$ (N$^3$LL) indicated in~\eqref{eq:syst}, 
and would have to be accounted for by including additional hard, 
potential and soft functions in~\eqref{eq:fact}.  As shown
in~\cite{Beneke:2009ye,Beneke:2010da}, these subleading effects vanish 
at NNLL accuracy due to rotational invariance, so that the
${\cal O}(\beta)$ corrections are absent and factorization holds in the
simpler form~\eqref{eq:fact}. Starting from N$^3$LL order, such
power-suppressed corrections in $\beta$ should be expected, requiring 
the introduction of new soft and potential functions, 
as is also
indicated by the massive two-loop soft anomalous dimension for 
non-threshold kinematics~\cite{Ferroglia:2009ep,Ferroglia:2009ii}, 
which exhibits a more complicated colour structure.

\subsection{Inputs to the resummation formula}
\label{sec:input}
From~\eqref{eq:syst} and the factorization formula~\eqref{eq:fact}, it
can be seen that the ingredients required for NNLL resummation are
given by the resummation functions in the
exponent~\cite{Becher:2007ty,Beneke:2010da}, the fixed-order one-loop
soft~\cite{Beneke:2009rj} and hard~\cite{Czakon:2008cx}
functions\footnote{The ${\cal O}(\alpha_s)$ constant contribution
appearing at NNLL in~\eqref{eq:syst} has been included in some
earlier NLL predictions \cite{Bonciani:1998vc} in an approximate way.} 
and the NLO-resummed
potential functions~\cite{Beneke:1999qg}.  Once the only non-trivial
process-dependent quantities, the one-loop hard functions, are known,
the formalism can be immediately applied to perform an NNLL
resummation for other processes, such as the production of coloured
supersymmetric particles. In this subsection we collect the required
ingredients.

\subsubsection{Hard functions}
\label{sec:hard}

The hard functions are related to the amplitudes of 
partonic top production processes $p p'\to t\bar t$ 
directly at threshold. Their perturbative expansion in 
the QCD coupling in the $\overline{\rm MS}$ scheme can be written as
\begin{equation}
\label{eq:hard-def}
   H^{R}_{pp'}(\mu_h)= H^{R(0)}_{pp'}(\mu_h)
   \left[1+\frac{\alpha_s(\mu_h)}{4\pi}h^{R(1)}_{pp'}(\mu_h)
     +\mathcal{O}(\alpha_s^2)\right] \, .
\end{equation}
The leading-order hard functions are related to the partonic
Born cross section at threshold in the
colour channel $R$, $\hat\sigma^{(0)}_{pp',R}$, according to
\begin{equation}
\label{eq:sigma-hard}
  \hat\sigma_{pp',R}^{(0)}(\hat s)
\underset{\hat s\to 4m_t^2}{\approx}
\frac{\beta m_t^2}{2\pi} H^{R(0)}_{pp'} \, . 
\end{equation}
The NLO coefficients $h_{pp'}^R$ at the hard matching scale $\mu_h$ can be
obtained by comparing the analytical result for the threshold 
expansion of the NLO top-pair cross section~\cite{Czakon:2008cx} to the NLO
expansion of the resummed cross section~\cite{Beneke:2010da}. Subtracting 
the NLO contribution from the Coulomb and soft function, we obtain 
\begin{equation}
\label{eq:hard-nlo}
\begin{aligned}
h^{{\bf 8}(1)}_{q\bar q}(\mu_h)=& -2 C_F\left[ 
\ln^2\left(\frac{\mu_h^2}{4m_t^2}\right)
+3\ln\left(\frac{\mu_h^2}{4m_t^2}\right)- \frac{7 \pi^2}{6}+16\right]\\
 &+\frac{2C_A}{3}\left[8\ln\left(\frac{\mu_h^2}{4m_t^2}\right)+\frac{100}{3}
-\frac{3\pi^2}{2}+4\ln 2 \right]-
\frac{20}{3}\ln\left(\frac{\mu_h^2}{4m_t^2}\right)-\frac{44}{3}\,,\\[0.2cm]
 h^{{\bf 1}(1)}_{gg}(\mu_h)=& -
 2C_A\left[\ln^2\left(\frac{\mu_h^2}{4m_t^2}
 \right)
  -\frac{2\pi^2}{3}-2\right]
+ 2C_F \left(\frac{\pi^2}{2}-10\right)\,,\\[0.2cm]
h^{{\bf 8}(1)}_{gg}(\mu_h)=& -2C_A\left[
  \ln^2\left(\frac{\mu_h^2}{4m_t^2}
  \right)
+ \ln\left(\frac{\mu_h^2}{4m_t^2}\right)
  +\frac{5\pi^2}{12}+4\right]
+ 2C_F \left(\frac{\pi^2}{2}-10\right)\,,
\end{aligned}
\end{equation}
in the $\overline{\rm MS}$ scheme with five active quark flavours. 
Here $R={\bf 1}$ stands for the colour-singlet and 
$R={\bf 8}$ for the colour-octet channel, respectively.
The colour factors are $C_A=N_C=3$, $C_F=(N_C^2-1)/(2 N_C)=4/3$. 
While strictly speaking the leading term in 
the threshold expansion of the Born cross section appears in the 
hard function~\eqref{eq:sigma-hard}, in our numerical results below 
we use the exact expression for the Born cross section, that can be 
found e.g. in~\cite{Moch:2008qy}. 

The choice $\mu_h\approx 2m_t$ eliminates the logarithmic terms in 
the hard function at the matching scale, but the soft and Coulomb 
functions naturally live at parametrically smaller scales, of the order of
$m_t\beta^2$ and $m_t\beta$, respectively. The solution to the renormalization 
group equation that sums logarithms of $\ln(\mu_h/\mu)$ is given by
\begin{equation}\label{eq:coeff_resummed}
H^R_{pp'}(\mu)=\exp[4 S(\mu_h,\mu)-2a_i^{V}(\mu_h,\mu)]
\left(\frac{4 m_t^2}{\mu_h^2}\right)^{-2a_\Gamma(\mu_h,\mu)} 
H^R_{pp'}(\mu_h)\, ,
\end{equation}
with the functions $S$, $a_i^{V}$ and $a_\Gamma$ defined
as \cite{Becher:2007ty}
\begin{eqnarray} 
\label{eq:res_funct_def}
S(\mu_h,\mu) &=& -\int_{\alpha_s(\mu_h)}^{\alpha_s(\mu)}d \alpha_s 
\frac{\Gamma^r_{\text{\text{cusp}}}(\alpha_s)
  +\Gamma^{r'}_{\text{\text{cusp}}}(\alpha_s)}{2\beta(\alpha_s)}
\int_{\alpha_s(\mu_h)}^{\alpha_s}\frac{d \alpha_s^{\prime}}
{\beta(\alpha_s^{\prime})} 
\, ,\nonumber\\
a_{\Gamma}(\mu_h,\mu) &=& -\int_{\alpha_s(\mu_h)}^{\alpha_s(\mu)} d \alpha_s 
\frac{\Gamma^{r}_{\text{\text{cusp}}}(\alpha_s)
+\Gamma^{r'}_{\text{\text{cusp}}}(\alpha_s)}{2\beta(\alpha_s)}
\, , \nonumber\\
a^{V}_i (\mu_h,\mu) &=& -\int_{\alpha_s(\mu_h)}^{\alpha_s(\mu)} d \alpha_s 
\frac{\gamma_i^{V}(\alpha_s)}{\beta(\alpha_s)}.
\end{eqnarray}
The labels $r, r'$ denote the colour representation of the initial-state 
partons $p,p'$, while the label $i$ refers to the colour of the initial-state 
partons and the representation $R$ of the $t\bar t$ pair.   
Explicit results for the $\overline{\rm MS}$ $\beta$-function, 
$a_\Gamma$ and the Sudakov
exponent $S$ up to N$^3$LL order
can be found in~\cite{Becher:2007ty}. The relevant anomalous dimension 
coefficients as needed for NNLL accuracy are collected 
in~\cite{Beneke:2009rj}. Starting from the two-loop order the evolution 
equation of the hard function is modified because of additional IR divergences, 
that are related to UV divergences of the
potential function due to insertions of non-Coulomb potentials.
In this work we will not consider these contributions (see end of 
section~\ref{sec:coulomb} below).

\subsubsection{Soft functions}
The resummed soft function in the momentum-space 
formalism~\cite{Becher:2006nr,Becher:2006mr,Becher:2007ty} can be written as
\begin{equation}
\label{eq:w-resummed}
W^{R,\text{res}}_{i}(\omega,\mu)=
\exp[-4 S(\mu_s,\mu)+2 a^{R}_{W,i}(\mu_s,\mu)]\,
\tilde{s}_{i}^{R}(\partial_\eta,\mu_s) 
\frac{1}{\omega} \left(\frac{\omega}{\mu_s}\right)^{2 \eta} \theta(\omega)
\frac{e^{-2 \gamma_E \eta}}{\Gamma(2 \eta)}\, .
\end{equation}
The auxiliary variable $\eta$  is set
to $\eta = 2 a_{\Gamma}(\mu_s,\mu)$ after performing the 
derivatives. The function $a^{R}_{W,i}$ is defined, analogously to 
$a^{V}_i$ in~\eqref{eq:res_funct_def}, in terms of  the anomalous 
dimension $\gamma^{R}_{W,i}$, first obtained up to the two-loop level 
in~\cite{Beneke:2009rj}. The function $\tilde s_i^{R}(\rho)$
is the Laplace transform of the soft function. For resummation at 
NNLL accuracy we require the NLO soft function, which 
reads~\cite{Beneke:2009rj,Beneke:2010da}
\begin{equation}
 \tilde  s^{R}_{i}(\rho,\mu)=1 +\frac{\alpha_s}{4\pi} 
\left[\left(C_r+C_{r'}\right)\left(   \rho^2+\frac{\pi^2}{6}\right)
- 2C_{R}\left( \rho-2\right)  \right]+\mathcal{O}(\alpha_s^2)\, ,
\label{eq:soft-laplace}
\end{equation}
where $C_r$ is the quadratic Casimir operator in the colour representation 
$r$ of the initial state parton $p$. As mentioned above, the soft 
matching scale $\mu_s$ should be chosen of the order $m_t\beta^2$ in 
order to resum large logarithms in the soft function. The precise 
choice we adopt in our numerical analysis  will be discussed in 
Section~\ref{sec:implementation}.

\subsubsection{The NLO Coulomb function}
\label{sec:coulomb}
The potential function is related to the imaginary part of
the zero-distance Coulomb Green function of the Schr{\"o}dinger
operator $-\vec{\nabla}^{\,2}/m_t-(-D_{R})\,\alpha_s/r$.
The coefficients of the Coulomb potential in the colour-singlet and 
colour-octet configuration are given by
\begin{equation}
D_1=-C_F=-\frac{N_C^2-1}{2N_C}\;,\quad
D_8=-\left[C_F-\frac{C_A}{2}\right]=  \frac{1}{2N_C} \, .
\end{equation}
For NNLL resummation we require the Coulomb Green function up to 
next-to-leading order, which includes one insertion of the 
NLO Coulomb potential
\begin{equation}
\delta\tilde{V}(\bff{p},\bff{q}) = 
\frac{4\pi D_{R}\alpha_s(\mu)}{\bff{q}^2}  \frac{\alpha_s(\mu)}{4\pi} 
\left(a_1-\beta_0\ln\frac{\bff{q}^2}{\mu^2}\right)\,,
\label{delV}
\end{equation}
where $\beta_0 = \frac{11}{3} C_A-\frac{4}{3} n_l T_f$ is
the one-loop beta-function coefficient, and $a_1 =\frac{31}{9}
C_A-\frac{20}{9} n_l T_f$. The insertion of 
$\delta\tilde{V}(\bff{p},\bff{q})$ yields a factor $\alpha_s \times 
\alpha_s/\beta$, 
which according to (\ref{eq:syst}) produces a NNLL correction.
The potential function up to next-to-leading order can be written as
\begin{equation}
J_{R}(E)=2 \,\mbox{Im} \left[\,
G^{(0)}_{C,R}(0,0;E) \,\Delta_{\rm nC}(E) + G^{(1)}_{C,R}(0,0;E) + 
\ldots\right] \, , 
\label{JRal}
\end{equation}
where $G^{(0)}_{C,R}$ is the solution to the Schr\"odinger equation with the
leading Coulomb potential, resumming all $(\alpha_s/\beta)^n$
corrections, while $G^{(1)}_{C,R}$ sums $\alpha_s \times (\alpha_s/\beta)^n$
corrections by solving perturbatively the Schr\"odinger equation with one  
insertion of $\delta\tilde{V}(\bff{p},\bff{q})$~\cite{Beneke:1999qg}. 
The explicit expressions are given in Appendix~\ref{app:coulomb}.

The NLO potential function 
is independent of the scale $\mu$, though a residual 
scale dependence, formally of higher order, remains. We use a 
$\beta$-dependent scale~\cite{Beneke:2010da},
\begin{equation}
\label{eq:mu-c} 
\tilde \mu_C= \text{max}\{2 m_{t} \beta, C_F m_{t}\alpha_s(\tilde \mu_C)\} \, ,
\end{equation}
in the numerical evaluation. With this choice logarithms of $\beta$ 
in the potential function arise only from the scale of the strong 
coupling. The lower limit in (\ref{eq:mu-c}) is chosen to coincide with 
the Bohr scale of the lowest-lying bound state in the colour singlet 
channel, which is indeed the relevant scale for $\beta \to 0$ in this case, 
as can be seen from the argument of the logarithm in~\eqref{eq:j0} below.
There are no bound states in the repulsive colour-octet channel, but 
since the Coulomb effects are small in this case, the precise choice 
of scale is not important.

Further NNLL effects from the non-relativistic dynamics arise from 
NNLO terms in the heavy-quark potential not related to the Coulomb potential,
\begin{equation}
\delta\tilde{V}_{\rm NNLO}(\bff{p},\bff{q}) =
\frac{4\pi D_{R}\alpha_s(\mu^2)}{\bff{q}^2} 
\left[\frac{\pi\alpha_s(\mu^2)|\bff{q}|}{4 m}
\left(\frac{D_{R}}{2}+C_A \right)
+ \frac{\bff{p}^2}{m^2}+\frac{\bff{q}^2}{m^2} \,v_{\rm spin}\right]\,,
\label{delV2}
\end{equation}
where $v_{\rm spin} = 0$ and $-2/3$ for a $t\bar t$ pair in a 
spin-singlet and spin-triplet state, respectively. The evaluation of 
an insertion of the non-Coulomb potential results in UV divergences 
that cancel against IR divergences in the two-loop hard function, as 
mentioned in Section~\ref{sec:hard}. After this cancellation a 
logarithm of $\beta$ remains at NNLO, which counts as NNLL. The 
divergence also produces an additional contribution to the anomalous 
dimension $\gamma_i^V$ of the hard function starting from the 
two-loop order. Since as mentioned above, we do not consider this 
additional contribution in the NNLL resummed hard function, the 
non-Coulomb corrections will be included only at fixed order in the 
number of non-relativistic logarithms.  
The corresponding NNLO correction for a final state in an arbitrary colour 
representation is given  by~\cite{Beneke:2009ye}
\begin{equation}
\sigma_{X|\rm nC} = \sigma_X^{(0)} 
\,\alpha_s^2(\mu)\,\ln\beta \,\left[-2 D_{R}^2 \,(1+v_{\rm spin}) 
+ D_{R} C_A\right]\,.
\label{eq:non-Coulomb}
\end{equation}
Here $ \sigma_X^{(0)}$ is the Born cross section in the spin and 
colour channel $X$. For top quarks the Born cross section in the 
$q\bar q$ initiated channel 
is a pure colour-octet spin-triplet, whereas in gluon-gluon fusion 
the $t\bar t$ state is spin-singlet but colour-octet or singlet.
In our numerical results, the corrections~\eqref{eq:non-Coulomb} are
added to the potential function $J_R(E)$ in (\ref{JRal}) through the 
factor\footnote{For simplicity we implement the non-Coulomb 
correction factor only in the continuum, and not 
for the bound states, since the uncorrected bound-state effect 
is already very small; hence the theta-function in (\ref{eq:JRalnC}).}
\begin{equation}
\Delta_{\rm nC}(E) = 1+ 
\alpha_s^2(\mu_C)\,\ln\beta \,\left[-2 D_{R}^2 \,(1+v_{\rm spin}) 
+ D_{R} C_A\right] \theta(E)
\label{eq:JRalnC}
\end{equation}
multiplying the leading Coulomb function $G^{(0)}_{C,R}(0,0;E)$, 
with $\beta = (E/m_t)^{1/2}$. After convolution with the soft function 
this procedure yields the correct  $\alpha_s^2\ln\beta$ and 
$\alpha_s^3\ln^3\beta$ terms in the threshold expansion. 
The factorized form of 
the logarithmic non-Coulomb correction can be deduced from the 
results of \cite{Beneke:1999qg} (given explicitly in the appendix of 
\cite{Pineda:2006ri}) and~\cite{Beneke:2009ye}, and produces 
correctly the series $\alpha_s^2\log\beta\times (\alpha_s/\beta)^k$ 
of terms associated with the first non-Coulomb logarithm and any 
number $k$ of Coulomb exchanges.

The non-relativistic anomalous dimension produces a series 
$\alpha_s\times (\alpha_s\log\beta)^n\times (\alpha_s/\beta)^k$, 
of which (\ref{eq:JRalnC}) is only the first term ($n=1$, any $k$), 
that should also be summed for complete NNLL 
accuracy. This requires the generalization of results for top-quark pair 
production in $e^+ e^-$ collisions 
(see, e.g., \cite{Hoang:2001mm,Pineda:2006ri}) to the production of a
colour-octet final state. The numerical effect of the leading 
non-Coulomb terms is an enhancement of the top production cross section 
of about $0.5\%$ at Tevatron and LHC. The higher-order terms should 
therefore be very small, and we leave this resummation of 
non-relativistic logarithms to future work.

\subsection{Calculation of the convolutions}
\label{sec:conv}

The total hadronic cross section for the production of a $t\bar t+X$ 
final state in collisions of hadrons $N_{1,2}$ with centre-of-mass energy 
$s$ is obtained from
\begin{equation}
\sigma_{N_1 N_2\to t\bar t X}(s)=
\sum_{p,p'=q,\bar q,g}\,\int_{4 m_t^2/s}^1 \!d\tau\,L_{pp'}(\tau,\mu_f)
\,\hat\sigma_{pp'} (s \tau,\mu_f)\,,
\label{eq:sig-had}
\end{equation}
where the parton luminosity is defined in terms of the parton distributions 
functions~(PDFs)
\begin{equation}
\label{eq:lumi}
L_{p p^\prime}(\tau,\mu) = \int_0^1 dx_1
dx_2\,\delta(x_1 x_2 - \tau) \,f_{p/N_1}(x_1,\mu)f_{p^\prime/N_2}(x_2,\mu)\,.
\end{equation}
We briefly discuss some technical issues encountered when performing 
the convolution of the resummed soft and potential functions in the 
partonic cross section~\eqref{eq:fact}, and the subsequent convolution 
with the parton luminosity.

We convolute the NLO Coulomb Green function~\eqref{JRal} with the 
resummed soft function, including the (pseudo) bound-state contributions from 
energies below the nominal top anti-top threshold. We note that for integrated 
quantities, such as the total cross section, the top width can be set 
to zero in the Coulomb function. Thus, for the continuum
contributions, the integral in~\eqref{eq:fact} has to be evaluated for
$0<\omega<2E$. The bound-state contributions are located at values 
$\omega_n > 2 E$, and we discuss them in Appendix~\ref{app:bound}.  
Here we focus on the continuum. For the relevant hierarchy of 
scales, $\mu_s<\mu_f$, the anomalous dimension $\eta$ is 
negative and the expression~\eqref{eq:w-resummed} should be understood in the
distributional sense~\cite{Becher:2006mr}. 

Let us first discuss the simpler 
case where the non-Coulomb corrections to the potential function are 
neglected, i.e.~$\Delta_{\rm nC}=1$ in~\eqref{JRal}. The convolution 
of the soft and potential functions in the
factorization formula~\eqref{eq:fact} is analytically continued to
negative values of $\eta$ according to
\begin{eqnarray}
\int_0^{2E}\!\!\!\!\!d\omega \,J_R(E-\frac{\omega}{2})
 \left[\frac{1}{\omega}\left(\frac{\omega}{\mu_s}\right)^{2\eta}\right]_*
&=&\int_0^{2E}\frac{d\omega}{\omega} 
\left[J_R(E-\frac{\omega}{2})-J_R(E)+\frac{\omega}{2} J_R'(E) \right]
\left(\frac{\omega}{\mu_s}\right)^{2\eta}\nonumber\\
&& 
+\,\left[\frac{J_R(E)}{2\eta}
-\frac{J_R'(E)E}{2\eta+1}\right]\left(\frac{2E}{\mu_s}\right)^{2\eta}\, .
\label{eq:star-dist}
\end{eqnarray}
The derivatives of the potential function appearing in~\eqref{eq:star-dist}  
have been evaluated by expressing the hypergeometric function in terms of 
harmonic sums using methods from~\cite{Bierenbaum:2008yu}, as described in 
Appendix~\ref{app:coulomb}. The double subtraction in~\eqref{eq:star-dist} 
is sufficient to render the $\omega$ integral finite for $\eta >-1$.
However, for $\eta <-1/2$ the convolution of the first term in 
the second-line square brackets with the parton luminosity 
in~\eqref{eq:sig-had} requires regularization at the partonic 
threshold $\tau\to 4m_t^2/s\equiv \tau_0$, if the potential function is 
non-vanishing at $E=0$, as is the case for the attractive colour-singlet 
Coulomb potential. To see this, note that in the region 
$\tau \sim \tau_0$ we can approximate $E=\sqrt{s\tau}-2m_t 
\approx \frac{1}{2} \frac{s}{2m_t}(\tau-\tau_0)$. With $L_{pp'}(\tau,\mu_f)$ 
and $J(E)$ approaching constants as $\tau \to \tau_0$, the 
$\tau$-integrand behaves as $(\tau-\tau_0)^{2\eta}$. Thus the $\tau$-integral 
for the colour-singlet cross section is analytically continued to
 $\eta<-1/2$ according to
\begin{multline}
\label{eq:subt-lumi}
\int_{\tau_0}^1d\tau L_{p p^\prime}(\tau) J_{\bf 1}(E)
\left(\frac{2E}{\mu_s}\right)^{2\eta}
= \frac{1}{2\eta+1}
L_{p p^\prime}(\tau_0)J_{\bf 1}(0)(1-\tau_0)
\left(\frac{s-4m_t^2}{2m_t\mu_s}\right)^{2\eta}\\
+\int_{\tau_0}^1d\tau 
\left[L_{p p^\prime}(\tau)   J_{\bf 1}(E)\left(\frac{2E}{\mu_s}\right)^{2\eta}-
  L_{p p^\prime}(\tau_0)J_{\bf 1}(0)
\left(\frac{s\tau-4m_t^2}{2m_t\mu_s}\right)^{2\eta}
\right ] \, .
\end{multline}
The pole at $\eta=-\frac{1}{2}$ in the 
first line cancels with the overall gamma function in the
resummed soft function~\eqref{eq:w-resummed}, so that the cross 
section~\eqref{eq:sig-had} is now well-defined for $\eta>-1$. In order to 
avoid large numerical cancellations, it is practical to split 
the integration interval into two regions and perform the analytical 
continuation~\eqref{eq:subt-lumi} only in a small region 
$\tau\in [\tau_0,\tau_0+\Delta\tau]$. The potential function in the
singlet channel at $E=0$, $J_{\bf 1}(0)$, is given in~\eqref{eq:j0}.
The derivative of the singlet potential function at $E=0$ vanishes only at 
leading order so that further subtractions would be 
required when $\mu_s$ and $\mu_f$ are such that $\eta<-1$.

Alternatively, the left-hand side of~(\ref{eq:star-dist}) can be 
analytically continued to negative $\eta$ 
by performing a partial integration with respect to $\omega$.
After one partial integration we obtain 
\begin{eqnarray}
\int_0^{2E} d\omega \,J_R(E-\frac{\omega}{2})
\left[\frac{1}{\omega}\left(\frac{\omega}{\mu_s}\right)^{\!2\eta}\right]_*
&=&
\frac{1}{2\eta}\left(\frac{2 E}{\mu_s}\right)^{\!2\eta}
J_R(0)\nonumber \\
&&
-\frac{1}{2\eta}\int_0^{2E} d\omega \,
\frac{d J_R(E-\frac{\omega}{2})}{d \omega}
\left(\frac{\omega}{\mu_s}\right)^{\!2\eta} \, .
\label{eq:part-int}
\end{eqnarray}
The integral in the second line is well defined for $\eta> -1/2$.  
By applying a second partial integration, it can be continued 
to values $\eta>-1$, as in (\ref{eq:star-dist}), generating 
boundary terms with derivatives of the Coulomb function. The boundary terms 
at $\omega=0$ are zero in the analytic continuation from positive $\eta$. 
For those at 
$\omega=2 E$ only the singlet channel contributes, since the octet Coulomb 
function $J_{\bf 8}(0)$ and its derivatives at $E=0$ vanish. The 
convolution of the boundary terms with the parton luminosity can be 
treated in the same way as the terms in the second line of 
(\ref{eq:star-dist}), leading to expressions similar to~(\ref{eq:subt-lumi}). 
Note that for the LO Coulomb function only $J_{\bf 1}(0)$ is non-zero, 
whereas all higher derivatives vanish.

Including the non-Coulomb correction~(\ref{eq:JRalnC}) in the potential
function~\eqref{JRal} makes the analytic continuation more complicated
due to the additional term $J_R^{(0)}(E)\log(E/m_t)$, where
$J^{(0)}_R(E)=\mbox{Im} \,\big[\, G^{(0)}_{C,R}(0,0;E)\big]$.  The
octet case does not pose a problem, since the octet Coulomb function 
and its derivatives vanish faster than any power as $E\rightarrow 0$. 
But for the singlet $J_{\bf 1}^{(0)}(E)\ln(E/m_t)$ diverges in this limit, 
and we have to introduce an additional subtraction, 
\begin{eqnarray}
\int_0^{2E} d\omega \,J_{\bf 1}^{(0)}(E-\frac{\omega}{2})
 \ln\Bigl(\frac{E-\frac{\omega}{2}}{m_t}\Bigr)
 \left[\frac{1}{\omega}\left(\frac{\omega}{\mu_s}\right)^{2\eta}\right]_*
&=&
 J_{\bf 1}^{(0)}(0)\int_0^{2E} d\omega \,
 \ln\Bigl(\frac{E-\frac{\omega}{2}}{m_t}\Bigr)
 \frac{1}{\omega}\left(\frac{\omega}{\mu_s}\right)^{\!2\eta}
 \nonumber \\[0.2cm]
&&\hspace{-8cm}+\int_0^{2E} d\omega \,
 \Bigl(J_{\bf 1}^{(0)}(E-\frac{\omega}{2})-J_{\bf 1}^{(0)}(0)\Bigr)
 \ln\Bigl(\frac{E-\frac{\omega}{2}}{m_t}\Bigr) 
 \frac{1}{\omega}\left(\frac{\omega}{\mu_s}\right)^{\!2\eta} \, .
\label{eq:part-int2}
\end{eqnarray}
The first integral on the right-hand side of (\ref{eq:part-int2})
can be calculated analytically, whereas the second integral 
can be continued to lower values of $\eta$ either using  partial 
integration, as in (\ref{eq:part-int}), or as in~(\ref{eq:star-dist}), 
with the replacement 
$J_{\bf 1}(E)\to (J_{\bf 1}^{(0)}(E)-J_{\bf 1}^{(0)}(0))\ln(E/m)$.

The convolution of the luminosity function $L_{p p^\prime}(\tau)$ with 
the second integral on the right-hand side of (\ref{eq:part-int2})
is finite, once the $\omega$ integration has been continued to 
negative values of $\eta$. On the contrary, the first integral, 
\begin{equation}
 J_{\bf 1}^{(0)}(0)\int_0^{2E} d\omega \,
 \ln\Bigl(\frac{E-\frac{\omega}{2}}{m_t}\Bigr)
 \frac{1}{\omega}\left(\frac{\omega}{\mu_s}\right)^{\!2\eta}
 =  \frac{J_{\bf 1}^{(0)}(0)}{2 \eta}  
\left(\frac{2 E}{\mu_s}\right)^{\!2 \eta} 
\left\{ \ln \left(\frac{E}{m_t} \right)-\gamma_E-\psi(1+2 \eta) \right\} \, , 
\end{equation}
contains 
a non-integrable singularity at $E=0$ for $\eta<-1/2$. This can be 
subtracted and analytically continued to $\eta>-1$, similarly 
to (\ref{eq:subt-lumi}).

\section{Expansion to $\alpha_s^3$}
\label{sec:n3lo}

The expression for the resummed cross section~\eqref{eq:fact}
can be expanded to order $\alpha_s^n$ in the strong coupling,
providing an approximation to the full $\mathcal{O}(\alpha_s^n)$ QCD
calculation. The expansion of the NLL and NNLL resummed corrections up
to order $\alpha_s^1$ and $\alpha_s^2$, respectively, has been given
in~\cite{Beneke:2009ye,Beneke:2010da}. Here we provide the
corresponding expansion of the NNLL result up to order
$\alpha_s^3$. In this way, we generate the first terms in the threshold 
expansion of the third-order fixed-order cross section, whose size 
should be indicative of the quality of the fixed-order expansion. 
If there are no cancellations, integrating the threshold 
expansion gives a rough estimate of the correction that may be 
expected on top of the exact NNLO result, once it is known.

To generate the threshold expansion from the resummation formula, 
the scales have to be chosen as $\mu_s=k_s
m_t\beta^2$, $\mu_C=k_Cm_t\beta$, $\mu_h=k_h m_t$, where the 
$k_X$ are numbers of order one. Those terms in the threshold expansion 
that are completely included at NNLL are independent of the 
$k_X$. The expansion of the resummation formula also 
includes terms that are beyond the NNLL accuracy, which may 
depend on the arbitrary constants $k_X$, and are presently not 
completely known. Adopting the notation 
\begin{equation}
\label{eq:sigma-series}
\hat \sigma_{pp',R}(\beta,\mu_f) = \hat \sigma^{(0)}_{pp',R} \Bigg\{ 1
+ \sum_n\sum_{m=0}^n\left(\frac{\alpha_s(\mu_f)}{4\pi}\right)^n
f^{(n,m)}_{pp'(R)} \ln^m\left(\frac{\mu_f}{m_t}\right)
\Bigg\}
\end{equation}
for the fixed-order expansion in the strong coupling of the partonic cross section in  
colour channel $R$,  
the result for the different production channels can be written as
\begin{subequations}
\label{eq:N3LO}
\begin{align}
f^{(3,0)}_{q\bar q(8)}=&
12945.4 \ln^6\beta-37369.1
   \ln^5\beta+27721.4 \ln^4\beta+41839.4 \ln
   ^3\beta\nonumber\\
   &+\frac{1}{\beta}\left[-2994.51 \ln^4\beta+2804.73 \ln^3\beta+
     3862.46 \ln^2\beta- 6528.61 \ln\beta\right]\nonumber\\
&+\frac{1}{\beta^2} \left[153.93 \ln^2\beta+122.866 \ln \beta-144.996\right]
+\tilde f^{(3,0)}_{q\bar q(8)} \, ,\\
f^{(3,0)}_{gg(1)}=&147456. \ln^6\beta-59065.6
   \ln^5\beta-286099. \ln^4\beta+349463. \ln^3\beta\nonumber\\
&+\frac{1}{\beta}\left[121278. \ln^4\beta+103557. \ln^3\beta-164944. \ln^2\beta
  + 56418.5\ln \beta\right]\nonumber\\
&+\frac{1}{\beta^2}\left[  22166. \ln^2\beta+39012.1 \ln\beta-2876.61\right]
+\tilde f^{(3,0)}_{gg(1)}\, ,\\
f^{(3,0)}_{gg(8)}=&147456. \ln^6\beta-169658.
   \ln^5\beta-140834. \ln^4\beta+524210. \ln^3\beta
   \nonumber\\
& +\frac{1}{\beta}  \left[-15159.7 \ln^4\beta-5364.82 \ln^3\beta
     +19598.9 \ln^2\beta-17054.7 \ln\beta\right]   \nonumber\\
&+\frac{1}{\beta^2}\left[346.343 \ln^2\beta+522.978 \ln\beta-71.7884\right]
+\tilde f^{(3,0)}_{gg(8)} \, .
\end{align}
\end{subequations}
The terms given explicitly in~\eqref{eq:N3LO} are those N$^3$LO terms,
which are predicted correctly by the NNLL approximation,  
and which belong to the NNLL terms according to~\eqref{eq:syst}. 
The remainder functions $\tilde f^{(3,0)}_i$, containing N$^3$LL and 
higher-order terms, are collected in Appendix~\ref{app:expansions}, 
together with the coefficients of the $\ln(\mu_f/m_t)$ terms. 
There we also summarize the  ${\cal O}(\alpha_s^2)$ terms 
from~\cite{Beneke:2009ye} and the $k_X$-dependent constants not 
given in~\cite{Beneke:2009ye}, which are generated from 
the expansion of the NNLL resummation formula, but are formally beyond 
NNLL.

We note that while the Coulomb terms are generically 
${\cal O}((\alpha_s/\beta)^n)$, there is no $1/\beta^{\,3}$ term 
at third order. Instead a bound-state contribution is present, 
that in the fixed-order expansion would appear as a 
$\delta(\beta)$ term, which is not included in~\eqref{eq:N3LO} but 
can be deduced from the expressions given in Appendix~\ref{app:bound}. 
The absence of a $1/\beta^{\,3}$ term implies that the convolution 
of the partonic cross section with the parton densities is still 
well-defined, since $d\hat \sigma_{p p'}/d\beta \propto \ln^2\beta$ for 
small $\beta$.  This ceases to be true at N$^4$LO, where a 
$1/\beta^{\,4}$ term is present in $f^{(4,0)}_i$. 
At this order the fixed-order 
expansion of heavy particle pair production breaks down, and the 
summation of Coulomb terms is mandatory even for the total cross 
section, which is not necessarily threshold-dominated after the 
resummation has been implemented.

\begin{figure}[t]
\begin{center}
\includegraphics[width=.55\textwidth]{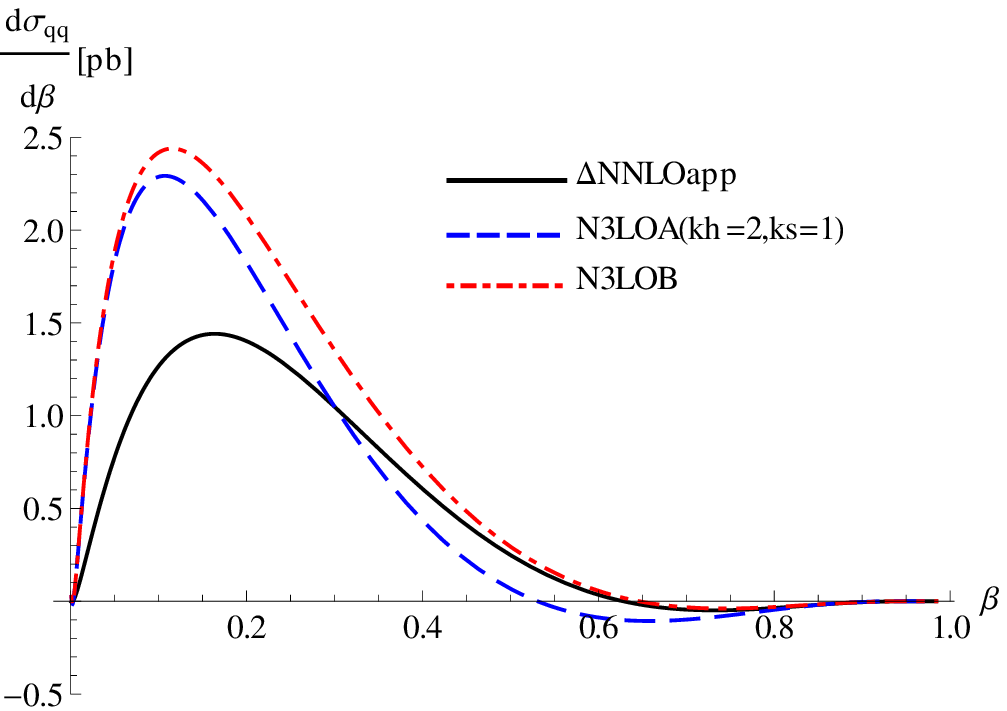}  
\includegraphics[width=.55\textwidth]{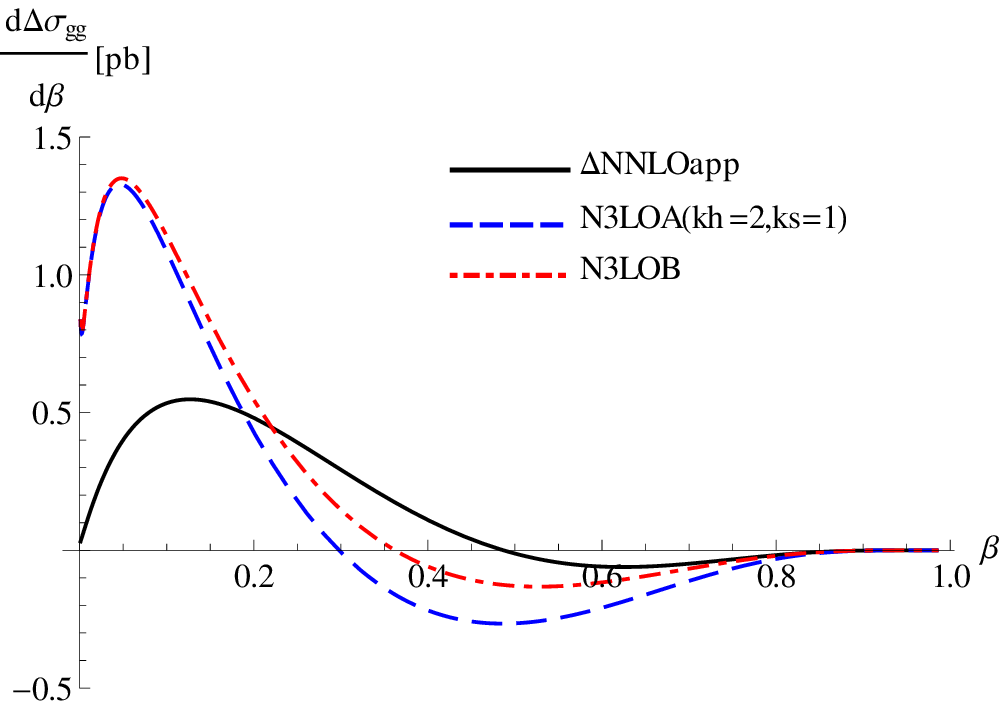}  
\end{center}
        \caption{\sf Partonic higher-order cross sections for $q\bar q\to
          t\bar t$ (top) and $gg\to t\bar t$ (below) at the Tevatron,
          multiplied with the parton luminosities. Black (solid):
          $\Delta$NNLO$_{\text{app}}$,
          blue (dashed): N$^3$LO$_{\text A}$ (all terms in~\eqref{eq:N3LO} 
          for $k_s=k_C=1$, $k_h=2$), red (dot-dashed): N$^3$LO$_{\text B}$.}
\label{fig:N3LO}
\end{figure}

\begin{figure}[t]
  \begin{center}
      \includegraphics[width=.55\textwidth]{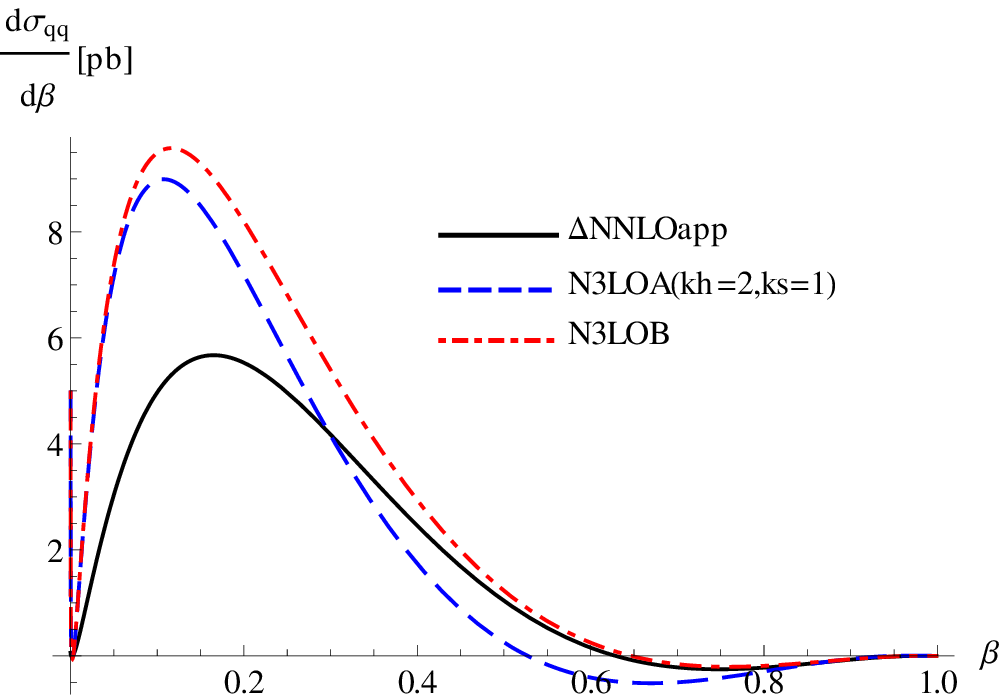}  
        \includegraphics[width=.55\textwidth]{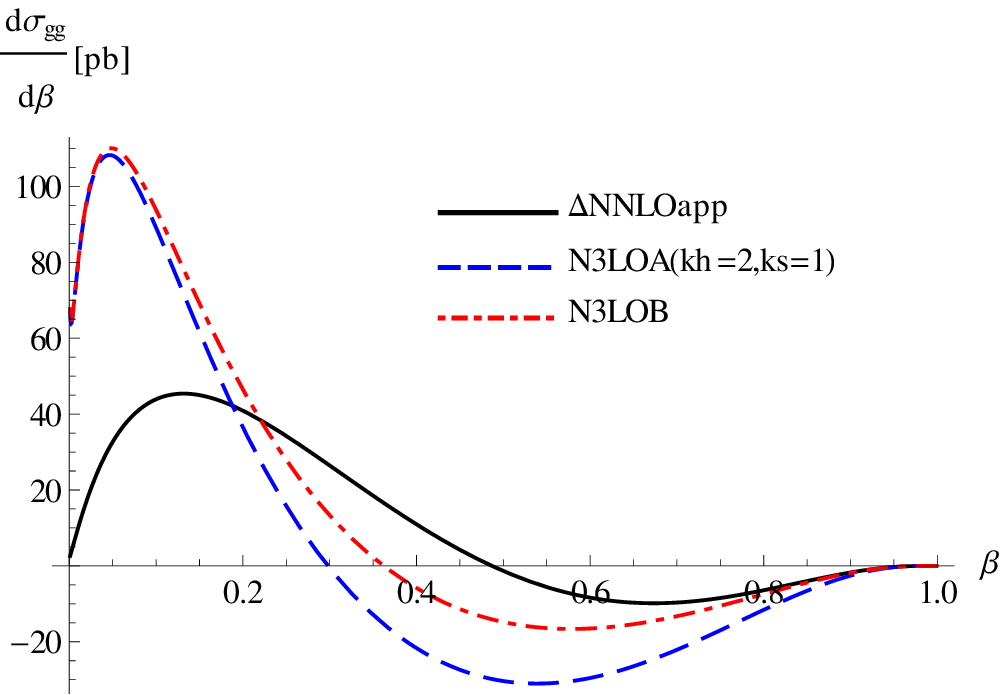}  
        \end{center}
        \caption{\sf Partonic higher-order cross sections for $q\bar q\to
          t\bar t$ (top) and $gg\to t\bar t$ (below) at the LHC 
          ($\sqrt{s} = 7\,$TeV),
          multiplied with the parton luminosities. Black (solid):
          $\Delta$NNLO$_{\text{app}}$,
          blue (dashed): N$^3$LO$_{\text A}$ (all terms in~\eqref{eq:N3LO} 
          for $k_s=k_C=1$, $k_h=2$), red (dot-dashed): N$^3$LO$_{\text B}$.}
\label{fig:N3LOLHC7}
\end{figure}

In order to judge the magnitude of the $\mathcal{O}(\alpha_s^3)$-terms, 
we plot the corrections to the partonic cross section in 
Figures~\ref{fig:N3LO} and~~\ref{fig:N3LOLHC7}.  We consider three
approximations: 
\begin{itemize}
\item $\Delta$NNLO$_{\text{app}}$, which consists of the singular terms 
of the  $\mathcal{O}(\alpha_s^2)$ 
correction as $\beta\to 0$~\cite{Beneke:2009ye}.  
\item N$^3$LO$_{\text A}$, which adds to the above the  
$\mathcal{O}(\alpha_s^3)$ terms from~\eqref{eq:N3LO}, including 
the remainder functions. 
\item N$^3$LO$_{\text B}$, which adds to 
$\Delta$NNLO$_{\text{app}}$ the $\mathcal{O}(\alpha_s^3)$
terms from~\eqref{eq:N3LO} 
with the remainder functions set to zero, i.e. the NNLL terms only. 
\end{itemize}
The difference between the first and the other two approximations 
provides an estimate of the importance of the third-order terms, 
and hence the convergence of the fixed-order expansion. The difference 
between the two N$^3$LO approximations represents an ambiguity 
in our estimate of the third-order terms.\footnote{For this estimate 
we choose $k_s=k_C=1$ and $k_h=2$, which differs from our  
canonical choice $k_s=k_C=k_h=2$ discussed below. The reason for 
this is that the remainder functions turn out to be small for the 
canonical choice, so the difference between  N$^3$LO$_{\text A}$ and 
N$^3$LO$_{\text B}$ would not provide a measure of the ``natural'' size 
of the sub-leading singular terms at N$^3$LO.}

In the figures we plot the integrand of the convolution in the 
formula for the hadronic cross section~\eqref{eq:sig-had}, i.e.  
the product of the partonic cross
section and the parton luminosity, as a function of $\beta$, including
the Jacobian $\partial \tau/\partial\beta$:
\begin{equation}
\label{eq:sigma-beta}
\frac{d \Delta\sigma_{pp'\to t\bar t}}{d\beta}=
\frac{8\beta m_{t}^2}{s(1-\beta^2)^2}
L_{pp'}(\beta,\mu_f)\Delta\hat\sigma_{pp'\to t\bar t}(\beta,\mu_f) \, .
\end{equation}
We show the partonic cross sections at the Tevatron (Figure~\ref{fig:N3LO}) 
and LHC ($\sqrt{s}=7\,$TeV, Figure~\ref{fig:N3LOLHC7}), using the
MSTW2008NNLO PDF set~\cite{Martin:2009iq} and $\mu_f=m_t$. The
third-order corrections are not negligible. 
The strong increase at small 
$\beta$ should be expected, since for such small 
values of $\beta$ the perturbative expansion breaks down and resummation
should be performed. However, we note that for the gluon 
channel the magnitude of the third-order terms can be comparable 
to the second-order ones up to $\beta\approx 0.6$ (N$^3$LO$_{\text A}$). Unless there are 
cancellations with higher-order terms in the $\beta$ expansion, 
which need not be small at $\beta\approx 0.5$, this may indicate a 
poor convergence of the fixed-order perturbative expansion. 

The area 
under the curves in the figures gives directly the correction to the 
total hadronic cross section that needs to be added to the NLO prediction. 
Adopting the top mass value $m_t=173.3\,$GeV, 
we obtain an additional $0.60\,$pb for 
$\Delta$NNLO$_{\text{app}}$, $0.93\,$pb ($0.71\,$pb) for 
N$^3$LO$_{\text B}$ (N$^3$LO$_{\text A}$) at the Tevatron and $12.1\,$pb for 
$\Delta$NNLO$_{\text{app}}$, $17.1\,$pb ($7.8\,$pb) for 
N$^3$LO$_{\text B}$ (N$^3$LO$_{\text A}$) at the LHC with $\sqrt{s}=7\,$TeV. 
The third-order term alone may therefore amount to up to $4\%$ of the cross
section at Tevatron and LHC. This might be considered as an estimate 
of the intrinsic uncertainty of the fixed-order prediction after 
the exact NNLO result is known. Below we shall see that 
resumming the threshold expansion to all orders in the NNLL approximation 
leads to a smaller effect from the sum of all higher-order terms. 

\section{Implementation of resummation and uncertainties}
\label{sec:implementation}

In the implementation of the resummation formalism 
one encounters several ambiguities: kinematical ambiguities, 
the precise matching to the fixed-order calculation, the choice of the 
scales $\mu_h$ and $\mu_s$ appearing in the momentum-space approach to 
threshold resummation, and the scale $\mu_C$ in the Coulomb function. We 
here discuss the choices made and the procedures to estimate 
the resulting theoretical uncertainties.

\paragraph{Kinematic ambiguities.} If the total hadronic cross
section~\eqref{eq:sig-had} is computed by convoluting the factorized
partonic cross section~\eqref{eq:fact} with the parton luminosity for
arbitrary values of $\tau$, the application of the factorization
formula is extended outside its domain of validity $\hat s\approx
4m_t^2$. As a result, there are kinematic ambiguities, since
expressions that agree at partonic threshold can show numerical
differences for $\beta \to 1$.
One such ambiguity is related to the expression used for the hard
function. As already mentioned in Section~\ref{sec:hard}, we use the
exact Born cross section, instead of the threshold limit, in the
determination of the leading hard function~\eqref{eq:sigma-hard}, since
this choice is observed to lead to a better agreement of the threshold
approximation with the exact NLO result. Another ambiguity arises from
the argument of the potential function $J_R$ in~\eqref{eq:fact}. The
derivation of the factorization formula~\cite{Beneke:2010da}
results in the dependence on the exact energy relative to the nominal
production threshold, $E=\sqrt{\hat s}-2m_t = 2m_t\times (1/\sqrt{1-\beta^2} - 
1)$. However, the threshold expansion is customarily formulated as an 
expansion in $\beta$. In the threshold limit the two choices are 
equivalent, since $E\approx m_t\beta^2$, and hence one may replace 
$E$ by $m_t\beta^2$ in the potential function. We have implemented both 
choices but use the second one as our default,
since in this case the expansion parameter $E/m_t\to\beta^2$ is always 
smaller than $1$, whereas it can grow to large values
when the exact expression for $E$ is used. The difference between
the two implementations is used as a numerical
estimate of the ambiguities in the resummation procedure.

\paragraph{Matching to the exact NLO result.}
Since the  total production cross section is not numerically dominated 
by the threshold region, where the factorization formula~\eqref{eq:fact} 
provides an accurate description of the cross section, we match the NNLL 
calculation to the exactly known NLO result. We consider two 
options. In the first, the NNLL cross section is expanded to NLO  
and subtracted from the resummed result. The resulting higher-order 
corrections are added to the full NLO result:
\begin{equation}
\label{eq:match1}
  \hat\sigma^{\text{NNLL}}_{pp',\text{matched}, 1}(\hat s)
  =\left[\hat\sigma^{\text{NNLL}}_{pp'}(\hat s)-
    \hat\sigma^{\text{NNLL}(1)}_{pp'}(\hat s)\right]
+ \hat\sigma^{\text{NLO}}_{pp'}(\hat s) \, .
\end{equation}
This matching prescription will be denoted by NNLL$_1$ in the following.
In this way, there is no double counting of $\mathcal{O}(\alpha_s)$
corrections. The expansion of the NNLL-resummed cross section to
$\mathcal{O}(\alpha_s)$, $\hat\sigma^{\text{NNLL}(1)}_{pp'}$, coincides
with the approximate NLO cross section given e.g.~in (B.6)
of~\cite{Beneke:2009ye}. For the NLO cross section we have
implemented both the analytical result~\cite{Czakon:2008ii} 
and the parameterization given in~\cite{Aliev:2010zk}.
For the former, 
we used the program provided in~\cite{Gehrmann:2001pz} for
numeric evaluation.

The matched cross section~\eqref{eq:match1} contains a
constant term at NNLO, i.e.~a $\beta$-independent
correction multiplying the Born cross section. This includes the
product of one-loop hard functions with the constant terms in the
one-loop soft function~\eqref{eq:soft-laplace}, as well as terms
related to ambiguities in the choice of the various scales. Since the
two-loop hard and soft functions have not been computed yet, the
constant NNLO term is not known completely, so the inclusion of a
partial result is a matter of choice. We therefore consider a
second matching option, denoted by NNLL$_2$, that matches to the NLO result by
subtracting the expansion of the NNLL corrections to
$\mathcal{O}(\alpha_s^2)$ and adding back the NNLO$_{\text{app}}$
corrections:~\cite{Beneke:2009ye} with the unknown constant set to
zero\footnote{For consistency the constant terms are set to zero also 
in the coefficients $f^{(2,2)}_X$ and $f^{(2,1)}_X$ of the 
factorization-scale dependent terms, even though they are known 
in this case, see Appendix~\ref{app:expNNLO}.}
\begin{equation}
\label{eq:match2}
  \hat\sigma^{\text{NNLL}}_{pp'\text{matched}, 2}(\hat s)
  =\left[\hat\sigma^{\text{NNLL}}_{pp'}(\hat s)-
    \hat\sigma^{\text{NNLL}(2)}_{pp'}(\hat s)\right]
+ \hat\sigma^{\text{NLO}}_{pp'}(\hat s)
+ \hat\sigma^{\text{NNLO}}_{\text{app},pp'}(\hat s) \, .
\end{equation}
The expansion of the NNLL correction to ${\cal O}(\alpha_s^2)$,
$\hat\sigma^{\text{NNLL}(2)}_{pp'}$, is given in
Appendix~\ref{app:expansions}.  The numerical difference between the
two implementations~\eqref{eq:match1} and~\eqref{eq:match2} can be
considered as an estimate of the unknown constant term at NNLO. We
discuss the issue of estimating the unknown constant further below.

For comparison, in the following we also present results for resummed
cross sections with NLL accuracy. In addition to making the appropriate
truncations for the resummation functions~\eqref{eq:res_funct_def},
in this approximation only the leading order terms in the
soft~\eqref{eq:soft-laplace} and hard function~\eqref{eq:hard-def} are
used, and the higher-order Green function $G_{C,R}^{(1)}$ and the
non-Coulomb corrections $\Delta_{\rm nC}$ are set to zero in the
potential function~\eqref{JRal}.  In contrast to the NNLL predictions,
bound-state corrections will not be included in our NLL results
presented in this paper.  Here the two options to implement the
matching are given by NLL$_1$, defined by
\begin{equation}
\label{eq:NLL1}
  \hat\sigma^{\text{NLL}}_{pp',\text{matched},1}(\hat s)
  =\left[\hat\sigma^{\text{NLL}}_{pp'}(\hat s)-
    \hat\sigma^{\text{NLL}(0)}_{pp'}(\hat s)\right]
  + \hat\sigma^{\text{LO}}_{pp'}(\hat s)\, ,
\end{equation}
and NLL$_2$, defined by
\begin{equation}
\label{eq:NLL2}
 \hat\sigma^{\text{NLL}}_{pp',\text{matched},2}(\hat s)
  =\left[\hat\sigma^{\text{NLL}}_{pp'}(\hat s)-
    \hat\sigma^{\text{NLL}(1)}_{pp'}(\hat s)\right]
+ \hat\sigma^{\text{NLO}}_{pp'}(\hat s) \, .
\end{equation}
Because of our choice of the hard function, the expansion of the NLL
correction to order $\alpha_s^0$, $\hat\sigma^{\text{NLL}(0)}_{pp'}$,
coincides with the exact LO cross section $
\hat\sigma^{\text{LO}}_{pp'}(\hat s)$, so that $
\hat\sigma^{\text{NLL}}_{pp',\text{matched},1}(\hat s)
=\hat\sigma^{\text{NLL}}_{pp'}(\hat s)$.  The expansion of the NLL
correction to ${\cal O}(\alpha_s^1)$, denoted by
$\hat\sigma^{\text{NLL}(1)}_{pp'}$, is given
in~\cite{Beneke:2010da}. The NLL$_2$ matching is identical to the
prescription used for squark-antisquark production
in~\cite{Beneke:2010da}. Since the full NLO corrections to top-pair
production are known, this is the preferred implementation for
phenomenological results, whereas the NLL$_1$ option is used to
estimate ambiguities in the resummation procedure at NLL.  

\paragraph{Unknown constant at $\mathcal{O}(\alpha_s^2)$.}
As mentioned in the previous paragraph, the constant term in the
threshold approximation is unknown at $\mathcal{O}(\alpha_s^2)$.  As a
default, we do not include such a constant in the
results, unless it is generated by the resummation, as in
the NNLL$_{1}$-option discussed below~\eqref{eq:match1}. 
However, we include an estimate of the order of magnitude of
the constants in the determination of the uncertainty. 
Adopting the MSTW08NNLO PDFs, $\mu_f=m_t$ and $m_t=173.3$ GeV, which is our
standard choice, the numerical effect
of non-vanishing constants for the different partonic channels is
given by
\begin{align}
\label{eq:delta-const}
\text{Tevatron}&: &  \Delta\sigma_{tt}^{(2)}&=
\left[ 0.345 \,\left(\frac{C^{(2)}_{qq}}{1000}\right)
+0.024\left(\frac{C^{(2)}_{gg,{\bf 8}}}{1000}\right)+
0.008\left(\frac{C^{(2)}_{gg,{\bf 1}}}{1000}\right)\right]\text{pb} \, ,
\nonumber\\[0.2cm]
\text{LHC} (\sqrt{s}=7~\text{TeV})&:&   \Delta\sigma_{tt}^{(2)}&=
\left[ 1.70 \,\left(\frac{C^{(2)}_{qq}}{1000}\right)
+4.31\left(\frac{C^{(2)}_{gg,{\bf 8}}}{1000}\right)
+ 1.31\left(\frac{C^{(2)}_{gg,{\bf 1}}}{1000}\right)\right]\text{pb} \, ,
\nonumber\\[0.2cm]
\text{LHC} (\sqrt{s}=14~\text{TeV})&:&   \Delta\sigma_{tt}^{(2)}&=
\left[ 5.34 \,\left(\frac{C^{(2)}_{qq}}{1000}\right)
+27.14\left(\frac{C^{(2)}_{gg,{\bf 8}}}{1000}\right)+
 7.97\left(\frac{C^{(2)}_{gg,{\bf 1}}}{1000}\right)\right]\text{pb}\,.
\end{align}
Here our conventions are such that the correction to the partonic cross 
section due to the constant reads
\begin{equation}
   \Delta\hat \sigma_{pp',R}^{(2)} = 
   \hat\sigma^{(0)}_{pp',R}\;
    \left(\frac{\alpha_s}{4\pi}\right)^2 C^{(2)}_{pp',R} \,  ,
\end{equation}
where $\hat\sigma^{(0)}_{pp',R}$ is the (exact) Born cross section for
the $pp'$ initial state in the colour channel~$R$.

In order to get a crude estimate of the order of magnitude of the
unknown $C^{(2)}_{pp',R}$, we consider the square of the
corresponding one-loop constants. This method is
observed to give a conservative estimate in cases where the full
two-loop corrections are known. For top-pair production the one-loop
constants have been calculated analytically~\cite{Czakon:2008cx}. In our 
conventions, they are related to the hard functions~\eqref{eq:hard-nlo} by
\begin{equation}
C^{(1)}_{pp',R}= h^{R(1)}_{pp'}(m_t)+ 
4\,(C_r+C_{r'}) \bigg[9\ln^22-12\ln 2+8 -\frac{11 \pi^2}{24}\bigg] -
 12  C_{R} \left[\ln 2-1\right]\,.
\end{equation}
Numerically, 
\begin{equation}
   C^{(1)}_{q\bar q}=14.531\;,\quad C^{(1)}_{gg,{\bf 8}}=30.586\;, \quad 
   C^{(1)}_{gg,{\bf 1}}=14.026\,,
\end{equation}
hence from $|C^{(2)\text{est.}}_{pp',R}|=(C^{(1)}_{pp',R})^2$ we obtain 
the estimates
\begin{equation}
| C^{(2)\text{est.}}_{qq}|=
211.2\;,\quad |C^{(2)\text{est.}}_{gg,{\bf 8}}|=935.5
\;,\quad |C^{(2)\text{est.}}_{gg,{\bf 1}}|=196.7\,.
 \label{eq:constant}
\end{equation}
In our numerical results we estimate the uncertainty due to the unknown 
constant by adding and subtracting $|C^{(2)\text{est.}}_{pp',R}|$
in all partonic channels.

\paragraph{Choice of the soft scale.}
As discussed in Section~\ref{sec:basics}, the resummation of
threshold logarithms in the momentum-space approach amounts to
resumming logarithms of the ratios $\mu_h/\mu_f$ and $\mu_s/\mu_f$ of
the hard, soft and the factorization scale. While an all-order
solution to the evolution equations would be independent of these
scales, a residual dependence remains after truncation of the
perturbative series of the resummation functions $S$ and
$a$. Furthermore, the potential function depends on the Coulomb scale
$\mu_C$. Our default choices for the hard scale, $\mu_h=2 m_t$, and the
Coulomb scale~\eqref{eq:mu-c} have been explained in
Section~\ref{sec:input}. 

The proper choice of the soft scale requires
some discussion.  
For any given $\beta$, the idea of scale separation in the effective 
theory suggests the $\beta$-dependent scale $\mu_s \sim m_t\beta^2$.
Indeed, this choice has to be made to recover 
the threshold logarithms in fixed-order in $\alpha_s$ from the 
expansion of the resummed result. However, the use of a running scale
in the resummed partonic cross section is problematic, since it 
leads to an oscillating partonic cross section for small values of 
$\beta$. The problem arises from 
$\eta=\frac{\alpha_s\Gamma_{\text{cusp}}}{2\pi}
\log(\tfrac{\mu_s}{\mu_f})+\dots$ becoming increasingly negative, 
such that the factor $1/\Gamma(2\eta)$ in (\ref{eq:w-resummed}) 
changes sign whenever a pole of the Gamma function is crossed. 
This effect appears in resummation for all hadron-collider 
processes such as Drell-Yan and Higgs production. In the traditional 
approach to threshold resummation in Mellin space, a similar 
difficulty appears when performing the inverse Mellin transform (see
e.g.~\cite{Catani:1996dj}), since the convolution of the resummed partonic 
cross section 
with the parton luminosity
does not converge. The most widely used prescription to perform the
convolution was proposed in~\cite{Catani:1996dj}, and amounts to
performing the inverse Mellin transform at the level of the hadronic
cross section. Other prescriptions have been suggested to perform the
inverse Mellin transform directly for the partonic cross section
but, in one way or another, employ a cutoff in the convolution with
the parton luminosity, see e.g.~\cite{Berger:1996ad,Bonvini:2010tp}.

We investigate two methods to determine the soft scale: (1) a
fixed soft scale determined according to a procedure proposed 
in~\cite{Becher:2007ty} and (2) a running scale frozen to a fixed value 
below a certain value of $\beta$.  
In our final analysis we use the second method by default.

\subsection{Method 1}
\label{sec:method1}

In~\cite{Becher:2006nr,Becher:2006mr,Becher:2007ty} it has been
advocated to choose a fixed soft scale $\mu_s$.  In this
approach one does not aim at
predicting the partonic cross section locally as a function of
$\beta$, but argues that logarithms in the \emph{hadronic} cross
section for $\tau_0=4m_t^2/s\to 1$ are resummed correctly by a fixed
soft scale $\mu_s=2 m_t k_s(1-\tau_0)$, where $k_s\sim 0.1$ due to the
steeply falling parton luminosity function~\cite{Becher:2007ty}.  The
partonic cross section is then predicted only in an average
sense.\footnote{See also~\cite{Ahrens:2011mw} for a discussion of the
differences between this approach and the use
of fixed-order expansions.}

In practice, the fixed soft scale is chosen such that  it minimizes the 
relative fixed-order one-loop soft correction to the \emph{hadronic} 
cross section.  
Following this 
procedure, we vary the scale in the PDFs and the soft correction and 
determine the value $\tilde \mu_s$ that minimizes the relative soft 
corrections:
\begin{equation}
\label{eq:def-mus}
0 =\tilde\mu_s\frac{d}{d\tilde\mu_s}
\sum_{p,p'}\int_{4m_{t}^2/s}^1 d\tau \,L_{pp'}(\tau,\tilde\mu_s)
\frac{\hat\sigma^{(1)}_{pp',\text{soft}} (\tau s,\tilde\mu_s)}{
\sigma^{(0)}_{N_1N_2}(s,\tilde \mu_s)}\,.
\end{equation}
Here the fixed-order NLO soft correction,
$\hat\sigma^{(1)}_{pp'\text{soft}}$, is obtained from the threshold
expansion of the NLO cross section (see e.g.~(D.3)
in~\cite{Beneke:2010da}) by setting the Coulomb correction and the
hard corrections $h_i$ to zero. In the denominator we divide by the
leading-order hadronic cross section. 

In this approach the value of the soft scale depends on the interplay
of coefficients of the logarithms and constants in the one-loop soft
function~\eqref{eq:soft-laplace}, which might be considered as going
against the philosophy of resumming only large logarithms.
Furthermore, the reduction of the soft scale by a factor of $10$
due to the behaviour of the PDFs compared to the naive expectation $
\sim 2 m_t (1-\tau_0)$, has been estimated~\cite{Becher:2007ty} to be
effective for $\tau_0>0.2$, while for $\tau_0 \to 0$ the relevant soft
scale is expected to be of the order $\mu_s \sim m_t$.  For top-quark
production at the Tevatron and the LHC with centre-of-mass energies up
to $14$ TeV the relevant range is $\tau_0=0.03-6\times 10^{-4}$, so it
cannot be expected {\it a priori} that the effect of the higher-order
$\log\beta$ terms in the partonic cross section can be properly
incorporated in an average sense through a fixed $\mu_s$. 

As discussed previously, there is a kinematic ambiguity 
from expressing the one-loop soft corrections in terms of 
$\log(E/\mu_s)$ or of $\log(m_t\beta^2/\mu_s)$. For a 
process dominated by threshold dynamics, the effect of this ambiguity
on the soft scale determined through~\eqref{eq:def-mus} should be
small. However, for the case of top-quark production, we obtain noticeable
differences between the soft scales determined using the  
two parameterizations of the soft function:
\begin{equation}
\begin{array}{llll}
\log(E/\mu_s): \hspace*{0.2cm} & 
\tilde\mu_s =52\, \text{GeV}\, (\text{Tevatron}),\hspace*{0.2cm}&
99\,\text{GeV}\, (\text{LHC}7),\hspace*{0.2cm}&
120\,\text{GeV}\, (\text{LHC}14),\\[0.2cm]
\log(m_t\beta^2/\mu_s):\hspace*{0.2cm}& 
\tilde\mu_s= 35\text{GeV} (\text{Tevatron}),&
58\,\text{GeV}\, (\text{LHC}7),&
65\,\text{GeV} \,(\text{LHC}14) .
\end{array}
\label{eq:mus-1}
\end{equation}
This difference, up to a factor of almost two, is a result of the fact
that the total cross section is dominated by values $\beta\gtrsim 0.3$, and is
another hint that the use of a fixed soft scale is somewhat
problematic for top quark production.\footnote{For the production of
  heavier particles, where the threshold region is more important,
  this ambiguity is indeed smaller.  For instance, for $m_t=1\,$TeV
  and the LHC with $\sqrt s=7\,$TeV we find $\tilde\mu_s= 188\,$GeV
  expressing the soft function in terms of $\log(E/\mu_s)$, and
  $\tilde\mu_s= 149\,$GeV using $\log(m_t\beta^2/\mu_f)$, so the
 difference is reduced to less than $30\%$.}
 
We nevertheless present results obtained using a fixed soft scale, in 
order to discuss the numerical impact of the potential problems in practice.
We show numbers for both the 
(N)NLL$_1$ and (N)NLL$_2$ matching prescriptions, and 
use $E \to m_t \beta^2$. Note that for a fixed
soft scale not all the logarithms of $\beta$ are resummed
locally in the partonic cross section, so the implementations
(N)NLL$_1$ and (N)NLL$_2$ differ not only by a constant but also by
logarithmic terms at $\mathcal{O}(\alpha_s^2)$. 
The soft scale 
is chosen as in the second line of~\eqref{eq:mus-1}.
For the factorization scale we take $\tilde \mu_f=m_t$, for the hard scale 
$\tilde \mu_h=2 m_t$, as discussed in Section~\ref{sec:hard}. The 
Coulomb scale $\tilde \mu_C$ is defined in~\eqref{eq:mu-c}. 
We estimate the theoretical uncertainty as follows:
\begin{description}
\item[Scale uncertainty:] We vary all scales $\mu_i$ (the
  factorization, soft, hard and Coulomb scales) in the interval
  $[\tilde\mu_i/2,2\tilde \mu_i]$ around their central values
  $\tilde\mu_i$. $\mu_s$ and $\mu_C$ are varied while keeping
  the other scales fixed. $\mu_h$ and $\mu_f$ are allowed to
  vary simultaneously,
  imposing the additional constraint $1\le \mu_h/\mu_f \le 4$.
  The errors from varying $\{\mu_f,\mu_h\}$,
  $\mu_s$ and $\mu_C$ are obtained by taking the respective
  maximum/minimum, and are added in quadrature.
\item[Resummation ambiguities:] We estimate ambiguities in the resummation 
procedure by taking the difference between the default  
$E\to m_t\beta^2$ and $E=\sqrt{\hat s}-2m_t$.  With the second 
expression for $E$ the soft scales in the first 
line of~\eqref{eq:mus-1} are used. 
\item[NNLO-constant:] All constants $C^{(2)}_{pp',R}$ 
in~\eqref{eq:delta-const} are varied by
$\pm|C^{(2)\text{est.}}_{pp',R}|$.
\item[PDF$+\alpha_s$ uncertainty:] We estimate the combined error of 
the NNLL results due to uncertainties in the PDFs and the strong coupling 
using the $90\%$ confidence level eigenvector set of the MSTW08NNLO 
PDFs~\cite{Martin:2009iq} and the five sets for variations of $\alpha_s$ 
provided in~\cite{Martin:2009bu}, which corresponds to $\alpha_s(M_Z) = 
0.1171 \pm 0.0034$. For the NLL prediction the 
corresponding NLO PDFs are used.
\end{description}

For the top quark pole mass $m_t=173.3\,$GeV, we obtain the 
total top pair production cross sections displayed in
Table~\ref{tab:method1}, where the four errors refer to the four 
sources of uncertainty detailed above. 
Both, at the NLL and NNLL order, the central values for the two matching
procedures lie inside the common error band. The total relative theory error,
given by the sum in quadrature of the scale, resummation and constant
uncertainties, decreases from NLL to NNLL, as one would expect
from including higher orders in the logarithmic expansion, and
increases from Tevatron to LHC, consistent with the fact that the
threshold kinematics is parametrically more dominant at Tevatron, and
thus matching and resummation ambiguities are expected to be
smaller.

\begin{table}[!t]
\newcommand{\m}{\hphantom{$-$}}
\newcommand{\cc}[1]{\multicolumn{1}{c}{#1}}
\renewcommand{\tabcolsep}{0.8pc} 
\renewcommand{\arraystretch}{1.0} 
\begin{center}
\begin{tabular}{@{}llllll}
\hline  \vspace{-4mm} \\  
     & \hspace{-4mm} \mbox{Tevatron} 
     & \hspace{-3mm} LHC ($\sqrt{s}=7\,$TeV)  
     & \hspace{-3mm} LHC ($\sqrt{s}=14\,$TeV)  \\
\hline   \\
\hspace{-7mm}
\phantom{ab} ${\text{NLL$_{1}$}}$
     & \hspace{-5mm} $\phantom{0}6.60^{\,+0.43}_{\,-0.65}{}^{\,+0.00}_{\,-0.61}
        {}^{\,+0.10}_{\,-0.10}{}^{\,+0.50}_{\,-0.44}$
     & \hspace{-5mm} $\phantom{0}129.9^{\,+26.3}_{\,-30.2}{}^{\,+0.00}_{\,-15.6}
       {}^{\,+4.7}_{\,-4.7}{}^{\,+10.2}_{\,-9.7}$
     & \hspace{-5mm} $\phantom{0}692^{\,+169}_{\,-180}{}^{\,+~0}_{\,-85}
       {}^{\,+28}_{\,-28}{}^{\,+46}_{\,-41}$
 \vspace{2mm} \\
 \vspace{-2mm} \\
\hspace{-7mm}
\phantom{ab} ${\text{NLL$_{2}$}}$     
     & \hspace{-5mm} $\phantom{0}6.90^{\,+0.32}_{\,-0.41}{}^{\,+0.00}_{\,-0.04}
        {}^{\,+0.10}_{\,-0.10}{}^{\,+0.52}_{\,-0.47}$
     & \hspace{-5mm} $\phantom{0}157.6^{\,+23.2}_{\,-20.2}{}^{\,+2.3}_{\,-0.00}
       {}^{\,+4.7}_{\,-4.7}{}^{\,+13.5}_{\,-12.8}$
     & \hspace{-5mm} $\phantom{0}876^{\,+135}_{\,-113}{}^{\,+14}_{\,-~0}
       {}^{\,+28}_{\,-28}{}^{\,+64}_{\,-56}$
 \vspace{2mm} \\
 \vspace{-2mm} \\
\hspace{-7mm}
\phantom{ab} ${\text{NNLL$_{1}$}}$     
     & \hspace{-5mm} $\phantom{0}6.87^{\,+0.31}_{\,-0.40}{}^{\,+0.00}_{\,-0.02}
        {}^{\,+0.10}_{\,-0.10}{}^{\,+0.65}_{\,-0.50}$
     & \hspace{-5mm} $\phantom{0}151.8^{\,+9.4}_{\,-4.9}{}^{\,+5.5}_{\,-0.0}
       {}^{\,+4.7}_{\,-4.7}{}^{\,+13.7}_{\,-13.2}$
     & \hspace{-5mm} $\phantom{0}837^{\,+69}_{\,-28}{}^{\,+36}_{\,-~0}
       {}^{\,+28}_{\,-28}{}^{\,+57}_{\,-57}$
 \vspace{2mm} \\
 \vspace{-2mm} \\
\hspace{-7mm}
\phantom{ab} ${\text{NNLL$_{2}$}}$     
     & \hspace{-5mm} $\phantom{0}7.08^{\,+0.15}_{\,-0.28}{}^{\,+0.07}_{\,-0.00}
        {}^{\,+0.10}_{\,-0.10} {}^{\,+0.69}_{\,-0.53}$
     & \hspace{-5mm} $\phantom{0}157.4^{\,+8.8}_{\,-3.6}{}^{\,+5.4}_{\,-0.0}
       {}^{\,+4.7}_{\,-4.7}{}^{\,+14.5}_{\,-13.9}$
     & \hspace{-5mm} $\phantom{0}868^{\,+63}_{\,-21}{}^{\,+29}_{\,-~0}
       {}^{\,+28}_{\,-28}{}^{\,+61}_{\,-60}$
 \vspace{2mm} \\
\hline
\end{tabular}\\[2pt]
\end{center}
\caption{ \sf Results for Method 1, $m_t=173.3\,$GeV, all numbers in pb. 
The four errors refer to scale variation, 
resummation ambiguity, the NNLO constant term, and the PDF$+\alpha_s$ 
uncertainty.}
\label{tab:method1}
\end{table} 

The difference of the NLL$_1$ results, both to the NNLL$_1$
and the complete fixed-order NLO predictions, (see
Tables~\ref{Tab:CentT} and~\ref{Tab:CentL} in Section \ref{sec:results}) 
are large, especially at the LHC, 
since in NLL$_1$ the fixed-order NLO corrections are not included through
matching.
The difference between NLL$_2$ and the fixed order NLO results, as
well as the two NNLL implementations, is small in comparison.
Since part of the difference is due to the use of different PDFs (NLO
vs.~NNLO), it is instructive to compare predictions using NNLO PDFs
also at NLL in order to ascertain the genuine effect of the higher-order
corrections. For the NLL$_{2}$ approximation we then find the central
values $\sigma^{\text{NLL}_2}_{t\bar t}=6.67$ pb at the Tevatron and
$\sigma^{\text{NLL}_2}_{t\bar t}=148.6$ pb at the LHC with
$\sqrt{s}=7$ TeV ($\sigma^{\text{NLL}_2}_{t\bar t}=830$ pb 
with $\sqrt{s}=14$ TeV). Since in both the NLL$_2$ and  NNLL$_1$ prescriptions 
the resummed cross section is matched to the fixed-order NLO result, 
their difference shows that the genuine effect of going from NLL to
NNLL is moderate, and leads to an enhancement of about  $3\%$ at the 
Tevatron and below $1\%$ at the LHC with $\sqrt{s}=14$ TeV.
The difference of the two NNLL implementations is of the order of
$3-4\%$, and is an estimate for the effect of the constant terms at
NNLO and the difference of including all $\log\beta$ terms 
exactly at NNLO or approximately due to the use of a fixed $\mu_s$.

The total error, obtained by summing in quadrature the four sources of
uncertainty, is dominated  by the scale uncertainty and the error 
associated with PDF and $\alpha_s$ variation. Despite the ambiguity in the
determination of the soft scale (\ref{eq:mus-1}), the error associated
to resummation is small at both Tevatron and LHC.
The scale uncertainty decreases from NLL to NNLL, and is
systematically smaller for the second matching option compared to the first.

\subsection{Method 2}
\label{sec:method2}

For the reasons explained in Section~\ref{sec:method1}, the use of a 
fixed soft scale may be problematic for top-quark
production at the Tevatron or LHC, where the total cross section is not
dominated by the threshold region, but where it is nevertheless useful
to include logarithmic higher-order corrections. We therefore
propose a procedure that uses a running scale and fixes the soft
scale only in the low-$\beta$ region. We divide the convolution of
the partonic cross section with the parton
luminosity~\eqref{eq:sig-had} into two regions using a parameter
$\beta_{\text{cut}}$, chosen such that on the one hand the perturbative
expansion in $\alpha_s$ is not spoiled by large logarithms in the
upper interval $\beta>\beta_{\text{cut}}$, and on the other hand
ambiguities in the threshold approximation are small in the lower
interval $\beta<\beta_{\text{cut}}$.  The partonic cross section is
then treated differently in the two regions:

\paragraph{$\beta < \beta_{\text{cut}}$}
For $\beta_{\text{cut}}$ chosen sufficiently small (say
$\beta_{\text{cut}}<0.5$) the threshold expansion in $\beta$ 
is convergent in the lower interval and the use of the factorization
formula~\eqref{eq:fact} is justified. In this region, the logarithms
of $\beta$ become large and need to be resummed for a reliable
prediction. 
The threshold-enhanced
contributions are numerically dominant and ambiguities in the resummation
procedure, like the kinematical ambiguities or the use of the
different matching prescriptions~\eqref{eq:match1}
and~\eqref{eq:match2}, should not lead to large numerical
differences. For the reasons discussed before we use the 
NNLL$_2$ implementation defined in~\eqref{eq:match2}, with $E\to
m_t\beta^2$ and a fixed soft scale
\begin{equation}
\label{eq:mus-low} 
\mu_s^<= k_s m_t\beta_{\text{cut}}^2\,.
\end{equation}
The choice of the constant $k_s$ is discussed below.

\paragraph{$\beta > \beta_{\text{cut}}$} In the upper interval, the
use of the threshold approximation cannot be justified a
priori. Nevertheless we here adopt the point of view that the
inclusion of a subset of higher-order corrections is useful even in
this region. For $\beta_{\text{cut}}$ chosen large enough, so that the
perturbative expansion is not spoiled by large logarithms, the
numerical difference of the NNLL resummed cross section and the
expansion to NNLO and N$^3$LO accuracy is expected to be small, so that
the result should not depend too strongly on the choice of one of
these approximations.  By default we use again the implementation
NNLL$_{\text{2}}$ with $E\to m_t\beta^2$. Since
for $\beta\to 1$ it is not possible to argue that a fixed soft scale
correctly includes dominant contributions to the hadronic cross
section, we choose a running soft scale, 
\begin{equation}
\label{eq:mus-up}
\mu_s^>= k_s m_t\beta^2 \,  ,
\end{equation}
in order to include all NNLL contributions to the \emph{partonic}
cross section correctly.  In order to estimate the ambiguities in this
treatment, we compare it to the approximate fixed-order cross 
sections NNLO$_{\text{app}}$ \cite{Beneke:2009ye}, as well as 
N$^3$LO$_{\text A}$ and N$^3$LO$_{\text B}$ defined in
Section~\ref{sec:n3lo}, in each case including all lower-order terms 
up to the indicated order. 

There is a certain tension in the requirements on $\beta_{\text{cut}}$. 
If chosen too large, the cross section result becomes too sensitive 
to the ambiguities in the matching
procedure and the resulting constant terms at
$\mathcal{O}(\alpha_s^2)$. Furthermore, a larger $\beta_{\text{cut}}$
implies a larger $\mu_s^<$, so that resummation becomes ineffective.
On the other hand, for $\beta_{\text{cut}}$ too small the difference
among successive orders in the perturbative expansion in the upper
interval becomes too large (c.f. Figure~\ref{fig:N3LO}). This suggests
the following procedure to determine $\beta_{\text{cut}}$: we consider
the cross section 
\begin{equation}
\label{eq:sigma-up-down}
  \hat \sigma_{t\bar t}(A_<,B_>,\beta_\text{cut})= 
\hat \sigma_{t\bar t}^{A_<}\;\theta(\beta_\text{cut}-\beta)
+\hat \sigma_{t\bar t}^{B_>}\;\theta(\beta-\beta_\text{cut}) \, ,
\end{equation}
defined using one of the appropriate approximations for the upper and
lower intervals, $A_<\in\{\text{NNLL}_1,\text{NNLL}_2\}$ and
$B_>\in\{\text{NNLL}_2,\text{NNLO}_{\text{app}},
\text{N$^3$LO}_{\text{A}},\text{N$^3$LO}_{\text{B}}\}$,
and determine $\beta_\text{cut}$ such that the difference between the
eight different implementations $\hat \sigma_{t\bar
  t}(A_<,B_>,\beta_\text{cut})$ becomes minimal. 

We repeat this procedure for values $k_s=1,2,4$ in the 
soft scales~\eqref{eq:mus-low}
and~\eqref{eq:mus-up} in the two intervals. As default we adopt $k_s=2$,
and obtain the following values for  $\beta_{\text{cut}}$ and the 
corresponding fixed soft scale in the lower interval:
\begin{equation}
\label{eq:betacut}
\begin{array}{llll}
\beta_{\text{cut}} (\text{NNLL}):\quad& 
0.35 \,(\text{Tevatron}),& 
0.54\,(\text{LHC}7),&
0.55 \,(\text{LHC}14),\\[0.1cm]
\mu_s^<=2m_t\beta^2_{\text{cut}}:&  
42\,\text{GeV}\,(\text{Tevatron}),\quad&
101\,\text{GeV}\, (\text{LHC}7),\quad&
105\,\text{GeV}\, (\text{LHC}14).
\end{array}
\end{equation}
It is seen that the $\beta_{\text{cut}}$ values are in a
region where we expect the $\beta$-expansion and the perturbative
expansion in $\alpha_s$ both to be reasonably reliable, so they satisfy
the requirements discussed above.
We treat the difference to the results
for $k_s=1,4$ as another source of the theoretical uncertainty. 
More details on this procedure and the  $\beta_{\text{cut}}$ values
for $k_s=1,4$ are given in Appendix~\ref{app:method2}.

To summarize, our default implementation is given by the option
NNLL$_2$ defined in~\eqref{eq:match2} using the soft
scales~\eqref{eq:mus-low} and~\eqref{eq:mus-up}, with $k_s=2$ in both
intervals and the $\beta_{\text{cut}}$-values given 
in~\eqref{eq:betacut}. For the
remaining scales we again take the default values $\tilde \mu_f=m_t$,
$\tilde \mu_h=2 m_t$ and the Coulomb scale as defined
in~\eqref{eq:mu-c}. We estimate the remaining uncertainties as
following:
\begin{description}
\item[Scale uncertainty:] as for Method 1, excluding the soft scale. 
  The latter is effectively replaced by the variation of $k_s$.
\item[Resummation ambiguities:] we consider three different sources
  for ambiguities: i) The difference between the default setting 
  $E = m_t\beta^2$ compared to $E=\sqrt{\hat s}-2m_t$ in the
  NNLL$_2$ implementation, ii) the difference between the NNLL$_2$
  implementation for the soft scale choices $k_s=1,4$ (and the
  corresponding $\beta_{\text{cut}}$ values) to the default choice
  $k_s=2$, iii) the envelope of the 8-different
  approximations~\eqref{eq:sigma-up-down} for variations of
  $\beta_{\text{cut}}$ by $20\%$ around the default
  values~\eqref{eq:betacut} for $k_s=2$. The resulting errors are
  added in quadrature.
\item[NNLO-constant, PDF$+\alpha_s$ uncertainty:] estimated as in  Method 1.
\end{description}

\begin{table}[t]
\newcommand{\m}{\hphantom{$-$}}
\newcommand{\cc}[1]{\multicolumn{1}{c}{#1}}
\renewcommand{\tabcolsep}{0.8pc} 
\renewcommand{\arraystretch}{1.0} 
\begin{center}
\begin{tabular}{@{}llllll}
\hline  \vspace{-4mm} \\  
     & \hspace{-4mm} \mbox{Tevatron} 
     & \hspace{-3mm} LHC ($\sqrt{s}=7\,$TeV) 
     & \hspace{-3mm} LHC ($\sqrt{s}=14\,$TeV) \\
\hline   \\
\hspace{-7mm}
\phantom{ab} ${\text{NLL$_{2}$}}$     
     & \hspace{-5mm} $\phantom{0}7.31^{\,+0.25}_{\,-0.03}{}^{\,+0.30}_{\,-0.53}
        {}^{\,+0.10}_{\,-0.10}{}^{\,+0.57}_{\,-0.54}$
     & \hspace{-5mm} $\phantom{0}172.8^{\,+14.8}_{\,-~0.8}{}^{\,+13.0}_{\,-14.7}
        {}^{\,+4.7}_{\,-4.7}{}^{\,+15.9}_{\,-14.6}$
     & \hspace{-5mm} $\phantom{0}954^{\,+85}_{\,-~5}{}^{\,+65}_{\,-71}
       {}^{\,+28}_{\,-28}{}^{\,+74}_{\,-66}$
 \vspace{2mm} \\
 \vspace{-2mm} \\
\hspace{-7mm}
\phantom{ab} ${\text{NNLL$_{2}$}}$
     & \hspace{-5mm} $\phantom{0}7.22^{\,+0.21}_{\,-0.41}{}^{\,+0.20}_{\,-0.21}
        {}^{\,+0.10}_{\,-0.10}{}^{\,+0.71}_{\,-0.55}$
     & \hspace{-5mm} $\phantom{0}162.6^{\,+4.2}_{\,-1.9}{}^{\,+3.9}_{\,-5.6}
        {}^{\,+4.7}_{\,-4.7}{}^{\,+15.4}_{\,-14.7}$
     & \hspace{-5mm} $\phantom{0}896^{\,+22}_{\,-~5}{}^{\,+18}_{\,-23}
       {}^{\,+28}_{\,-28}{}^{\,+65}_{\,-64}$
 \vspace{2mm} \\
\hline
\end{tabular}\\[2pt]
\end{center}
\caption{ \sf Results for Method 2, $m_t=173.3\,$GeV, all numbers in pb. 
The four errors refer to scale variation, 
resummation ambiguity, the NNLO constant term, and the PDF$+\alpha_s$ 
uncertainty.}
\label{tab:method2}
\end{table} 

Our results for Method 2 are shown in Table~\ref{tab:method2}, 
where the four errors refer to the four 
sources of uncertainty as detailed above. 
The separate sources of resummation ambiguity are shown in more
detail in Table \ref{Tab:ErrorRes}, from which one can see that the variation
of $k_s$ and $\beta_{\text{cut}}$ (i.e. the soft-scale variation)
dominate the error,
while the ambiguity due to the use of $E=m_t \beta^2$ is negligible. 
To see the genuine effect of NNLL resummation it is again useful to
consider the NLL$_2$ results using the NNLO PDFs for which we obtain
$\sigma^{\text{NLL}_2}_{t\bar t}=7.02$ pb at the Tevatron
and $\sigma^{\text{NLL}_2}_{t\bar t}=162.2$ pb at the LHC with
$\sqrt{s}=7$ TeV ($\sigma^{\text{NLL}_2}_{t\bar t}=901$ pb at the LHC
with $\sqrt{s}=14$ TeV). Therefore the effect of the NNLL corrections
on the central values is small, about $3\%$ at the Tevatron and in the
per-mille range at the LHC, showing a good convergence of successive
orders of the logarithmic approximations.

\begin{table}[!t]
\newcommand{\m}{\hphantom{$-$}}
\newcommand{\cc}[1]{\multicolumn{1}{c}{#1}}
\renewcommand{\tabcolsep}{0.8pc} 
\renewcommand{\arraystretch}{1.0} 
\begin{center}
\begin{tabular}{@{}llllll}
\hline  
      Collider
     & \phantom{ab} i) $E$
     & \phantom{ab} ii) $k_s$ & \phantom{ab} iii) $\beta_{\text{cut}}$  \\
\hline  \\
                Tevatron
              & $\phantom{0}^{+0.01}_{-0.00}~~\{^{+0.1\%}_{-0.0\%}\}$
              & $\phantom{0}^{+0.10}_{-0.10}~~\{^{+1.4\%}_{-1.4\%}\}$
              & $\phantom{0}^{+0.17}_{-0.18}~~\{^{+2.4\%}_{-2.5\%}\}$
 \vspace{2mm} \\
                LHC ($\sqrt{s}=7$ TeV)
              & $\phantom{0}^{+1.0}_{-0.0}~~~\{^{+0.6\%}_{-0.0\%}\}$
              & $\phantom{0}^{+0.8}_{-2.2}~~~\{^{+0.5\%}_{-1.4\%}\}$
              & $\phantom{0}^{+3.7}_{-5.1}~~~\{^{+2.3\%}_{-3.1\%}\}$
 \vspace{2mm} \\
                LHC ($\sqrt{s}=14$ TeV)
              & $\phantom{0}^{+5}_{-0}\phantom{.}~~~~\{^{+0.6\%}_{-0.0\%}\}$
              & $\phantom{0}^{+~0}_{-10}\phantom{.}~~~\{^{+0.0\%}_{-1.1\%}\}$ 
              & $\phantom{0}^{+17}_{-21}\phantom{.}~~~\{^{+1.9\%}_{-2.3\%}\}$
 \vspace{2mm} \\ 
\hline
\end{tabular}\\[2pt]
\caption{ \sf Different resummation ambiguities of the $\text{NNLL$_{2}$}$
  cross section for $m_{t}=173.3$ GeV.
   Absolute errors are in pb. Relative errors are given in brackets.}
\label{Tab:ErrorRes}
\end{center}
\end{table} 
 
The NNLL$_2$ results in Table~\ref{tab:method2} are about $3\%$ larger
than those obtained in Table~\ref{tab:method1} using Method 1, while
for the NLL$_2$ results the differences are of the order of $10\%$.
However the results from the two methods agree within the estimated
resummation ambiguity. Since the soft scales corresponding to
$\beta_{\text{cut}}$ in~\eqref{eq:betacut} lie in the range of  the fixed
soft scales used for Method 1~\eqref{eq:mus-1}, 
the differences between the two methods are smaller than might have
been expected, and serve as an estimate of the effect
of treating logarithmic corrections exactly for large $\beta$, as in
Method 2, or approximately through a fixed soft scale, as in Method~1. 
Since the results of both (very different) methods for the
treatment of the soft scale agree within errors, we are confident that
we are realistically estimating the resummation ambiguities.

The total relative theoretical error (not including the PDF$+\alpha_s$
uncertainty) is shown in Table~\ref{Tab:RelErrLHC} for the NLL$_2$ and
NNLL$_2$ cross sections, and for the NNLO$_{\text{app}}$ result (see
Section \ref{sec:results}, Tables \ref{Tab:CentT} and~\ref{Tab:CentL},
for the absolute values).  Both for the Tevatron and the LHC, the
theoretical error decreases when higher-order logarithmic corrections
are added to
NLL$_2$. However at the Tevatron the smallest uncertainty is obtained
for NNLO$_{\text{app}}$, while NNLL$_2$ yields the smallest error
estimate for LHC.  To understand this somewhat surprising behaviour at
the Tevatron, in Table~\ref{tab:channels} we study separately the
results for the quark-antiquark and gluon-gluon channels. It is seen
that the increase of the error estimate for the quark-antiquark
induced channel from NNLO$_{\text{app}}$ to NNLL$_2$ occurs both at
the Tevatron and the LHC. 
On the contrary, at the LHC the error of the
dominant gluon channel is reduced at NNLL$_2$, which leads to the
observed reduction of the total theoretical uncertainty.  At the Tevatron a
reduction of the error in the gluon channel is observed as well, if the
same value $\beta_{\text{cut}}=0.54$ as for LHC is used.
\footnote{This observation could suggest a separate
  $\beta_{\text{cut}}$ choice for both channels, however due to the
  small effect on the central value we continue to use the same
  values.}
It is interesting to recall that a comparison of the
NLO singular terms with the exact NLO results (see
e.g.~\cite{Ahrens:2010zv,Beneke:2010da}) shows that, while the exact
partonic cross section for the $gg$ channel is approximated reasonably
well by its threshold approximation over the whole $\beta$ range, for the $q \bar{q}$ channel the
approximation breaks down at $\beta \sim
0.3$. This points to a more problematic behaviour of the
quark-antiquark channel.  
Therefore one might conclude that the small
error of the NNLO$_{\text{app}}$ prediction at the Tevatron is an
underestimate of the true uncertainty, due to the particular behaviour of
the dominant quark-antiquark channel.\footnote{Another indication 
of this comes
  from a study of the choice $\tilde \mu_r=2 m_t$ for the central
  value of the renormalization scale in NNLO$_{\text{app}}$, that is
  identical to the central value of $\mu_h$ in the NNLL results. For
  this choice, the scale-uncertainty increases moderately to
  ${}^{+7.3\%}_{-8.2\%}$ for NNLO$_{\text{app}}$ at the LHC with $\sqrt
  s=7$~TeV, while the error at the Tevatron increases to
  ${}^{+4.9\%}_{-6.8\%}$ and becomes larger than for NNLL$_2$.}
\begin{table}[!t]
\newcommand{\m}{\hphantom{$-$}}
\newcommand{\cc}[1]{\multicolumn{1}{c}{#1}}
\renewcommand{\tabcolsep}{0.8pc} 
\renewcommand{\arraystretch}{1.0} 
\begin{center}
\begin{tabular}{@{}llllll}
\hline  \vspace{-4mm} \\  
     & \hspace{-4mm} \mbox{Tevatron} 
     & \hspace{-3mm} LHC ($\sqrt{s}=7\,$TeV) 
     & \hspace{-3mm} LHC ($\sqrt{s}=14\,$TeV) \\
\hline   \\
 NLL$_2$
     & 
$\{+5.5\%, -7.4\%\}$& $\{+11.7\%,-8.9\%\}$&  $\{+11.6\%,-8.0\%\}$  
\vspace{2mm} \\
     NNLO$_{\text{app}}$&
     $\{+3.8\%, -4.9\%\}$& $\{+7.7\%,-7.4\%\}$&$\{+8.6\%,-7.7\%\}$ 
     \vspace{2mm} \\
     NNLL$_2$
     & 
$\{+4.2\%, -6.5\%\}$& $\{+4.6\%,-4.6\%\}$&  $\{+4.5\%,-4.1\%\}$
 \vspace{2mm} \\ \hline
\end{tabular}\\[2pt]
\end{center}
\caption{ \sf Relative error of NLL$_2$, NNLO$_{\text{app}}$ 
         and NNLL$_{2}$
         at the LHC. Scale uncertainty, 
         resummation ambiguity and the error from the NNLO constant are added in 
         quadrature.}
\label{Tab:RelErrLHC}
\end{table}

\begin{table}[t]
\newcommand{\m}{\hphantom{$-$}}
\newcommand{\cc}[1]{\multicolumn{1}{c}{#1}}
\renewcommand{\tabcolsep}{0.8pc} 
\renewcommand{\arraystretch}{1.0} 
\begin{center}
\begin{tabular}{@{}llllll}
\hline  \vspace{-4mm} \\
     & \hspace{-4mm} \mbox{Channel}
     & \hspace{-3mm} \mbox{Tevatron}
     & \hspace{-3mm} LHC ($\sqrt{s}=7\,$TeV)  \\
\hline   \\[-0.2cm]
\hspace{-7mm}
\phantom{ab} ${\text{NNLO$_{\text{app}}$}}$
     &  $q\overline{q}$
     & \hspace{-5mm} $\phantom{0}6.18^{\,+0.21}_{\,-0.27}{}^{\,+0.07}_{\,-0.07}
        ~\Bigl\{{}^{\,+3.6\%}_{\,-4.5\%}\Bigr\}$
     & \hspace{-5mm} $\phantom{0}28.7^{\,+2.6}_{\,-2.9}{}^{\,+0.4}_{\,-0.4}
        ~\Bigl\{{}^{\,+9.2\%}_{\,-10.2\%}\Bigr\}$
 \vspace{2mm} \\
 \vspace{-2mm} \\
\hspace{-7mm}
\phantom{ab} ${\text{NNLL$_{2}$}}$
     &  $q\overline{q}$
     & \hspace{-5mm} $\phantom{0}6.31^{\,+0.23}_{\,-0.41}{}^{\,+0.12}_{\,-0.20}
       {}^{\,+0.07}_{\,-0.07}
        ~\Bigl\{{}^{\,+4.3\%}_{\,-7.3\%}\Bigr\}$
     & \hspace{-5mm} $\phantom{0}28.7^{\,+2.4}_{\,-3.0}{}^{\,+0.3}_{\,-1.9}
       {}^{\,+0.4}_{\,-0.4}
        ~\Bigl\{{}^{\,+8.5\%}_{\,-12.5\%}\Bigr\}$
 \vspace{2mm} \\
 \hline
 \vspace{-2mm} \\
\hspace{-7mm}
\phantom{ab} ${\text{NNLO$_{\text{app}}$}}$
     &  $gg$
     & \hspace{-5mm} $\phantom{0}0.95^{\,+0.10}_{\,-0.09}{}^{\,+0.02}_{\,-0.02}
        ~\Bigl\{{}^{\,+10.7\%}_{\,-9.7\%}\Bigr\}$
     & \hspace{-5mm} $\phantom{0}133.2^{\,+17.4}_{\,-13.8}{}^{\,+4.3}_{\,-4.3}
        ~\Bigl\{{}^{\,+13.5\%}_{\,-10.9\%}\Bigr\}$
 \vspace{2mm} \\
 \vspace{-2mm} \\
\hspace{-7mm}
\phantom{ab} ${\text{NNLL$_{2}$}}$
     &  $gg$
     & \hspace{-5mm} $\phantom{0}0.98^{\,+0.02}_{\,-0.08}{}^{\,+0.12}_{\,-0.05}
       {}^{\,+0.02}_{\,-0.02}
        ~\Bigl\{{}^{\,+12.6\%}_{\,-9.8\%}\Bigr\}$
     & \hspace{-5mm} $\phantom{0}134.7^{\,+8.2}_{\,-12.5}{}^{\,+3.6}_{\,-3.9}
       {}^{\,+4.3}_{\,-4.3}
        ~\Bigl\{{}^{\,+7.4\%}_{\,-10.2\%}\Bigr\}$
 \vspace{2mm} \\
 \vspace{-2mm} \\
\hspace{-7mm}
\phantom{ab} {\text{NNLL$_{2}$}}
     &  $gg$
     & \hspace{-5mm} $\phantom{0}0.99^{\,+0.06}_{\,-0.08}{}^{\,+0.05}_{\,-0.06}
       {}^{\,+0.02}_{\,-0.02}
        ~\Bigl\{{}^{\,+8.1\%}_{\,-10.3\%}\Bigr\}$
     & \hspace{-5mm} $\phantom{0}134.7^{\,+8.2}_{\,-12.5}{}^{\,+3.6}_{\,-3.9}
       {}^{\,+4.3}_{\,-4.3}
        ~\Bigl\{{}^{\,+7.4\%}_{\,-10.2\%}\Bigr\}$
 \vspace{-2mm} \\
{\small ($\beta_{\text{cut}}=0.54$)}&&&\\[0.2cm]
\hline
\end{tabular}\\[2pt]
\end{center}
\caption{ \sf Results for the $q\overline{q}$- and $gg$-channel, Method 2,
$m_t=173.3\,$GeV, all numbers in pb.
The errors refer to scale variation,
resummation ambiguity (for NNLL$_2$ only) and the NNLO constant term. 
The numbers in brackets
denote the total relative error. Unless stated
otherwise, the default values of $\beta_{\text{cut}}$, Eq. (\ref{eq:betacut}), are used.}
\label{tab:channels}
\end{table}

To summarize, by using a running soft scale within Method 2, we
implement the principle of scale separation in effective theories at
the level of the partonic cross section, rather than at the level of
the hadronic cross section as in Method 1. Threshold logarithms are
resummed locally rather than in an average sense in the region above
$\beta_{\text{cut}}$, while problems related to very small soft scales
for small $\beta$ are avoided.  The sensitivity to the precise value
of $\beta_{\text{cut}}$ is moderate and included in our error
estimate. In Section \ref{sec:results} we use this method 
as our default implementation. 

\subsection{Effect of Coulomb resummation}
\label{sec:Coulombeffect}

In this section we would like to comment on the size of
the bound-state contributions and Coulomb resummation,
which were not (or only partially) included in previous works
on $t \bar{t}$ threshold resummation. 

Toponium-like bound states may form below the nominal production
threshold due to the strong Coulomb attraction in the colour-singlet 
channel. In practice, the bound-states are smeared into a broad
resonance due to the rapid $t\to bW$ decay. For an inclusive
observable such as the total cross section the resonance effect 
can be computed by neglecting the top decay width and summing 
instead over the series of would-be toponium bound-state poles 
in the Coulomb Green function. The technical aspects are discussed in detail 
in Appendix \ref{app:bound}. In Table \ref{Tab:bstcb}
we show the corresponding contribution to the cross section, denoted
by BS, which are always included in the NNLL results we present. 
The first error denotes the sum in quadrature of scale and resummation
uncertainties, determined as from Method 2, 
and the second one the PDF$+\alpha_s$ error. The bound-state
contributions, which contribute first at $O(\alpha_s^5)$, are rather 
small and make up less than $0.5\%$ of the total cross 
section. The large theoretical error might be due to the more singular 
behaviour of the convolution of the bound-state correction with the soft 
function and the PDFs (c.f. (\ref{eq:BS})), 
which makes the cross section more sensitive to variations 
of $\mu_f$ and $k_s$.

Next, we discuss the effects due to the resummation of Coulomb 
corrections in the continuum ($E>0$). To switch off the Coulomb
effects in the resummation formula, we set  the Coulomb function 
(\ref{JRal}) in Eq. \ref{eq:fact}  to its first-order term in the 
expansion in $\alpha_s$ (see~(\ref{GCR})),
\begin{eqnarray}
 J^{\text{triv}}_R(E)=\frac{m_t^2}{2\pi} \sqrt{\frac{E}{m_t}}~.
 \label{Jctriv}
\end{eqnarray}
We then consider the quantities 
\begin{eqnarray}
 \delta \text{Cb}_1 &=&
   \sigma^{\text{NNLL$_1$}}
  -\sigma^{\text{NNLL$_1$}}_{\text{triv}} \, ,\nonumber\\
   \delta \text{Cb}_2 &=&
   \sigma^{\text{NNLL$_2$}}
  -\sigma^{\text{NNLL$_2$}}_{\text{triv}}~ , 
\end{eqnarray}
where $\sigma^{\text{NNLL$_i$}}$ denotes the total resummed cross
section with the full Coulomb function (but excluding bound-state
effects), matched according to prescription (\ref{eq:match1}) or
(\ref{eq:match2}), and $\sigma^{\text{NNLL$_i$}}_{\text{triv}}$ the
analogous quantity for the trivial potential function (\ref{Jctriv}). 
In this case, the expansion of $\sigma^{\text{NNLL}}_{\text{triv}}$ to
NLO or NNLO order is subtracted in the matching prescriptions 
(\ref{eq:match1}) or (\ref{eq:match2}).
As a consequence of the matching to the fixed-order results, $\delta
\text{Cb}_1$ contains only terms of order $\alpha_s^4$ or higher. In
particular, it contains all the purely Coulomb contributions, and the
interference of Coulomb corrections with hard and soft terms, at
NNLO. $\delta \text{Cb}_2$ contains only terms of order $\alpha_s^5$ or
higher, and gives a measure of the effect of Coulomb resummation
beyond NNLO.
  
In Table \ref{Tab:bstcb} we show the central values for $\delta
\text{Cb}_1$ and $\delta \text{Cb}_2$. We do not show errors, which
are large, due to $\delta \text{Cb}_{1,2}$
being defined as a difference. From the numbers in the table, it is 
clear that the dominant effect 
of Coulomb resummation is accounted for by the terms already 
included in NNLO$_\text{app}$, which coincide with the difference 
$\delta \text{Cb}_1-\delta \text{Cb}_2$. They give a non-negligible 
correction at both Tevatron ($\sim 2\%$) and LHC ($\sim 1-2 \%$).
Beyond NNLO, Coulomb corrections, and their interference with soft and
hard resummation, is very small ($\lesssim 0.5 \%$), as can be 
inferred from the numbers for $\delta \text{Cb}_2$.   

\begin{table}[!t]
\newcommand{\m}{\hphantom{$-$}}
\newcommand{\cc}[1]{\multicolumn{1}{c}{#1}}
\renewcommand{\tabcolsep}{0.8pc} 
\renewcommand{\arraystretch}{1.0} 
\begin{center}
\begin{tabular}{@{}llllll}
\hline  
     & ~Tevatron 
     & ~LHC ($\sqrt{s}=7$ TeV)
     & ~LHC ($\sqrt{s}=14$ TeV) \\
\hline  \\
                ${\text{BS}}$
              & $\phantom{0-}{0.014^{\,+0.011+0.005}_{\,-0.009-0.004}}$
              & $\phantom{0-}{0.67^{\,+0.33+0.12}_{\,-0.31-0.10}}$
              & $\phantom{0-}{3.1^{\,+1.7+0.5}_{\,-1.6-0.4}}$
 \vspace{2mm} \\
                $\delta {\text{Cb}}_1$
              & $\phantom{0}{-0.140}$
              & $\phantom{0+}{2.81}$
              & $\phantom{0+}{7.8}$
 \vspace{2mm} \\
                $\delta {\text{Cb}}_2$
              & $\phantom{0}{-0.052}$
              & $\phantom{0+}{0.13}$
              & $\phantom{0}{-0.3}$
 \vspace{2mm}              \\
\hline
\end{tabular}\\[2pt]
\end{center}
\caption{ \sf Bound-state and Coulomb contributions for $m_{t}=173.3$
  GeV, Method 2. All numbers in pb.}
\label{Tab:bstcb}
\end{table} 

\subsection{Comparison to other NNLO$_{\text{app}}$ and NNLL
  predictions}

Previous results for the total top cross section beyond NLO and 
NLL \cite{Aliev:2010zk,Beneke:2010fm} use theoretical input equivalent
to the NNLO$_{\text{app}}$ result discussed in the present paper (and 
presented in detail in the subsequent section), but the 
complete NNLL threshold resummation (including Coulomb effects) 
performed here is new. 

There exists a complementary approach to resummation, which is 
based on resumming logarithms in various kinematical limits 
different from the partonic threshold $\beta\to 0$ such as 
pair invariant mass~(PIM)
kinematics~\cite{Ahrens:2009uz,Ahrens:2010zv} to compute the invariant
mass distribution, and one-particle inclusive~(1PI)
kinematics~\cite{Kidonakis:2010dk,Ahrens:2011mw} for rapidity and
transverse-momentum distributions.  Results for the total cross
section are then obtained by integrating PIM or 1PI differential cross
sections, and do not include the higher-order Coulomb corrections, 
which, as discussed in section~\ref{sec:Coulombeffect},
amounts to an effect of $-0.13$ pb at the Tevatron and $+3.5$ pb ($+11$
pb) at the LHC with $\sqrt{s}=7$ TeV ($\sqrt{s}=14$ TeV). 
On the other hand, these calculations keep
certain sets of power-suppressed terms in $\beta$ that can be
non-negligible, since the total cross section is dominated by
contributions with $\beta>0.3$. Therefore these approaches complement
our treatment using the partonic threshold limit, and differences in the
predictions can indicate systematical uncertainties of resummation.
In Table~\ref{tab:comparison} we compare the results from these 
approaches, which are either based on NNLO$_{\text{app}}$ expansions
or on NNLL resummations to ours.

The approximations based on 1PI and PIM
kinematics are themselves subject to large ambiguities, analogous to
the ambiguity between $E=\sqrt{\hat s}-2m_t$ and $m_t\beta^2$ in the
threshold limit. 
In~\cite{Ahrens:2010zv,Ahrens:2011mw} it is argued by comparison to
the known fixed-order NLO results that a particular form of the
singular distributions arising in the SCET-based resummation formalism
is preferred over those used conventionally
(e.g.~\cite{Kidonakis:2001nj,Kidonakis:2010dk}). This ambiguity
between the conventional 1PI approximation and the 1PI$_{\text{SCET}}$
implementation of~\cite{Ahrens:2011mw} is reflected in the different predictions in the first two rows of Table~\ref{tab:comparison}. 
It is interesting to note that these kinematical ambiguities appear to be larger in 1PI and PIM kinematics 
than in the threshold approximation where we only found a small ambiguity (c.f. Table~\ref{Tab:ErrorRes}).

\begin{table}[t]
\newcommand{\m}{\hphantom{$-$}}
\newcommand{\cc}[1]{\multicolumn{1}{c}{#1}}
\renewcommand{\tabcolsep}{0.8pc} 
\renewcommand{\arraystretch}{1.0} 
\begin{center}
\begin{tabular}{@{}llccc}
\hline  \vspace{-4mm} \\  
   &  & \hspace{-4mm} \mbox{Tevatron} 
     & \hspace{-3mm} LHC ($\sqrt{s}=7\,$TeV)  
     & \hspace{-3mm} LHC ($\sqrt{s}=14\,$TeV)  \\
\hline   \\
\hspace{-7mm}
\phantom{ab} $\text{NNLO}_{\text{app}}^{1\text{PI}}$&(Ref.~\cite{Kidonakis:2010dk})
 &$7.08^{+0.20}_{-0.24}$  & $163^{+7}_{-5}$& $920^{+60}_{-39}$  \vspace{2mm} \\
$\text{NNLO}_{\text{app}}^{1\text{PI}_{\text{SCET}}}$&(Ref.~\cite{Ahrens:2011mw})
 &$6.63^{+0.00}_{-0.27}$  & $155^{+3}_{-2}$& $851^{+25}_{-5}$  \vspace{2mm} \\
$\text{NNLO}_{\text{app}}^{\text{PIM}_{\text{SCET}}}$& (Ref.~\cite{Ahrens:2010zv})&
$6.62^{+0.05}_{-0.40}$ &$155^{+8}_{-8}$& $860^{+46}_{-43}$\\\hline \\
$\text{NNLL}^{1\text{PI}_{\text{SCET}}}$&(Ref.~\cite{Ahrens:2011mw}) &
$6.55^{+0.16}_{-0.14}$&$150^{+7}_{-7}$ &$824^{+41}_{-44}$\vspace{2mm}\\
$\text{NNLL}^{\text{PIM}_{\text{SCET}}}$& (Ref.~\cite{Ahrens:2010zv})&
$6.46^{+0.18}_{-0.19}$&$147^{+7}_{-6}$&$811^{+45}_{-42}$\vspace{2mm} \\
\hline\\
$\text{NNLL}_2$&this work &$7.22^{+0.31}_{-0.47}$&$163^{+7}_{-8}$&
$896^{+40}_{-37}$ \vspace{2mm} \\
\hline
\end{tabular}\\[2pt]
\end{center}
\caption{ \sf Comparison to previous NNLO$_{\text{app}}$ and NNLL predictions for the central value $\mu_f=m_t$ and  MSTW08NNLO PDFs, all numbers in pb. In the results of~\cite{Kidonakis:2010dk} $m_t=173\,\text{GeV}$, in those of~\cite{Ahrens:2010zv,Ahrens:2011mw} $m_t=173.1\text{GeV}$. Only theory errors, excluding PDF and $\alpha_s$ errors, are shown. The NNLL$_2$ theory errors from table~\ref{tab:method2} have been added in quadrature.}
\label{tab:comparison}
\end{table} 

From the results in Table~\ref{tab:comparison} it is seen that our
results agree within the quoted errors with those of
Kidonakis~\cite{Kidonakis:2010dk} in 1PI kinematics. While the tension
with the predictions of Ahrens
et. al.~\cite{Ahrens:2010zv,Ahrens:2011mw} is larger, the results for
the LHC agree within the given errors, if the size of the Coulomb
corrections included in our results is taken into account. The size of
the theory uncertainties quoted by the different groups are comparable
at the LHC (except for the very small error assigned to
$\text{NNLO}_{\text{app}}^{1\text{PI}_{\text{SCET}}}$), however our
theory error includes an estimate of the missing NNLO constant and the
resummation ambiguities, which are not included by the other groups.
At the Tevatron, the agreement with the results
of~\cite{Ahrens:2010zv,Ahrens:2011mw} is worse, in particular for the
resummed results. In this case our estimate of the theory error is
more conservative than that of the other groups.\footnote{The quoted
  results from~\cite{Kidonakis:2010dk} for the Tevatron include an
  independent variation of $\mu_f$ and $\mu_r$ while the
  NNLO$_{\text{app}}$ results from~\cite{Ahrens:2010zv,Ahrens:2011mw}
  set $\mu_f=\mu_r$. Adding separate variations of both scales in
  quadrature and averaging the $1\text{PI}_{\text{SCET}}$ and
  PIM$_{\text{SCET}}$ kinematics, Ahrens et.al.~\cite{Ahrens:2011px}
  recently obtained a slightly increased error estimated
  $\sigma_{t\bar t}^{\text{NNLO}_{\text{app}}}=6.63^{+0.07}_{-0.41}$
  pb. It is interesting to note that the difference between the two
  kinematics is larger for the resummed cross sections than for the
  NNLO$_{\text{app}}$-results combined in the prediction
  from~\cite{Ahrens:2011px}.} The larger discrepancies between
results obtained using different kinematical approximations might
again point to a worse behaviour of the threshold approximation for the
quark-antiquark channel dominant at the Tevatron, as discussed already
at the end of Section~\ref{sec:method2}.


\section{Detailed numerical results}
\label{sec:results}

In this section we present
numerical results for the combined NNLL resummation of soft and
Coulomb effects for the total $t\overline{t}$ cross section at hadron
colliders.  This requires a choice of the parton distribution
functions used in the convolution of the partonic cross section with
the parton luminosity.  Available fits for PDFs at NNLO accuracy are
the sets of MSTW2008NNLO~\cite{Martin:2009iq},
JR09~\cite{JimenezDelgado:2008hf}, ABKM09~\cite{Alekhin:2009ni},
HERAPDF1.0~\cite{Aaron:2009wt} and NNPDF2.1~\cite{Ball:2011uy}. Recent
discussions of the impact of the differences among the different PDF fits
on the top-pair cross section can be found
in~\cite{Watt:2011kp,Ball:2011uy}, where it was shown that the MSTW and
NNPDF sets yield consistent results, whereas there are larger
differences between the other sets, partly due to different values of the
strong coupling constant obtained by the different groups.  For our
results we use the MSTW2008NNLO PDF sets at $90\%$
confidence level and the associated value of the strong coupling constant
\cite{Martin:2009iq},
\begin{eqnarray}
 \alpha_s(M_Z^2)=0.1171^{+0.0034}_{-0.0034}~. \label{alsval}
\end{eqnarray}
Besides numbers for the NNLL cross section, we also present the NLO 
result~\cite{Nason:1987xz,Czakon:2008ii} and NNLO$_{\text{app}}$, 
which adds to NLO the threshold expansion\footnote{Note that we apply a strict
  threshold approximation for $\text{NNLO}_{\text{app}}$, i.e.  we
  discard all terms scaling as $\mathcal{O}(\alpha_s^2 \beta^0)$ or
  higher with respect to the leading Born cross section. Consequently,
  our central value for $\text{NNLO}_{\text{app}}$ differs, albeit
  slightly, from the one obtained e.g. using {\sf HATHOR},
  \cite{Aliev:2010zk}, since there ${\cal O}(\alpha_s^2 \beta^2)$
  terms from the $gq$--channel are included. Furthermore, terms of the
  form $\text{const.}\times\ln\left({\mu_f}/{m_t}\right)$,
  although known exactly at NNLO, are not included, in the spirit of a
  strict threshold expansion.} of the 
NNLO correction\cite{Beneke:2009ye} (see Appendix~\ref{app:expansions}
for a summary of the analytical expressions) to illustrate the 
significance of higher-order results. For NLO we use the 
NLO PDFs and the associated value
$\alpha_s(M_Z^2)=0.1202^{+0.0032}_{-0.0039}$.  In all three cases, we
determine the joint PDF$+\alpha_s$ uncertainty using the method
described in \cite{Martin:2009bu}.  Following~\cite{Cacciari:2008zb} the theoretical
uncertainty of the fixed-order NLO and NNLO$_{\text{app}}$
approximations is determined by setting the central values of the
renormalization and factorization scales,
$\{\tilde{\mu}_r,~\tilde{\mu}_f\}$, to $m_t$ and varying both scales
simultaneously in the interval $[\tilde{\mu}_i/2,2\tilde{\mu}_i]$, imposing
the additional constraint $1/2\le \mu_r/\mu_f \le 2$.

For the NNLL resummed cross section we choose the $\beta$-dependent
soft scale (\ref{eq:mus-low}), (\ref{eq:mus-up}), and the
matching condition \eqref{eq:match2}, denoting the corresponding 
result by NNLL$_2$. The error
estimate follows the procedure detailed in
Section~\ref{sec:method2}. The errors from scale variation and
the resummation ambiguities are added in quadrature, while the uncertainty from
variation of the unknown NNLO constant is given as a separate error
both for NNLL$_2$ and NNLO$_{\text{app}}$.  The default value of the
pole mass of the top is chosen as $m_t=173.3$~GeV~\cite{TopMass2010}.
Results for $m_t=165\ldots 180$~GeV are given in Table
\ref{tab:mt-tev} for Tevatron, and in Tables \ref{tab:mt-lhc7} and
\ref{tab:mt-lhc14} for LHC. In Sections \ref{sec:Tevatron} and
\ref{sec:LHC} we also compare our results to recent measurements at
Tevatron and LHC, while in Section \ref{sec:mass} we illustrate the
impact on the extraction of the top mass from
measurements of the cross section.

\subsection{Tevatron}
\label{sec:Tevatron}

\begin{table}[p]
\newcommand{\m}{\hphantom{$-$}}
\newcommand{\cc}[1]{\multicolumn{1}{c}{#1}}
\renewcommand{\tabcolsep}{0.8pc} 
\renewcommand{\arraystretch}{1.0} 
\begin{center}
\begin{tabular}{@{}llllll}
\hline  \vspace{-4mm} \\  
       \phantom{ab} ${\text{NLO}}$
     & \phantom{ab} ${\text{NNLO}_{\text{app}}}$
     & \phantom{ab} ${\text{NNLL$_{2}$}}$  \\
\hline   \\
       $\phantom{0}{6.68^{\,+0.36+0.51}_{\,-0.75-0.45}}$
     & $\phantom{0}{7.06^{\,+0.25+0.10+0.69}_{\,-0.33-0.10-0.53}}$
     & $\phantom{0}{7.22^{\,+0.29+0.10+0.71}_{\,-0.46-0.10-0.55}}$
 \vspace{2mm} \\
\hline
\end{tabular}\\[2pt]
\caption{ \sf The total top-pair cross section (in pb) 
for $m_{t}=173.3$~GeV at the Tevatron. 
The first set of errors refers to scale variation (scale 
variation+resummation ambiguities for NNLL$_2$),
the last to PDF+$\alpha_s$ error. The second set of errors for 
NNLO$_\text{app}$/NNLL$_2$ arises from
variations of the unknown NNLO constant term.}
\label{Tab:CentT}
\end{center}
\end{table} 
\begin{figure}[p]
\begin{center}
\includegraphics[angle=0, width=.56\textwidth]{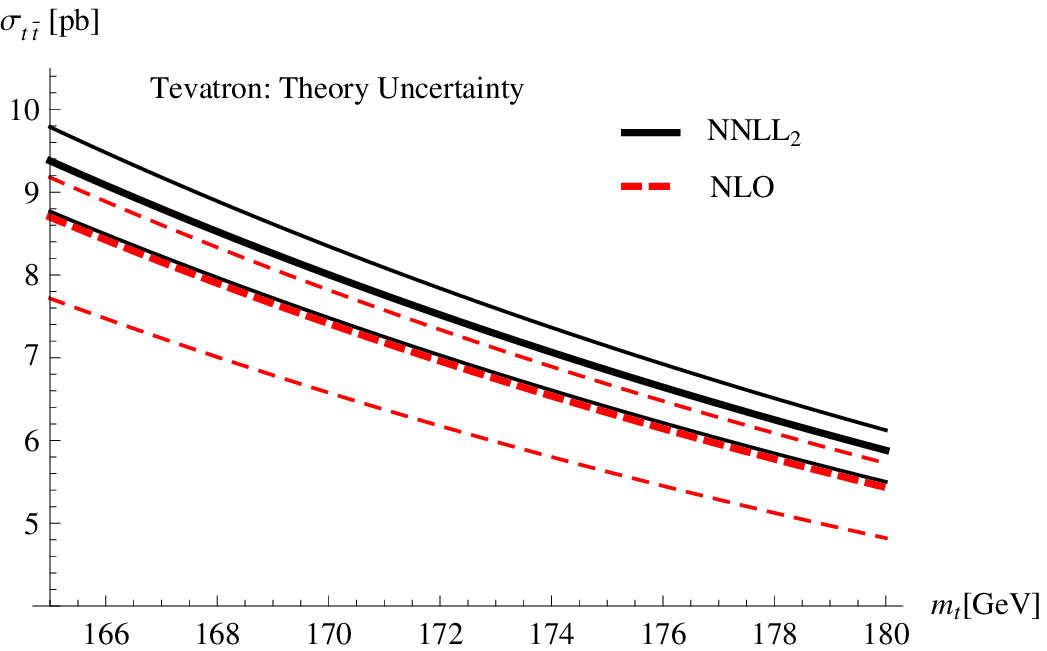}
\includegraphics[angle=0, width=.56\textwidth]{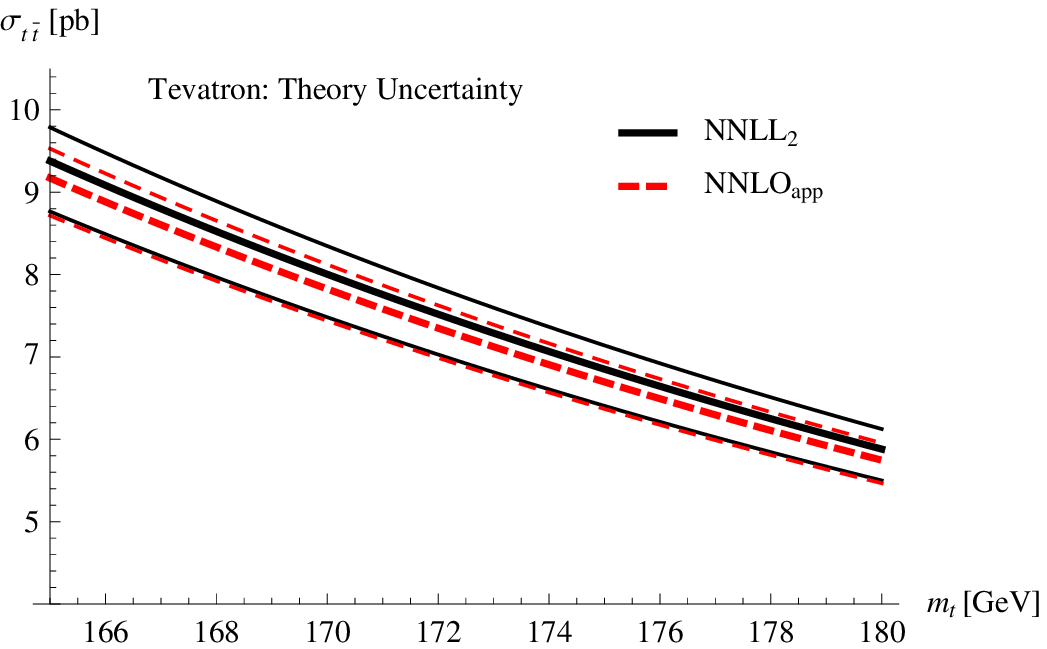}
\end{center}
\caption{\sf Dependence of the total cross section on the top mass at the 
Tevatron. The bands denote the total theory uncertainty, i.e. 
the sum (in quadrature) of the scale and resummation ambiguities, 
and the estimate of the NNLO constant but not the PDF + $\alpha_s$ 
error.}
\label{Fig:MTTev}
\end{figure}

The cross section results for Tevatron kinematics 
are summarized in Table~\ref{Tab:CentT}. The difference
between NNLL$_2$ and NLO amounts to about $8\%$ of the NLO
result. This correction is an interplay of genuine 
resummation effects, which amount to about  $12\%$ of the NLO result,
and of switching from NLO to NNLO PDFs, which lowers the NLO cross
section from the value given in the table to $6.46\, \text{pb}$. As
evident from Table~\ref{Tab:CentT}, the bulk of the resummation
corrections are accounted for by the fixed-order soft and Coulomb 
terms at $O(\alpha_s^4)$  (NNLO$_{\text{app}}$), which give a $9.5 \%$
correction to the NLO result, though higher-order contributions 
(from NNLO$_{\text{app}}$ to NNLL$_2$)
amount to a non-negligible 
$3.5 \%$. The total theoretical error of the cross section is
reduced at both NNLO$_\text{app}$ and NNLL$_2$ compared to the 
NLO result, and amounts to $\{+4.2\%,-6.5\%\}$ for the NNLL$_2$ result,
and $\{+3.8\%,-4.9\%\}$ for NNLO$_\text{app}$. As discussed in 
Section~\ref{sec:method2}, the counter-intuitive increase of the 
uncertainty of the resummed result compared to the fixed-order NNLO 
is related to the behaviour of the quark-antiquark induced partonic
subprocess. Interestingly, the
relative PDF$+\alpha_s$ uncertainty increases when going from NLO to
NNLO$_{\text{app}}$/NNLL$_2$, an
effect which is also observed at the LHC and has been found in
\cite{Watt:2011kp} as well.  The ambiguity arising from the unknown constant
(second error) is small compared to the scale, resummation and 
PDF+$\alpha_s$ uncertainties, which dominate the total error at NNLL$_2$. 

\begin{table}[!t]
\newcommand{\m}{\hphantom{$-$}}
\newcommand{\cc}[1]{\multicolumn{1}{c}{#1}}
\renewcommand{\tabcolsep}{0.8pc} 
\renewcommand{\arraystretch}{1.0} 
\caption{ \sf Total cross sections in pb at the Tevatron for 
$m_{t}=165\ldots 180$~GeV. The errors denote the scale variation 
(scale variation+resummation
ambiguities for NNLL$_2$), the NNLO constant variation 
(for NNLO$_{\text{app}}$ and NNLL$_2$) and the PDF$+\alpha_s$ error.}
\label{tab:mt-tev}
\begin{center}
\begin{tabular}{llll}
\hline  
$m_t$~[GeV]  &${\text{NLO}}$&${\text{NNLO}_{\text{app}}}$&${\text{NNLL$_{2}$}}$
\\\hline\\[-3mm] 
      165
     & $\phantom{0}{{8.70^{\,+0.48+0.66}_{\,-0.98-0.61}}}$
     & $\phantom{0}{{9.17^{\,+0.33+0.13+0.93}_{\,-0.44-0.13-0.71}}}$
     & $\phantom{0}{{9.38^{\,+0.39+0.13+0.96}_{\,-0.60-0.13-0.73}}}$
      \vspace{2mm} \\
      166
     & $\phantom{0}{{8.42^{\,+0.46+0.64}_{\,-0.95-0.58}}}$
     & $\phantom{0}{{8.88^{\,+0.32+0.13+0.89}_{\,-0.42-0.13-0.68}}}$
     & $\phantom{0}{{9.08^{\,+0.37+0.13+0.92}_{\,-0.58-0.13-0.71}}}$
      \vspace{2mm} \\
      167
     & $\phantom{0}{{8.16^{\,+0.45+0.62}_{\,-0.92-0.56}}}$
     & $\phantom{0}{{8.60^{\,+0.31+0.12+0.86}_{\,-0.41-0.12-0.66}}}$
     & $\phantom{0}{{8.80^{\,+0.36+0.12+0.89}_{\,-0.56-0.12-0.68}}}$
      \vspace{2mm} \\
      168
     & $\phantom{0}{{7.90^{\,+0.43+0.60}_{\,-0.89-0.54}}}$
     & $\phantom{0}{{8.33^{\,+0.30+0.12+0.83}_{\,-0.39-0.12-0.63}}}$
     & $\phantom{0}{{8.52^{\,+0.35+0.12+0.86}_{\,-0.54-0.12-0.66}}}$
      \vspace{2mm} \\
      169
     & $\phantom{0}{{7.65^{\,+0.42+0.58}_{\,-0.86-0.52}}}$
     & $\phantom{0}{{8.07^{\,+0.29+0.11+0.80}_{\,-0.38-0.11-0.61}}}$
     & $\phantom{0}{{8.26^{\,+0.34+0.11+0.83}_{\,-0.53-0.11-0.64}}}$
      \vspace{2mm} \\
      170
     & $\phantom{0}{{7.41^{\,+0.41+0.56}_{\,-0.84-0.51}}}$
     & $\phantom{0}{{7.82^{\,+0.28+0.11+0.78}_{\,-0.37-0.11-0.59}}}$
     & $\phantom{0}{{8.00^{\,+0.32+0.11+0.80}_{\,-0.51-0.11-0.61}}}$
      \vspace{2mm} \\
      171
     & $\phantom{0}{{7.18^{\,+0.39+0.54}_{\,-0.81-0.49}}}$
     & $\phantom{0}{{7.58^{\,+0.27+0.11+0.75}_{\,-0.36-0.11-0.57}}}$
     & $\phantom{0}{{7.76^{\,+0.31+0.11+0.77}_{\,-0.49-0.11-0.59}}}$
      \vspace{2mm} \\
      172
     & $\phantom{0}{{6.96^{\,+0.38+0.53}_{\,-0.78-0.47}}}$
     & $\phantom{0}{{7.35^{\,+0.26+0.10+0.72}_{\,-0.35-0.10-0.55}}}$
     & $\phantom{0}{{7.52^{\,+0.30+0.10+0.75}_{\,-0.48-0.10-0.57}}}$
      \vspace{2mm} \\
      173
     & $\phantom{0}{{6.74^{\,+0.37+0.51}_{\,-0.76-0.46}}}$
     & $\phantom{0}{{7.12^{\,+0.25+0.10+0.70}_{\,-0.34-0.10-0.53}}}$
     & $\phantom{0}{{7.29^{\,+0.29+0.10+0.72}_{\,-0.46-0.10-0.55}}}$
      \vspace{2mm} \\
      174
    & $\phantom{0}{{6.54^{\,+0.36+0.50}_{\,-0.74-0.44}}}$
     & $\phantom{0}{{6.91^{\,+0.24+0.09+0.67}_{\,-0.32-0.09-0.51}}}$
     & $\phantom{0}{{7.07^{\,+0.28+0.09+0.70}_{\,-0.45-0.09-0.53}}}$
      \vspace{2mm} \\
      175
     & $\phantom{0}{{6.34^{\,+0.34+0.48}_{\,-0.71-0.42}}}$
     & $\phantom{0}{{6.70^{\,+0.23+0.09+0.65}_{\,-0.31-0.09-0.49}}}$
     & $\phantom{0}{{6.85^{\,+0.27+0.09+0.67}_{\,-0.44-0.09-0.52}}}$
      \vspace{2mm} \\
      176
     & $\phantom{0}{{6.14^{\,+0.33+0.47}_{\,-0.69-0.41}}}$
     & $\phantom{0}{{6.49^{\,+0.22+0.09+0.63}_{\,-0.30-0.09-0.48}}}$
     & $\phantom{0}{{6.64^{\,+0.26+0.09+0.65}_{\,-0.42-0.09-0.50}}}$
      \vspace{2mm} \\
      177
     & $\phantom{0}{{5.96^{\,+0.32+0.45}_{\,-0.67-0.40}}}$
     & $\phantom{0}{{6.30^{\,+0.22+0.09+0.61}_{\,-0.30-0.09-0.46}}}$
     & $\phantom{0}{{6.44^{\,+0.26+0.09+0.63}_{\,-0.41-0.09-0.48}}}$
      \vspace{2mm} \\
      178
     & $\phantom{0}{{5.78^{\,+0.31+0.44}_{\,-0.65-0.38}}}$
     & $\phantom{0}{{6.11^{\,+0.21+0.08+0.59}_{\,-0.29-0.08-0.45}}}$
     & $\phantom{0}{{6.25^{\,+0.25+0.08+0.61}_{\,-0.40-0.08-0.46}}}$
      \vspace{2mm} \\
      179
     & $\phantom{0}{{5.60^{\,+0.30+0.43}_{\,-0.63-0.37}}}$
     & $\phantom{0}{{5.92^{\,+0.20+0.08+0.57}_{\,-0.28-0.08-0.43}}}$
     & $\phantom{0}{{6.06^{\,+0.24+0.08+0.59}_{\,-0.39-0.08-0.45}}}$
      \vspace{2mm} \\
      180
     & $\phantom{0}{{5.43^{\,+0.29+0.42}_{\,-0.61-0.36}}}$
     & $\phantom{0}{{5.75^{\,+0.19+0.08+0.55}_{\,-0.27-0.08-0.42}}}$
     & $\phantom{0}{{5.88^{\,+0.23+0.08+0.57}_{\,-0.37-0.08-0.43}}}$ \\
\hline
\end{tabular}\\[2pt]
\end{center}
\end{table} 

The dependence of the total cross section on $m_t$ is plotted in
Figure \ref{Fig:MTTev} and given explicitly in Table~\ref{tab:mt-tev}.
The relative theoretical errors in the mass range $m_t=165\ldots
180$~GeV differ only on the permille level from those for
$m_t=173.3$~GeV, while the PDF$+\alpha_s$ uncertainty decreases
slightly for larger masses, resulting in a practically constant
overall uncertainty for NNLL$_2$ of $^{+11.1\%}_{-10.2\%}$ at
$m_t=165$~GeV and $^{+10.5\%}_{-9.7\%}$ at $m_t=180$~GeV.

The NNLL$_2$ results are in good agreement with measurements performed
at the Tevatron. The D0 experiment obtains 
$\sigma_{t\bar t}= 7.56^{+0.63}_{-0.56}$ pb from combining
measurements of the dilepton 
and lepton plus jets final states with up to $5.6\text{fb}^{-1}$ of  
data~\cite{Abazov:2011cq}.
The CDF collaboration quotes  
$\sigma_{t\bar t}= 7.50^{+0.48}_{-0.48}$ pb from 
a combination~\cite{cdf:sigmatt} of the dilepton, the lepton plus jets and the all-hadronic channel using up to $4.6\text{fb}^{-1}$ of data.
A value of  $\sigma_{t\bar t}= 7.70^{+0.52}_{-0.52}$ pb has been obtained from a measurement of the ratio of the 
top-pair and $Z$-boson production cross sections~\cite{Aaltonen:2010ic}.
In these measurements a top mass $m_t=172.5$~GeV has been used in the 
kinematical reconstruction of the top-events.

\subsection{LHC}
\label{sec:LHC}

\begin{table}[!t]
\newcommand{\m}{\hphantom{$-$}}
\newcommand{\cc}[1]{\multicolumn{1}{c}{#1}}
\renewcommand{\tabcolsep}{0.8pc} 
\renewcommand{\arraystretch}{1.0} 
\begin{center}
\begin{tabular}{@{}llllll}
\hline  \vspace{-4mm} \\
        $\sqrt{s}$
     &  \phantom{ab} ${\text{NLO}}$
     & \phantom{ab} ${\text{NNLO}_{\text{app}}}$
     & \phantom{ab} ${\text{NNLL$_{2}$}}$  \\
\hline  \\
                7
              & $\phantom{0}{158.1^{\,+19.5+13.9}_{\,-21.2-13.1}}$
              & $\phantom{0}{161.1^{\,+11.4+4.7+15.2}_{\,-10.9-4.7-14.5}}$
              & $\phantom{0}{162.6^{\,+5.7+4.7+15.4}_{\,-5.9-4.7-14.7}}$
 \vspace{2mm} \\
                14
              & $\phantom{0}{884^{\,+107+65}_{\,-106-58}}$
              & $\phantom{0}{891^{\,+71+28+64}_{\,-63-28-63}}$
              & $\phantom{0}{896^{\,+29+28+65}_{\,-24-28-64}}$
 \vspace{2mm} \\
\hline
\end{tabular}\\[2pt]
\end{center}
\caption{ \sf Predictions for the total cross section (in pb) 
for $m_{t}=173.3$~GeV at the LHC. The errors are defined as 
in Table~\ref{Tab:CentT}.}
\label{Tab:CentL}
\end{table} 

\begin{figure}[!p]
\begin{center}
\includegraphics[angle=0, width=0.55\textwidth]{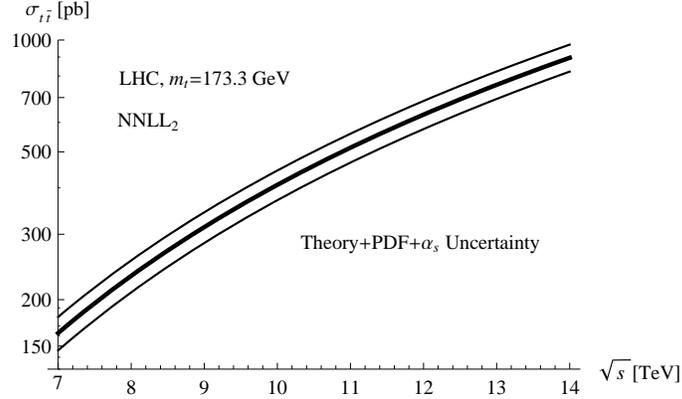}
\end{center}
\caption{\sf Dependence of the total cross section on $\sqrt{s}$ 
at the LHC. The central line is obtained with our NNLL$_2$ prediction,
while the band corresponds to the sum (in quadrature) of theory and PDF+$\alpha_s$ errors.}
\label{Fig:SLHCB}
\end{figure}

Our results for the LHC with centre-of-mass energies $\sqrt{s}=7$ and
$14$~TeV are shown in Table~\ref{Tab:CentL}. In Figure \ref{Fig:SLHCB} 
we plot the dependence on $\sqrt{s}$ of the NNLL$_2$ cross section. 
The difference between the NNLL$_2$ and NLO results is smaller than at
Tevatron, and amounts to $3\%$ ($1\%$) at $\sqrt{s}=7$ TeV ($\sqrt{s}=14$ TeV).
Again, the genuine effect of resummation is partly hidden by the switch to NNLO
PDFs for the resummed results. Using NNLO PDFs for the NLO
result would yield $148.9$ pb and $837$ pb for the NLO cross section at
7 and 14 TeV, which differ from the NNLL$_2$ results by  $9 \%$ and $7\%$,
respectively. As for the Tevatron, the bulk of the corrections beyond NLO come
from the ${\cal O}(\alpha_s^2)$ terms, with higher order contributions at 
the $1 \%$ level.

The total theoretical error is largely reduced from 
NLO to NNLO$_\text{app}$, and from NNLO$_\text{app}$ to NNLL$_2$ as
seen from Table~\ref{Tab:CentL}. For NNLL$_2$ it amounts to a
$\{+4.5\%,-4.6\%\}$ residual uncertainty at $\sqrt{s}=7$ TeV, 
and $\{+4.5\%,-4.1\%\}$ at 
$\sqrt{s}=14$ TeV. Contrary to the Tevatron, the
uncertainty arising from the variation of the unknown constant is
comparable to the remaining theoretical error. This is a consequence of our
estimate $|C_{pp',R}^{(2) \text{est.}}|=(C_{pp',R}^{(1)})^2$, see 
Eq. (\ref{eq:constant}), which
gives larger values at LHC, where the $gg$ channel is dominant. At
NNLL$_2$ the dominant source of uncertainty is the PDF+$\alpha_s$
error, which is $\pm 9\%$.
The total relative uncertainty, given by the sum in quadrature of 
all sources of error, is reduced from $\pm~15.5\%$ (NLO)
to approximately $\pm~10.5\%$ (NNLL$_2$) at $7$~TeV .  
A similar picture holds for 14 TeV, with a smaller total relative error,
$\pm~8.5\%$.

\begin{figure}[!p]
\begin{center}
\includegraphics[angle=0, width=0.56\textwidth]{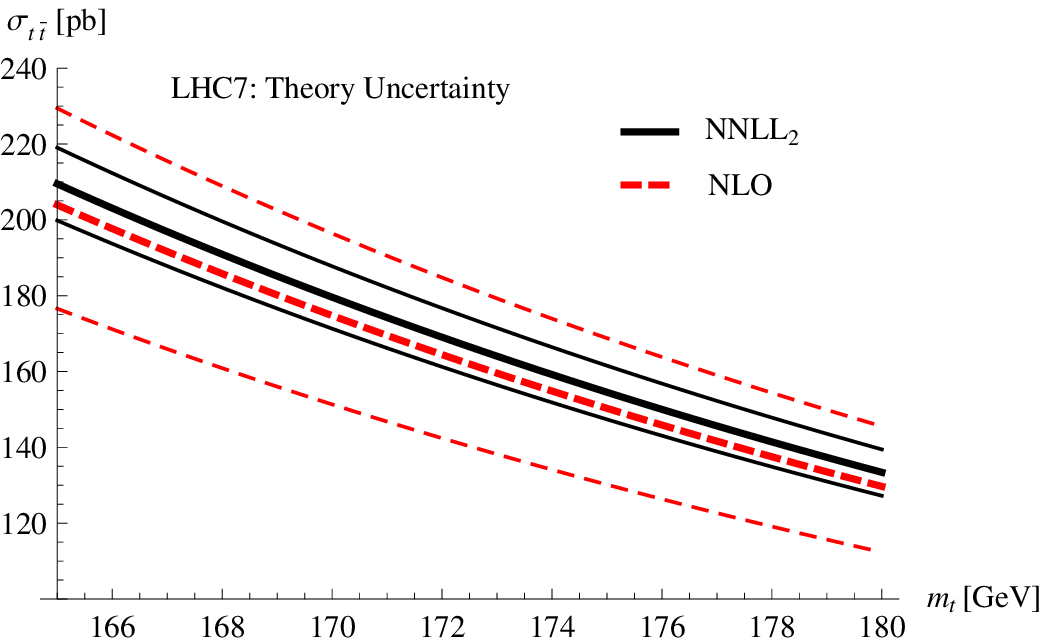}
\includegraphics[angle=0, width=0.56\textwidth]{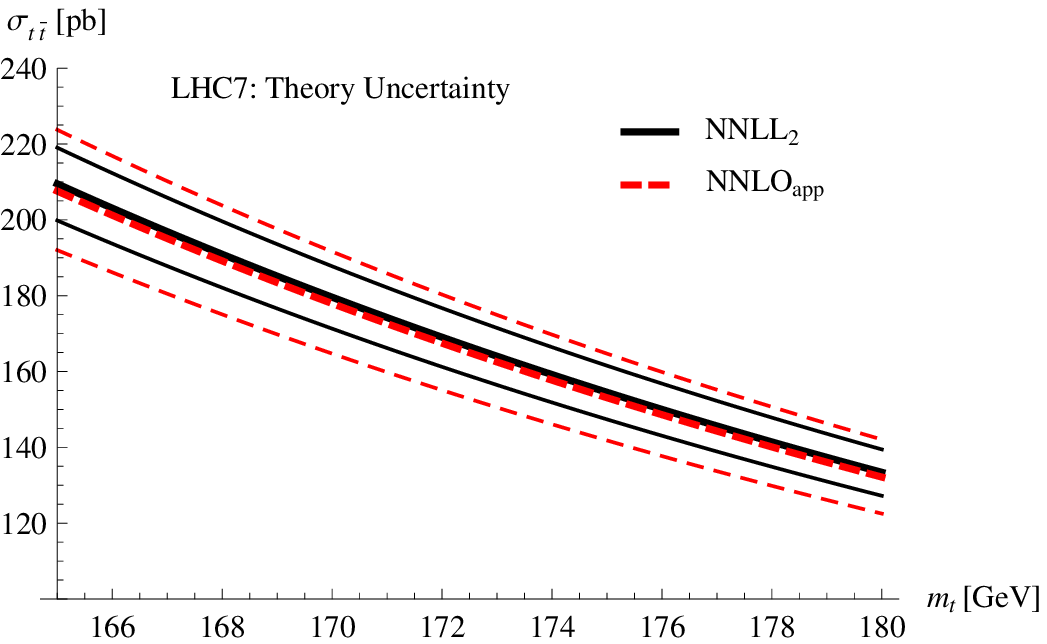}
\end{center}
\caption{\sf Dependence of the total cross section on the mass at the
  LHC with $\sqrt s=7$ TeV.  The bands denote the total theory
  uncertainty, i.e. the sum (in quadrature) of the scale and
  resummation ambiguities, and the estimate of the NNLO constant 
  but not the PDF + $\alpha_s$ error.}
\label{Fig:MTLHC7}
\end{figure}

The dependence  of the cross section on $m_t$ for $\sqrt{s}=7$~TeV 
in different approximations is plotted in Figure \ref{Fig:MTLHC7}. 
Additionally, numerical results for $m_t=165\ldots 180$~GeV for NLO,
NNLO$_{\text{app}}$ and NNLL$_2$ at the LHC can be found in Tables
\ref{tab:mt-lhc7} and \ref{tab:mt-lhc14}.  As for the Tevatron, the
relative theory error is constant to a good accuracy in the considered
mass interval, while the PDF+$\alpha_s$ error increases slightly,
resulting in a practically constant overall uncertainty ranging
from ${}^{+10.4\%}_{-10.0\%}$ to ${}^{+10.6\%}_{-10.3\%}$ at $\sqrt{s}=7$~TeV 
and from ${}^{+8.4\%}_{-8.2\%}$ to ${}^{+8.6\%}_{-8.3\%}$ at $\sqrt{s}=14$~TeV.

\begin{table}[!tb]
\caption{ \sf Total cross sections in pb at the LHC ($\sqrt{s}=7$ TeV) for $m_{t}=165\ldots 180$~GeV. The errors denote the scale variation (scale variation+resummation
ambiguities for NNLL$_2$), the NNLO constant variation (for NNLO$_{\text{app}}$ and NNLL$_2$) and the PDF$+\alpha_s$ error.}
\label{tab:mt-lhc7}
\newcommand{\m}{\hphantom{$-$}}
\newcommand{\cc}[1]{\multicolumn{1}{c}{#1}}
\renewcommand{\tabcolsep}{0.8pc} 
\renewcommand{\arraystretch}{1.0} 
\begin{center}
\begin{tabular}{llll}
\hline  
$m_t$~[GeV]  &${\text{NLO}}$&${\text{NNLO}_{\text{app}}}$&${\text{NNLL$_{2}$}}$
\\\hline\\[-3mm] 
      165
     & $\phantom{0}{{203.9^{\,+25.5+17.8}_{\,-27.4-16.7}}}$
     & $\phantom{0}{{207.6^{\,+15.0+6.1+19.3}_{\,-14.3-6.1-18.6}}}$
     & $\phantom{0}{{209.5^{\,+7.3+6.1+19.6}_{\,-7.5-6.1-18.8}}}$
      \vspace{2mm} \\
      166
     & $\phantom{0}{{197.7^{\,+24.7+17.3}_{\,-26.5-16.2}}}$
     & $\phantom{0}{{201.2^{\,+14.5+5.9+18.8}_{\,-13.9-5.9-18.0}}}$
     & $\phantom{0}{{203.1^{\,+7.1+5.9+19.0}_{\,-7.3-5.9-18.2}}}$
      \vspace{2mm} \\
      167
     & $\phantom{0}{{191.6^{\,+23.9+16.8}_{\,-25.7-15.7}}}$
     & $\phantom{0}{{195.1^{\,+14.0+5.7+18.2}_{\,-13.4-5.7-17.5}}}$
     & $\phantom{0}{{196.9^{\,+6.9+5.7+18.5}_{\,-7.1-5.7-17.7}}}$
      \vspace{2mm} \\
      168
     & $\phantom{0}{{185.8^{\,+23.1+16.3}_{\,-24.9-15.3}}}$
     & $\phantom{0}{{189.2^{\,+13.5+5.5+17.7}_{\,-13.0-5.5-17.0}}}$
     & $\phantom{0}{{190.9^{\,+6.7+5.5+17.9}_{\,-6.9-5.5-17.2}}}$
      \vspace{2mm} \\
      169
     & $\phantom{0}{{180.2^{\,+22.4+15.8}_{\,-24.2-14.9}}}$
     & $\phantom{0}{{183.5^{\,+13.1+5.4+17.2}_{\,-12.6-5.4-16.5}}}$
     & $\phantom{0}{{185.2^{\,+6.5+5.4+17.4}_{\,-6.7-5.4-16.7}}}$
      \vspace{2mm} \\
      170
     & $\phantom{0}{{174.7^{\,+21.7+15.3}_{\,-23.4-14.4}}}$
     & $\phantom{0}{{177.9^{\,+12.7+5.2+16.7}_{\,-12.2-5.2-16.0}}}$
     & $\phantom{0}{{179.6^{\,+6.3+5.2+16.9}_{\,-6.5-5.2-16.2}}}$
      \vspace{2mm} \\
      171
     & $\phantom{0}{{169.5^{\,+21.0+14.9}_{\,-22.7-14.0}}}$
     & $\phantom{0}{{172.6^{\,+12.3+5.0+16.2}_{\,-11.8-5.0-15.6}}}$
     & $\phantom{0}{{174.2^{\,+6.1+5.0+16.5}_{\,-6.3-5.0-15.8}}}$
      \vspace{2mm} \\
      172
     & $\phantom{0}{{164.4^{\,+20.3+14.5}_{\,-22.0-13.6}}}$
     & $\phantom{0}{{167.5^{\,+11.9+4.9+15.8}_{\,-11.4-4.9-15.1}}}$
     & $\phantom{0}{{169.0^{\,+6.0+4.9+16.0}_{\,-6.1-4.9-15.3}}}$
      \vspace{2mm} \\
      173
     & $\phantom{0}{{159.6^{\,+19.7+14.0}_{\,-21.4-13.3}}}$
     & $\phantom{0}{{162.5^{\,+11.5+4.7+15.4}_{\,-11.0-4.7-14.7}}}$
     & $\phantom{0}{{164.0^{\,+5.8+4.7+15.6}_{\,-6.0-4.7-14.9}}}$
      \vspace{2mm} \\
      174
     & $\phantom{0}{{154.8^{\,+19.1+13.6}_{\,-20.7-12.9}}}$
     & $\phantom{0}{{157.7^{\,+11.1+4.5+14.9}_{\,-10.7-4.5-14.3}}}$
     & $\phantom{0}{{159.2^{\,+5.6+4.5+15.1}_{\,-5.8-4.5-14.4}}}$
      \vspace{2mm} \\
      175
     & $\phantom{0}{{150.3^{\,+18.5+13.3}_{\,-20.1-12.5}}}$
     & $\phantom{0}{{153.1^{\,+10.8+4.4+14.5}_{\,-10.4-4.4-13.9}}}$
     & $\phantom{0}{{154.5^{\,+5.5+4.4+14.7}_{\,-5.6-4.4-14.0}}}$
      \vspace{2mm} \\
      176
     & $\phantom{0}{{145.9^{\,+17.9+12.9}_{\,-19.5-12.2}}}$
     & $\phantom{0}{{148.6^{\,+10.4+4.3+14.1}_{\,-10.0-4.3-13.5}}}$
     & $\phantom{0}{{150.0^{\,+5.3+4.3+14.3}_{\,-5.5-4.3-13.6}}}$
      \vspace{2mm} \\
      177
     & $\phantom{0}{{141.6^{\,+17.4+12.5}_{\,-19.0-11.9}}}$
     & $\phantom{0}{{144.3^{\,+10.1+4.1+13.7}_{\,-9.7-4.1-13.1}}}$
     & $\phantom{0}{{145.6^{\,+5.2+4.1+13.9}_{\,-5.3-4.1-13.3}}}$
      \vspace{2mm} \\
      178
     & $\phantom{0}{{137.5^{\,+16.9+12.2}_{\,-18.4-11.5}}}$
     & $\phantom{0}{{140.1^{\,+9.8+4.0+13.4}_{\,-9.4-4.0-12.7}}}$
     & $\phantom{0}{{141.4^{\,+5.0+4.0+13.5}_{\,-5.2-4.0-12.9}}}$
      \vspace{2mm} \\
      179
     & $\phantom{0}{{133.5^{\,+16.3+11.8}_{\,-17.9-11.2}}}$
     & $\phantom{0}{{136.1^{\,+9.5+3.9+13.0}_{\,-9.2-3.9-12.4}}}$
     & $\phantom{0}{{137.4^{\,+4.9+3.9+13.2}_{\,-5.0-3.9-12.5}}}$
      \vspace{2mm} \\
      180
     & $\phantom{0}{{129.7^{\,+15.9+11.5}_{\,-17.3-10.9}}}$
     & $\phantom{0}{{132.2^{\,+9.2+3.8+12.6}_{\,-8.9-3.8-12.0}}}$
     & $\phantom{0}{{133.4^{\,+4.7+3.8+12.8}_{\,-4.9-3.8-12.2}}}$ \\
\hline
\end{tabular}\\[2pt]
\end{center}
\end{table} 

Recently the LHC collaborations presented measurements of the total
$t\bar t$ cross section that approach the experimental accuracy of
the Tevatron. Utilizing kinematic information of lepton plus jets events, the ATLAS experiment obtains~\cite{atlas:sigmattlatest}
$\sigma_{t\bar t}=179.0 \pm 9.8 \text{(stat+syst)}\pm
6.6\text{(lumi)}=179.0 \pm 11.8$ pb using up to $0.7 \text{fb}^{-1}$ of
data, while the CMS collaboration obtains~\cite{CMS:2011yy}
$\sigma_{t\bar t}=154\pm 17\text{(stat+syst)}\pm 6\text{(lumi)}$ pb
from a combination of lepton plus jet and
dileptonic decay channels, using $0.036 \text{fb}^{-1}$.  Both measurements are in agreement with
our predictions within the uncertainties.

\begin{table}[!tb]
\caption{ \sf Total cross sections in pb at the LHC ($\sqrt{s}=14$ TeV) for $m_{t}=165\ldots 180$~GeV. The errors denote the scale variation (scale variation+resummation
ambiguities for NNLL$_2$), the NNLO constant variation (for NNLO$_{\text{app}}$ and NNLL$_2$) and the PDF$+\alpha_s$ error.}
\label{tab:mt-lhc14}
\newcommand{\m}{\hphantom{$-$}}
\newcommand{\cc}[1]{\multicolumn{1}{c}{#1}}
\renewcommand{\tabcolsep}{0.8pc} 
\renewcommand{\arraystretch}{1.0} 
\begin{center}
\begin{tabular}{llll}
\hline  
$m_t$~[GeV]  &${\text{NLO}}$&${\text{NNLO}_{\text{app}}}$&${\text{NNLL$_{2}$}}$
\\\hline\\[-3mm] 
      165
     & $\phantom{0}{{1111^{\,+136+82}_{\,-133-73}}}$
     & $\phantom{0}{{1118^{\,+91+36+80}_{\,-80-36-78}}}$
     & $\phantom{0}{{1124^{\,+36+36+80}_{\,-30-36-79}}}$
      \vspace{2mm} \\
      166
     & $\phantom{0}{{1080^{\,+132+80}_{\,-129-71}}}$
     & $\phantom{0}{{1087^{\,+88+35+77}_{\,-78-35-76}}}$
     & $\phantom{0}{{1093^{\,+35+35+78}_{\,-29-35-77}}}$
      \vspace{2mm} \\
      167
     & $\phantom{0}{{1051^{\,+128+78}_{\,-126-69}}}$
     & $\phantom{0}{{1057^{\,+86+34+75}_{\,-76-34-74}}}$
     & $\phantom{0}{{1064^{\,+34+34+76}_{\,-28-34-75}}}$
      \vspace{2mm} \\
      168
     & $\phantom{0}{{1022^{\,+124+76}_{\,-122-67}}}$
     & $\phantom{0}{{1029^{\,+83+33+73}_{\,-73-33-72}}}$
     & $\phantom{0}{{1035^{\,+33+33+74}_{\,-27-33-73}}}$
      \vspace{2mm} \\
      169
     & $\phantom{00}{{994^{\,+121+73}_{\,-119-65}}}$
     & $\phantom{0}{{1001^{\,+81+32+72}_{\,-71-32-71}}}$
     & $\phantom{0}{{1007^{\,+32+32+72}_{\,-27-32-71}}}$
      \vspace{2mm} \\
      170
     & $\phantom{00}{{967^{\,+117+71}_{\,-116-63}}}$
     & $\phantom{00}{{974^{\,+79+31+70}_{\,-69-31-69}}}$
     & $\phantom{00}{{979^{\,+31+31+70}_{\,-26-31-69}}}$
      \vspace{2mm} \\
      171
     & $\phantom{00}{{941^{\,+114+70}_{\,-113-62}}}$
     & $\phantom{00}{{948^{\,+76+30+68}_{\,-67-30-67}}}$
     & $\phantom{00}{{953^{\,+31+30+69}_{\,-25-30-68}}}$
      \vspace{2mm} \\
      172
     & $\phantom{00}{{916^{\,+111+68}_{\,-110-60}}}$
     & $\phantom{00}{{922^{\,+74+29+66}_{\,-65-29-65}}}$
     & $\phantom{00}{{928^{\,+30+29+66}_{\,-25-29-66}}}$
      \vspace{2mm} \\
      173
     & $\phantom{00}{{892^{\,+108+66}_{\,-107-58}}}$
     & $\phantom{00}{{898^{\,+72+28+64}_{\,-64-28-64}}}$
     & $\phantom{00}{{903^{\,+29+28+65}_{\,-24-28-64}}}$
      \vspace{2mm} \\
      174
     & $\phantom{00}{{868^{\,+105+64}_{\,-104-57}}}$
     & $\phantom{00}{{874^{\,+70+28+63}_{\,-62-28-62}}}$
     & $\phantom{00}{{879^{\,+28+28+63}_{\,-23-28-63}}}$
      \vspace{2mm} \\
      175
     & $\phantom{00}{{845^{\,+102+63}_{\,-101-55}}}$
     & $\phantom{00}{{851^{\,+68+27+61}_{\,-60-27-60}}}$
     & $\phantom{00}{{856^{\,+27+27+62}_{\,-23-27-61}}}$
      \vspace{2mm} \\
      176
     & $\phantom{00}{{823^{\,+99+61}_{\,-98-54}}}$
     & $\phantom{00}{{829^{\,+66+26+60}_{\,-58-26-59}}}$
     & $\phantom{00}{{834^{\,+27+26+60}_{\,-22-26-59}}}$
      \vspace{2mm} \\
      177
     & $\phantom{00}{{801^{\,+96+59}_{\,-96-53}}}$
     & $\phantom{00}{{807^{\,+64+25+58}_{\,-57-25-57}}}$
     & $\phantom{00}{{812^{\,+26+25+59}_{\,-22-25-58}}}$
      \vspace{2mm} \\
      178
     & $\phantom{00}{{780^{\,+94+58}_{\,-93-51}}}$
     & $\phantom{00}{{786^{\,+62+25+57}_{\,-55-25-56}}}$
     & $\phantom{00}{{791^{\,+25+25+57}_{\,-21-25-57}}}$
      \vspace{2mm} \\
      179
     & $\phantom{00}{{760^{\,+91+56}_{\,-91-50}}}$
     & $\phantom{00}{{766^{\,+61+24+55}_{\,-54-24-55}}}$
     & $\phantom{00}{{770^{\,+25+24+56}_{\,-21-24-55}}}$
      \vspace{2mm} \\
      180
     & $\phantom{00}{{741^{\,+89+55}_{\,-89-49}}}$
     & $\phantom{00}{{746^{\,+59+23+54}_{\,-52-23-53}}}$
     & $\phantom{00}{{751^{\,+24+23+55}_{\,-20-23-54}}}$ \\
\hline
\end{tabular}\\[2pt]
\end{center}
\end{table}

\subsection{Top-quark mass determination}
\label{sec:mass}

The ATLAS experiment also extracted the top mass from the measurement
of the total cross section~\cite{atlas:mtt} by comparing the
NNLO$_{\text{app}}$ prediction of~\cite{Langenfeld:2009wd} to the
cross section measured using different top-mass values in the
kinematical reconstruction of the events.  With $35 \text{pb}^{-1}$ of
data and an experimental cross section $\sigma_{t\bar t}=186.3\text{
  pb} \pm 5\% \text{(stat)} \pm 12\% \text{(syst)} \pm 3\%
\text{(lumi)}$ for the reference top-mass $m_t=172.5$~GeV, the ATLAS
collaboration obtained the value $m_t=166.4^{+ 7.8}_{-7.3}$~GeV for
the pole mass. With a similar procedure, the D0 collaboration obtained
the pole mass~\cite{Abazov:2011pt} $m_t=167.5^{+ 5.2}_{-4.7}$~GeV from
the cross section measurement at the Tevatron.
 
It is interesting to speculate how the central value and the accuracy 
of the mass determination will change with the improved statistical accuracy of
the latest LHC measurements and our NNLL$_2$ predictions. Following the 
method described in~\cite{atlas:mtt} we define a likelihood function
\begin{equation} \label{eq:likelihood}
f(m_t) = \int f_{\text{th}}(\sigma|m_t) 
\cdot f_{\text{exp}}(\sigma|m_t) d \sigma \, ,
\end{equation} 
where $f_{\text{th}}$ is a normalized Gaussian 
distribution centred on the theoretical prediction of the cross section,
having a width equal to the total error of the theory prediction
\begin{equation}
f_{\text{th}}(\sigma|m_t) = \frac{1}{\sqrt{2 \pi} \Delta 
\sigma^\text{th}_{t \bar{t}}(m_t)} \exp \left[-\frac{\left( 
\sigma-\sigma^\text{th}_{t \bar{t}}(m_t)\right)^2}{2 (\Delta
\sigma^\text{th}_{t \bar{t}}(m_t))^2} \right] \, .
\end{equation}
$f_{\text{exp}}$ is defined in a similar way, with the central value and width of the Gaussian given by the measured value of the cross section, 
$\sigma^\text{exp}_{t \bar{t}}(m_t)$, and the
total experimental error, $\Delta \sigma^\text{exp}_{t \bar{t}}(m_t)$. 
The top-quark mass is then extracted from the maximum of the likelihood function (\ref{eq:likelihood}), 
with the error obtained from the $68\%$ area around the maximum. 

The theoretical cross section $\sigma_{t \bar{t}}^{\text{th}}(m_t)$ is obtained using the numbers for the NNLL$_2$
approximation given in Table \ref{tab:mt-lhc7}, interpolated by a function
of the form 
\begin{equation} \label{eq:theoryxs}
\sigma_{t \bar{t}}^{\text{th}}(m_t) = \left(\frac{172.5}{m_t}\right)^4 \left(c_0+c_1 (m_t-172.5)+c_2 (m_t-172.5)^2+c_3 (m_t-172.5)^3 \right) \, \text{pb}\, ,
\end{equation}
where all the masses are given in GeV.  The total theory error is
obtained by summing in quadrature the scale, resummation and the
NNLO-constant uncertainties.  For the PDF$+\alpha_s$ uncertainty, the
$68\%$ CL set was used in the Atlas note~\cite{atlas:mtt} and added
linearly to the theory error. We account for this by rescaling our
PDF$+\alpha_s$ error. Furthermore, when the errors in the positive and
negative directions differ from each other, the maximum of the two
values is used.  For the coefficients in the fit~\eqref{eq:theoryxs}
we find $c_0=166.496\pm 7.706\pm 7.464$, $c_1=-(1.15093\pm 0.05687\pm
0.03741)$, $c_2=(5.06265\pm 0.15147\pm 0.15886)\times 10^{-3}$,
$c_3=(8.53722\pm 12.1116\pm (-3.19098))\times 10^{-5}$ where the first
error denotes the total theory error and the second the $68\%$ CL
PDF$+\alpha_s$ error.
Using the theoretical prediction provided by our NNLL$_2$ approximation, the 
measured $t \bar{t}$ cross section for the $35 \text{pb}^{-1}$ data set given in Table 1
of~\cite{atlas:mtt}, and the extraction procedure described above, we obtain $m_t = 166.5^{+8.0}_{-7.0}$~GeV,  
which is in good agreement with the value extracted by the 
ATLAS collaboration using the NNLO$_{\text{app}}$ 
result from~\cite{Langenfeld:2009wd}.

\begin{figure}[t!]
\begin{center}
\includegraphics[width=0.65 \linewidth]{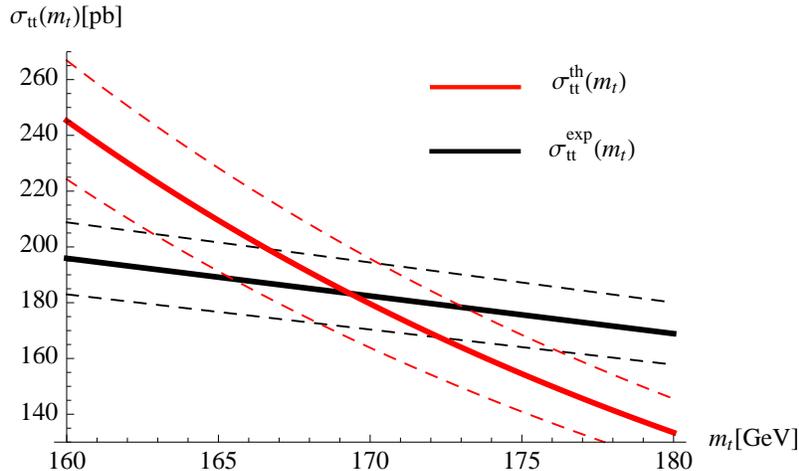}
\end{center}
\caption{\sf Mass dependence of the theoretical NNLL$_2$ cross section (red) and of the measured cross section (black), as obtained
from Ref. ~\cite{atlas:sigmattlatest}. The solid
lines represent the central values, while the total uncertainties of the theoretical and experimental results, determined as explained in the text, are
given by the external dashed lines.}
\label{fig:mass_determination}
\end{figure}

For the more recent measurement obtained by ATLAS with improved
statistics, the dependence of the measured cross section on the
reference top mass was found to be well described by a linear
fit~\cite{atlas:sigmattlatest}
\begin{equation} \label{eq:expxs}
\sigma_{t \bar{t}}^{\text{exp}}(m_t) = \left(411.9-1.35\, m_t \right) \, \text{pb}
\end{equation} 
in the interval $m_t=160-190$ GeV. The mass dependence of both experimental and theoretical
cross section is shown in Figure \ref{fig:mass_determination}.
Using the experimental result (\ref{eq:expxs}) and our 
theoretical prediction (\ref{eq:theoryxs}) for the evaluation of the likelihood function, we
extract the pole mass  
\begin{equation}\label{eq:mpole} 
m_t = (169.8^{+4.9}_{-4.7}) \, \text{GeV} \, ,
\end{equation}
where we have assumed that the total relative experimental error
does not depend on $m_t$, and is equal to the relative error at the
reference mass, $\Delta \sigma_{t \bar{t}}^{\text{exp}} /\sigma_{t \bar{t}}^{\text{exp}}=\pm 6.6\%$. 
It can be seen that the central value obtained from the $0.7\, \text{fb}^{-1}$ data set is 
higher than the one obtained in~\cite{atlas:mtt}, and the error is 
reduced to $\pm 3\%$.   
Also note that (\ref{eq:mpole})
agrees with the Tevatron measurement from direct
reconstruction, $m_t=173.3\pm1.1$~GeV, at better than 1$\sigma$ accuracy.

\section{Conclusion}

We calculated the total production cross section
of top-antitop pairs at the Tevatron and the LHC based on a combined
resummation of threshold logarithms and Coulomb corrections at NNLL
accuracy, using the formalism developed
in~\cite{Beneke:2009rj,Beneke:2010da}.  
We carefully assessed the ambiguities inherent in the 
resummation prescription,
and adopted a new procedure for choosing the soft scale in the
momentum-space approach to soft-gluon resummation in addition to the
fixed scale advocated in~\cite{Becher:2006mr,Becher:2007ty}, 
obtaining consistent results with both methods within the estimated 
uncertainties. We also
used the expansion of the NNLL result to obtain new approximate
results at N$^3$LO accuracy.

Our calculation accounts for bound-state corrections and 
higher-order Coulomb corrections not included in NNLL or 
NNLO$_{\text{app}}$ calculations based on summing threshold logarithms
at fixed invariant 
mass~\cite{Ahrens:2009uz,Ahrens:2010zv,Kidonakis:2010dk,Ahrens:2011mw}.
We find that the main effect of these corrections is included in
the NNLO$_{\text{app}}$ corrections obtained
in~\cite{Beneke:2009ye}, whereas the Coulomb corrections beyond NNLO
are in the sub-percent range.  The effect of NNLL resummation compared
to the NNLO$_{\text{app}}$ result of~\cite{Beneke:2009ye} is small,
about $2\%$ at the Tevatron and below $1\%$ at the LHC. While this
suggests that effects beyond NNLO are small, it is interesting to
note that a larger effect is obtained by adding the partial N$^3$LO
results obtained from the expansion of the NNLL prediction. This might
indicate that, from a certain perturbative order onwards, resummation 
should always be performed, even though the total effect is small.

Our results for a pole mass of $m_t=173.3$~GeV, and using the 
MSTW08NNLO PDFs, are given by
\begin{eqnarray}
&& \sigma_{t\bar t}(\mbox{Tevatron}) 
= (7.22^{+0.31+0.71}_{-0.47-0.55})\,\mbox{pb} \nonumber \\[0.2cm]
&& \sigma_{t\bar t}(\mbox{LHC,}\,\sqrt{s}=7\,\mbox{TeV}) = 
(162.6^{+7.4+15.4}_{-7.6-14.7})\,\mbox{pb}\\[0.2cm]
&& \sigma_{t\bar t}(\mbox{LHC,}\,\sqrt{s}=14\,\mbox{TeV}) = 
(896^{+40+65}_{-37-64})\,\mbox{pb}
\nonumber
\end{eqnarray}
where the first error denotes the combined theoretical uncertainty,
including the scale uncertainty, estimates of resummation
ambiguities and the unknown constant term in the threshold
expansion of the NNLO cross section, and the second the PDF+$\alpha_s$
uncertainty. The first error should be reduced once the 
full NNLO calculation of the top cross section is available. 
The reduction of the second one depends on the availability of a 
suitable reference cross section that depends on the gluon 
luminosity. 

Using our NNLL$_2$ prediction and recent ATLAS measurements, we
estimated that the top-quark mass could be extracted with an accuracy 
of $\pm 5$~GeV from the currently available LHC data on the total
cross section, and our result
\begin{equation}\label{eq:mpole1} 
m_t = (169.8^{+4.9}_{-4.7}) \, \text{GeV} 
\end{equation}
is compatible with the mass 
determination from direct reconstruction.

\subsubsection*{Acknowledgements}
We would like to thank M. Cacciari, M. Mangano and A. Mitov
for fruitful discussions during the completion of this manuscript.
M.B. also thanks the Kavli Institute for Theoretical Physics 
at UC Santa Barbara for hospitality while part of this work was done. 
The work of M.B. and S.K. is supported by the DFG
Sonder\-for\-schungs\-bereich/Trans\-regio~9 ``Computergest\"utzte
Theoreti\-sche Teilchenphysik''. P.F. acknowledges support by the ``Stichting voor Fundamenteel Onderzoek
der Materie (FOM)". This research was supported in part 
by the National Science Foundation under Grant 
No.~PHY05-51164.

\appendix
\section{NLO Coulomb function}
\label{app:coulomb}

The Coulomb Green functions at leading and next-to-leading order 
appearing in (\ref{JRal}) 
read~\cite{Beneke:1999zr,Beneke:1999qg}\footnote{The 
expression for the NLO Coulomb-Green
function obtained in~\cite{Beneke:1999qg} is quoted
e.g. in~\cite{Pineda:2006ri}.}
\begin{equation}
\begin{aligned}
\label{GCR}
G^{(0)}_{C,R}(E) &= 
\frac{m_t^2}{4 \pi} \,\bigg\{-\sqrt{\frac{-E}{m_t}}
    - (-D_{R}) \alpha_s \bigg[-L_E -\frac{1}{2} 
    +\hat \psi(1-\lambda)\bigg]\,\bigg\}\,, \\[0.2cm]
G^{(1)}_{C,R}(E) &=
\frac{m_t^2}{16 \pi^2} 
\,4\pi(-D_{R})\alpha_s\,\frac{\alpha_s}{4\pi}\,
\bigg\{a_1\left[L_E+j_0\right]+\beta_0\left[L_E^2 + 
2 j_0 L_E+j_1\right]\!\bigg\}\, ,
\end{aligned}
\end{equation}   
where we have introduced the modified Euler psi function, 
$\hat\psi(x) = \gamma_E+\psi(x)$, the short-hands 
\begin{equation}
\lambda = \frac{(-D_{R})\alpha_s}{2\sqrt{-E/m_t}} \qquad 
L_E = -\frac{1}{2}\ln\left(-\frac{4\,m_tE}{\mu^2}\right)\,,
\end{equation}
and the functions
\begin{eqnarray}
j_0&=&\lambda\psi^\prime(1-\lambda)-\hat\psi(1-\lambda) \,,
\\[0.2cm]
j_1&=& 4\,_4F_3(1,1,1,1;2,2,1-\lambda;1)
+\lambda\psi^{\prime\prime}(1-\lambda)
-2\lambda\hat \psi(1-\lambda)\psi^\prime(1-\lambda)
\nonumber\\
&& -\,3\psi^\prime(1-\lambda)
+\hat\psi(1-\lambda)^2 -\frac{\pi^2}{6}. \label{j1eq}
\end{eqnarray}
The coefficients $\beta_0=11-\frac{2}{3}n_l$ and $a_1 = 
\frac{31}{3}-\frac{10}{9}n_l$ ($n_l=5$ light quarks) denote 
the first coefficient of the QCD beta-function and the one-loop 
correction to the Coulomb potential, respectively.
The computation of the convolution (\ref{eq:star-dist}) requires
the numerical evaluation of the function $j_1$ and its
derivatives for complex argument $\lambda$, which is explained in 
Section \ref{sec:hyperg}. After the analytic continuation 
of the convolution of partonic cross sections with the parton luminosity, 
the potential function from (\ref{JRal}) with 
$\Delta_{\rm nC}$ set to one needs to be evaluated at $E=0$, 
see~(\ref{eq:subt-lumi}). For the colour-singlet 
configuration ($D_{R}=-C_F$) this is given by
\begin{equation}
\label{eq:j0}
J_{\bf 1}(0)=\frac{m_t^2 C_F}{2\pi} \left\{\pi \alpha_s -
    \alpha_s^2 \left[\frac{\beta_0}{2}\left(
         \ln\left(\frac{C_F\alpha_s m_t}{\mu}\right)
     + \gamma_E+1\right) -\frac{a_1}{4} 
      \right]\right\} \qquad (\mbox{for }\Delta_{\rm nC}=1)\,.\quad 
\end{equation}
The octet Coulomb function vanishes in this limit. 

\subsection{Evaluation of the NLO Coulomb function}
\label{sec:hyperg}

The only non-trivial term in the NLO Coulomb function is the hypergeometric 
function $_4F_3$ in (\ref{j1eq}), which needs to be evaluated 
for complex $\lambda$. We define
\begin{eqnarray}
F_{43}\equiv \,_4F_3(1,1,1,1;2,2,N;1)
&=& \sum_{i=0}^{\infty} \frac{\Gamma(i+1)^3\Gamma(N)}
{\Gamma(i+2)^2\Gamma(N+i)} \quad\mbox{with}\quad N=1-\lambda\,.
\quad
\label{HYP4F3}\end{eqnarray}
By assuming that $N$ is a positive integer, this sum is of the type 
which has been considered in~\cite{Bierenbaum:2008yu},
and can be expressed in terms of harmonic sums 
\cite{Vermaseren:1998uu,Blumlein:1998if},
\begin{eqnarray}
 F_{43} &=&
  (N-1)\,\Bigl\{\zeta_3-\zeta_2S_1(N-2)+S_{3}(N-2)+S_{1}(N-2)S_{2}(N-2)
  \nonumber\\ &&
     -S_{2,1}(N-2)\Bigr\}~. \label{4F3RES}
\end{eqnarray}
The single harmonic sums can be
continued analytically to complex $N$ by expressing them in
terms of the $\psi$-function and its derivatives
\begin{eqnarray}
S_1(N)  &=&\psi(N+1)+\gamma_E~,\label{s1psi}    \\
S_a(N)  &=&\frac{(-1)^{a-1}}{\Gamma(a)}\psi^{(a-1)}(N+1)
    +\zeta_a~, \quad\mbox{for}\quad a\in \mathbb{N}, a \ge 2\,.
\quad
\label{sapsi}
\end{eqnarray}
The continuation 
of the nested harmonic sum $S_{2,1}(N)$ is well-defined as well
and can be easily implemented, cf. e.g.~\cite{Blumlein:2009ta,Albino:2009ci}.
The result for $F_{43}$ in terms of $\lambda\in {\mathbb{C}}$ then reads 
\begin{equation}
 F_{43} = \zeta_2-S_{2}(-\lambda)
   -\lambda\,\Big[ \zeta_3
                   +S_{3}(-\lambda)
                   -S_1(-\lambda)\,
                    \big(\zeta_2-S_{2}(-\lambda)\big)
                   -S_{2,1}(-\lambda)\,
          \Big]\,. \label{4F3RES2}
\end{equation}
The computation of the convolution (\ref{eq:star-dist}) requires
the numerical evaluation of derivatives of the NLO Coulomb function
with respect to $\lambda$. Derivatives of the $\psi$-function and hence 
single harmonic sums are easily done, but the derivative of 
the $S_{2,1}$ requires more work. We use the package 
{\sf HarmonicSums}, \cite{Ablinger:2010kw} for this task 
and obtain for the first derivative of~$F_{43}$ 
\begin{eqnarray}
\frac{d}{d\lambda} F_{43}&=&
  -\lambda\frac{\zeta_2^2}{2}+\zeta_2S_{1}(-\lambda )
  +\lambda \zeta_2 S_{2}(-\lambda )-S_{1}(-\lambda )S_{2}(-\lambda )
\nonumber \\ &&
  -\lambda \frac{S^2_{2}(-\lambda )}{2}
  -3S_{3}(-\lambda )
  -2\lambda S_{1}(-\lambda )S_{3}(-\lambda )
  -5\lambda \frac{S_{4}(-\lambda )}{2}
\nonumber\\[0.1cm] &&
  +S_{2,1}(-\lambda )
  +2\lambda S_{3,1}(-\lambda )+\zeta_3
  +2\lambda S_{1}(-\lambda )\zeta_3~. \label{D4F3RES}
\end{eqnarray}
In~(\ref{D4F3RES}) an additional nested harmonic sum appears, 
$S_{3,1}$, which can again be continued to complex $\lambda$ using 
methods described in~\cite{Blumlein:2009ta,Albino:2009ci}.

\section{Bound-state corrections}
\label{app:bound}

Above the production threshold, $E>0$, the potential function $J_R$ is 
determined by the branch cut of the Coulomb Green function $G_{C,R}(E)$ 
along the real axis. Since we neglect the top decay width in the 
potential function, which is allowed when computing the total cross 
section, an imaginary part arises below threshold from 
isolated poles corresponding to $t\bar{t}$ bound-state production. The 
bound-state contribution to the potential function takes the form  
\begin{equation} 
\label{eq:J_bound}
 J_{R}(E)=2
\sum_{n=1}^\infty \delta(E-E_n)\,R_n \; \theta(-D_{R})
\,,\qquad E<0
\end{equation}
There are no bound states corrections when the Coulomb potential is 
repulsive ($D_R>0$), as is the case in the colour-octet production 
channel, as indicated by the step function $\theta(-D_R)$.
The leading-order bound-state energies and the residues read
\begin{equation}
E_n^{(0)} =-\frac{\alpha_s^2D^2_{R}m_t}{4n^2}, 
\qquad 
R_n^{(0)} = \left(\frac{m_{t} (-D_{R_\alpha})\alpha_s}{2 n}\right)^{\!3}\,.
\end{equation}
While these expressions are sufficient for NLL resummation, the NLO 
Coulomb Green function shifts the energy and residue of the 
bound states to 
\begin{equation}
E_n =E_n^{(0)}(1+\frac{\alpha_s}{4\pi}\,e_1)\,,
\qquad 
R_n =R_n^{(0)}(1+\frac{\alpha_s}{4\pi}\,\delta r_1)\,, 
\end{equation} 
with \cite{Pineda:1997hz,Beneke:2005hg}
\begin{eqnarray}
e_1 &=& 2 a_1 + 4 \beta_0 \left[S_1(n) - 
\ln \left(\frac{m_t \alpha_s (-D_R)}{n \mu}\right) \right],
\\[0.2cm] 
\delta r_1 &=& 3 a_1 +2 \beta_0\left[S_1(n) + 2 n S_2(n) - 1  
-\frac{n\pi^2}{3}- 
3 \ln \left(\frac{m_t \alpha_s (-D_R)}{n \mu}\right) \right] .
\end{eqnarray}

The bound-state contributions to the partonic cross section are 
convoluted with the parton luminosity, similarly to the procedure 
discussed in Section~\ref{sec:conv}. We mention here the relevant 
modifications. Recall that we do no apply the small correction 
$\Delta_{\rm nC}$ in (\ref{JRal}) to the already small bound-state 
contribution. Inserting (\ref{eq:J_bound}) into the resummation formula and 
the convolution with the parton luminosity, we find
\begin{eqnarray} 
\label{eq:BS}
\sigma^{BS}_{N_1 N_2 \rightarrow t \bar{t} X} &=& 
\sum_{p,p'=q,\bar{q},g} H_{p p'}^{\bf 1}(m_t,\mu) 
\int_0^1 d \tau L_{p p'}(\tau) \int_0^\infty d \omega
J_{\bf 1}(E-\omega/2) W^{\bf 1}(\omega,\mu) \theta(\omega/2-E)  
\nonumber\\
&& \hspace*{-2cm} 
= \sum_{p,p'=q,\bar{q},g} \frac{2 H_{p p'}^{\bf 1}(m_t,\mu)}{m_t} 
\exp[-4 S(\mu_s,\mu_f)+2 a^{\bf 1}_{W}(\mu_s,\mu_f)] 
\tilde{s}^{\bf 1} (\partial_\eta,\mu_s) 
\frac{e^{-2 \gamma_E \eta}}{\Gamma(2 \eta)} 
\left(\frac{2 m_t}{\mu_s}\right)^{2 \eta} \nonumber\\
&&\hspace*{-1.2cm}
\times  \sum_n  R_n
\int_{\tau_n}^1 d \tau L_{p p'}(\tau) 
\left(\frac{E-E_n}{m_t}\right)^{2 \eta-1} \,  ,
\end{eqnarray}
where $E=\sqrt{\tau s}-2 m_t$, and $H_{p p'}^{\bf 1}(m_t,\mu)$, 
$\tilde{s}^{\bf 1} (\partial_\eta,\mu_s)$ are the 
hard function and Laplace transform of the soft function for the singlet 
state. The integrals in the series have different lower 
integration limits, given by
\begin{equation}
\tau_n= \frac{1}{s} \left( 2 m_t+E_n \right)^2 < \tau_0 \equiv 
\frac{4 m_t^2}{s} \, .
\end{equation}

For $\eta<0$ the $\tau$ integrals have a non-integrable singularity 
at $\tau=\tau_n$ where $E-E_n\approx \sqrt{s/(4\tau_n)}\,(\tau-\tau_n)$.
This is analogous  to the convolution of the continuum cross section 
with the parton luminosity in Section~\ref{sec:conv}. Following the 
method discussed there, the analytic continuation to negative $\eta$ 
is constructed. However, here the integrand behaves 
$\sim (\tau-\tau_n)^{2\eta-1}$ rather than 
$\sim (\tau-\tau_0)^{2\eta}$, so one further subtraction is required.
To improve numerical stability we again introduce a parameter 
$\Lambda > \tau_0$ and split the integral into two integration 
domains,
\begin{eqnarray}
&& \sum_n  R_n \int_{\tau_n}^1 d \tau L_{p p'}(\tau) 
\left(\frac{E-E_n}{m_t}\right)^{2 \eta-1}
= \sum_n R_n
\left(\frac{\sqrt{s}}{m_t}\right)^{2\eta-1} \int_{\tau_n}^\Lambda d \tau L_{p p'}(\tau) \left(\sqrt{\tau}-\sqrt{\tau_n}
\right)^{2\eta-1} \nonumber\\
&& \hspace*{0.5cm}
+ \int_\Lambda^1 d \tau L_{p p'}(\tau) 
\left(\frac{\sqrt{s}}{m_t}\right)^{2\eta-1} 
\sum_nR_n \left(\sqrt{\tau}-\sqrt{\tau_n} \right)^{2\eta-1} \,.
\end{eqnarray}
In the second term we interchanged summation and integration. 
The second integral is free of singularities. Furthermore, 
the sum inside the integral is convergent. 
To isolate the singularities in the region $[\tau_n,\Lambda]$ we introduce 
the expansion of the integrand
around $\tau=\tau_n$ as
\begin{eqnarray}
\hspace*{-1.5cm}
\mathcal{T}_{\tau_n}^{(1)}
\left[L_{pp'}(\tau)(\sqrt{\tau}-\sqrt{\tau_n})^{2\eta-1}\right] &=&
(4\tau_n)^{1/2-\eta}
\bigg[L_{p p'}(\tau_n) (\tau-\tau_n)^{2\eta-1}  
\nonumber\\
&&
\,+\frac{4 \tau_n L^{'}_{p p'}(\tau_n)+(1-2 \eta) L_{pp'}(\tau_n)}{4 \tau_n} (\tau-\tau_n)^{2 \eta} \bigg]\,.
\end{eqnarray}
Then, for $\eta<0$, the integrals over the interval $[\tau_n,\Lambda]$ are 
analytically continued to $\eta>-1$ by the identity 
\begin{eqnarray}
&& \hspace*{-0.6cm}\int_{\tau_n}^\Lambda d \tau L_{p p'}(\tau) 
\left(\sqrt{\tau}-\sqrt{\tau_n}\right)^{2\eta-1} 
= \int_{\tau_n}^\Lambda
d \tau (1-\mathcal{T}_{\tau_n}^{(1)})
\left[L_{pp'}(\tau)(\sqrt{\tau}-\sqrt{\tau_n})^{2\eta-1}\right]
\nonumber\\
&&\hspace*{0.5cm} + \,(4\tau_n)^{1/2- \eta} (\Lambda-\tau_n)^{2 \eta}
\left[ \frac{L_{pp'}(\tau_n)}{2 \eta} 
+\frac{4 \tau_n L_{pp'}^\prime(\tau_n)+
(1-2 \eta) L_{pp'}(\tau_n)}{4 (1+2 \eta) \tau_n} (\Lambda-\tau_n)\right]. 
\,\qquad
\label{eq:BS_sub}
\end{eqnarray}
The poles at $\eta=0,-1/2$ are now manifest and cancelled by the 
factor $1/\Gamma(2 \eta)$ in~(\ref{eq:BS}). 
The remaining integral in~(\ref{eq:BS_sub}) is singularity-free if 
$\eta>-1$, and can be computed numerically. This requires the numerical 
evaluation of the first derivative of the parton-luminosity 
functions, $L_{pp'}^\prime$. The integral~(\ref{eq:BS_sub}) can be continued 
to arbitrary negative values of $\eta$ by performing more subtractions. 
However, for the applications presented in this paper, the continuation 
to $\eta>-1$ is sufficient. The convergence of the series in $n$, and 
the stability of the numerical integration,
depend on the choice of the separation parameter $\Lambda$. 
We adopt $\Lambda=1.002 \tau_0$, and truncate the series over bound states 
at level $n_{\text{max}}=100$. 

\section{Fixed-order expansions}
\label{app:expansions}

\subsection{Expansion to $\mathcal{O}(\alpha_s^2)$}
\label{app:expNNLO}

In this appendix we provide the NNLO corrections to the cross
section~\eqref{eq:sigma-series} resulting from expanding the NNLL
resummed result. 
The scale-independent terms for the three partonic channels are given by
\begin{align}
f^{(2,0)}_{q\bar q(8)}=&\,\frac{3.60774}{\beta^2}
+\frac{1}{\beta}\left(-140.368\ln^2\beta+32.106\ln\beta+3.95105\right)
\nonumber \\ &
+910.222\ln^4\beta-1315.53\ln^3\beta+592.292\ln^2\beta
+528.557\ln\beta+ \tilde f^{(2,0)}_{q\bar q(8)}\,,\\[0.2cm]
f^{(2,0)}_{gg(1)}=&\, \frac{230.896}{\beta ^2}+\frac{1}{\beta}
\left(2526.62 \ln^2\beta+1347.76 \ln\beta+66.1114\right)\nonumber\\&
+4608. \ln^4\beta-249.2 \ln^3\beta-2385.73
   \ln ^2\beta+1600.47 \ln\beta+\tilde f^{(2,0)}_{gg(1)}\,,\\[0.2cm]
f^{(2,0)}_{gg(8)}=&\, \frac{3.60774}{\beta^2}
+\frac{1}{\beta }\left(-315.827 \ln^2\beta-89.5134 \ln\beta-38.5162\right)
\nonumber\\&
+4608 \ln^4\beta-2553.2 \ln^3\beta-322.996 \ln^2\beta+2799.24 \ln\beta 
+ \tilde f^{(2,0)}_{gg(8)}\,.
\end{align}
The terms shown explicitly are known exactly and have first been
obtained fully in~\cite{Beneke:2009ye}. The functions $\tilde
f^{(2,0)}_X$ are constant terms in the threshold expansion 
that are currently not known
completely, since, for example, the two-loop hard functions contribute 
to them. At the NNLL level, they therefore show a residual dependence on 
the numerical factors $k_i$ in the scale choices $\mu_s=k_s m_t\beta^2$,
and $\mu_h=k_h m_t$. These terms are given by
\begin{align}
\tilde f^{\text{NNLL}(2,0)}_{q\bar q(8)}=&\,
-56.889 \ln^4k_h+240.692 \ln^3k_h
-116.423\ln^2k_h-456.803 \ln k_h 
\nonumber\\&
-56.889 \ln^4 k_s-164.441 \ln^3 k_s -52.0528 \ln^2 k_s +319.029 \ln k_s 
\nonumber\\&
+49.7741\,,\\[0.2cm]  
\tilde f^{\text{NNLL}(2,0)}_{gg(1)}=&\,-288. \ln^4 k_h +921.172 \ln^3 k_h
-8.19545 \ln^2k_h-1322.62 \ln k_h-288. \ln^4k_s
\nonumber\\&
-31.15 \ln^3k_s+1231. \ln^2k_s+1295.98 \ln k_s-328.235\,, \\[0.2cm]
\tilde f^{\text{NNLL}(2,0)}_{gg(8)}=&\,-288. \ln^4 k_h +633.172 \ln^3k_h
+543.378 \ln^2 k_h-1109.7 \ln k_h-288. \ln^4 k_s 
\nonumber\\&
-319.15 \ln^3 k_s+847.638 \ln^2k_s+1626.13 \ln k_s-43.6962\,.
\end{align}
The factorization-scale dependent terms for the partonic channels are
\begin{align}
f^{(2,1)}_{gg(1)}=&\,\frac{1}{\beta }\left(403.557-2526.62 \ln\beta\right)
-9216 \ln^3\beta+3567.82\ln^2\beta\nonumber\\&
+2639.47 \ln\beta+\tilde  f^{(2,1)}_{gg(1)}\,,
  \\[0.2cm]
f^{(2,2)}_{gg(1)}=&\,4608 \ln ^2\beta-3563.96 \ln \beta
  +\tilde  f^{(2,2)}_{gg(1)} ,\\
f^{(2,1)}_{gg(8)}=&\,\frac{1}{\beta }\left(315.827 \ln\beta-50.4446\right)
  -9216 \ln ^3\beta +5871.82 \ln ^2\beta\nonumber\\&
 -25.2877 \ln\beta+  \tilde  f^{(2,1)}_{gg(8)}\,,\\[0.2cm]
f^{(2,2)}_{gg(8)}=&\, f^{(2,2)}_{gg(1)}\,.
\end{align}
In this case the constant terms are known exactly:
\begin{align}
\tilde  f^{(2,1)}_{q\bar q(8)}&=327.048 ,&
\tilde  f^{(2,2)}_{q\bar q(8)}&=385.383 , \\
\tilde  f^{(2,1)}_{gg(1)}&=-487.213,&
\tilde  f^{(2,2)}_{gg(1)}&=-417.165,\\
\tilde  f^{(2,1)}_{gg(8)}&=-283.914 ,&
\tilde  f^{(2,2)}_{gg(8)}&=\tilde  f^{(2,2)}_{gg(1)}.
\end{align}

\subsection{Expansion to $\mathcal{O}(\alpha_s^3)$}

Here we provide those parts of the N$^3$LO corrections to the cross
section~\eqref{eq:sigma-series} that have not been given already in
Section~\ref{sec:n3lo}, i.e. the scale dependent terms $f^{(3,i)}_X$ for
$i=1,2,3$ and the NNLL approximations to the functions 
$\tilde f^{(3,i)}_X$, that are beyond the NNLL accuracy of the 
threshold expansion. The expressions that follow are 
generated from expanding the resummation formula as given in 
Eqs.~(3.48) and (3.49) of~\cite{Beneke:2010da} (appropriately adapted
to the case of top-quark production).

\paragraph{\it Quark-antiquark channel.} 
The scaling functions of the scale-dependent contributions in the N$^3$LO 
expansion of the quark-antiquark production channel are given by
\begin{align}
f^{(3,1)}_{q\bar q(8)}=&
\frac{1}{\beta^2}\left(-153.93\ln\beta+56.8546\right)
+\frac{1}{\beta}
\left(5989.02 \ln ^3\beta-7733.19 \ln ^2\beta-2669.2 \ln \beta\right)
\nonumber\\
&-38836.1 \ln^5\beta+109310. \ln^4\beta
-78403.7 \ln ^3\beta+\tilde f^{(3,1)}_{q\bar q(8)}\,,\\[0.2cm]
f^{(3,2)}_{q\bar q(8)}=&\frac{1}{\beta }
\left(-2994.51 \ln^2\beta+5287.18 \ln\beta\right)+38836.1 \ln^4\beta
-111164\ln^3\beta+\tilde f^{(3,2)}_{q\bar q(8)} \,,\\[0.2cm]
f^{(3,3)}_{q\bar q(8)}=&
-12945.4 \ln^3\beta
+\tilde f^{(3,3)}_{q\bar q(8)}\,,
\end{align}
where all the terms shown explicitly are exact. These are the only
terms included in the N$^3$LO$_{\text B}$ approximation, while the
N$^3$LO$_{\text A}$ approximation includes in addition the NNLL
approximation to the $\tilde f^{(3,i)}$-functions.  Most of the terms 
in these functions can only be predicted exactly from resummation 
beyond NNLL accuracy. At the NNLL level, they show a residual 
dependence on the numerical
factors $k_i$ in the scale choices $\mu_s=k_s m_t\beta^2$, $\mu_h=k_h
m_t$ and $\mu_C=k_C m_t\beta$.  For the quark-antiquark channel these
functions read
\begin{align}
\tilde f^{\text{NNLL}(3,0)}_{q\bar q(8)}=&\,\frac{1}{\beta}
\left(773.485 \ln ^2k_C-335.834 \ln k_C+187.157 \ln ^4k_h-791.846 \ln^3k_h
\right.\nonumber\\
&\left.   +383.015 \ln ^2k_h+1502.82 \ln k_h+187.157 \ln ^4k_s+
81.5526 \ln^3k_s\right.\nonumber\\
&\left. -1199.28 \ln ^2k_s -2174.59 \ln k_s+749.55\right)+
\ln \beta \left(-2331.66 \ln k_C\right.\nonumber\\
&\left.+1172.51 \ln ^4k_h-4960.79 \ln^3k_h+2399.53 \ln ^2k_h+
9414.94 \ln k_h\right.\nonumber\\
&\left. +4661.69 \ln ^4k_s+4398.91 \ln^3k_s-27481 \ln ^2k_s 
-60914.9 \ln k_s+11370.2\right)\nonumber\\
&\left.+\ln ^2\beta \left(-2427.26 \ln^4k_h+10269.5 \ln ^3k_h\right.
-4967.37 \ln^2k_h\right.\nonumber\\
&\left.-19490.3 \ln k_h-2427.26 \ln^4k_s-7016.16 \ln ^3k_s
-2220.92 \ln^2k_s\right.\nonumber\\
&\left. +13611.9 \ln k_s-65560\right)-404.543 \ln ^6k_h+3730.45 \ln ^5k_h
\nonumber\\
&-10015.2 \ln^4k_h+5148.7 \ln ^3k_h+9150.36 \ln^2k_h-5460.36 \ln k_h
\nonumber\\
&+404.543 \ln^6k_s+2917.1 \ln ^5k_s+1022.1 \ln^4k_s-12319 \ln ^3k_s
\nonumber\\
&-15042.1 \ln^2k_s+5890.83 \ln k_s+702.307 \, , \\[0.2cm]
    \tilde f^{\text{NNLL}(3,1)}_{q\bar q(8)}=&\,
    \frac{2272.92}{\beta}-38012.7 \ln^2\beta+\ln\beta
    \left(2427.26 \ln^4k_h-10269.5 \ln^3k_h\right.\nonumber\\
    &\left. +4967.37 \ln^2k_h +19490.3 \ln k_h+2427.26 \ln ^4k_s
      +7016.16 \ln ^3k_s\right.\nonumber\\
    &\left. +2220.92 \ln^2k_s-13611.9 \ln k_s+67825.2\right)
    -3323.77 \ln ^4k_h+14062.6 \ln^3k_h\nonumber\\
    &-6802.08 \ln^2k_h-26436.3 \ln k_h-3323.77 \ln^4k_s-5069.58 \ln^3k_s
\nonumber\\
    &+12832\ln^2k_s+36025.7 \ln k_s+327.661\, , \\[0.2cm]
\tilde f^{\text{NNLL}(3,2)}_{q\bar q(8)}=&\,
-\frac{1005.67}{\beta }+78770 \ln ^2\beta-3383.32 \ln\beta-3697.7 \, , \\[0.2cm]
\tilde f^{\text{NNLL}(3,3)}_{q\bar q(8)}=&\,
39223.6 \ln^2\beta-28867.1 \ln\beta+1140.45 \, .
\end{align}

\paragraph{\it Gluon fusion colour-singlet channel.} 
The scaling functions for the factorization-scale dependent terms are
\begin{align}
f^{(3,1)}_{gg(1)}=&\,\frac{1}{\beta^2}\left(
- 22166\ln \beta-8283.49\right)
+\frac{1}{\beta}\left(-242555\ln^3\beta-51902.1 \ln^2\beta
+217024\ln \beta\right)\nonumber\\
&
-442368\ln^5\beta+300977 \ln^4\beta+563908\ln ^3\beta+\tilde f^{(3,1)}_{gg(1)}
\,,\\[0.2cm]
f^{(3,2)}_{gg(1)}=&\,\frac{1}{\beta }
\left(121278\ln^2\beta-58112.2 \ln\beta\right)
+442368\ln^4\beta-448309\ln^3\beta+\tilde f^{(3,2)}_{gg(1)} \, , \\[0.2cm]
f^{(3,3)}_{gg(1)}=&\,
- 147456\ln^3\beta
+\tilde f^{(3,3)}_{gg(1)}\,.
\end{align}
The NNLL approximations for the terms in the scaling
functions not exactly known read 
\begin{align}
\tilde f^{(3,0)}_{gg(1)}=&\,
\frac{1}{\beta}\left(-6187.88 \ln^2k_C+2686.67 \ln k_C
-7579.86  \ln^4k_h+24244.3 \ln^3k_h\right.\nonumber\\
&\left.-215.695 \ln^2k_h-34809.9 \ln k_h-7579.86 \ln^4k_s
+17787.4 \ln^3k_s\right.\nonumber\\
&\left.+49075.2 \ln^2k_s+10292.7 \ln k_s-14467\right)
+\ln\beta  \left(36589.2 \ln k_C\right.\nonumber\\
&\left.-2196.4 \ln^4k_h+7025.22 \ln^3k_h-62.5017 \ln^2k_h
-10086.8 \ln k_h\right.\nonumber\\
&\left.+15467.6 \ln^4k_s-101709 \ln^3k_s-211987 \ln^2k_s
-10792.4 \ln k_s+112623.\right)\nonumber\\
&+\ln^2\beta  \left(-27648 \ln^4k_h+88432.5 \ln^3k_h
-786.763 \ln^2k_h-126971 \ln k_h\right.\nonumber\\
&\left.-27648 \ln^4k_s-2990.4 \ln^3k_s+118176 \ln^2k_s
+124414 \ln k_s-146090\right)\nonumber\\
&-4608 \ln^6k_h+27996.1 \ln^5k_h-42685.9 \ln^4k_h-17071.1 \ln^3k_h
\nonumber\\
&+45365.4 \ln^2k_h+13255 \ln k_h+4608 \ln^6k_s+6635.6 \ln^5k_s
-62517.1 \ln^4k_s\nonumber\\
&-79096.6 \ln^3k_s+37998.9 \ln^2k_s+37910.9 \ln k_s-9739.8 \, , 
\\[0.2cm]
    \tilde f^{\text{NNLL}(3,1)}_{gg(1)}=&\,
-\frac{29709.3}{\beta }-618319 \ln^2\beta+\ln\beta
   \left(27648 \ln^4 k_h-88432.5 \ln^3 k_h \right.\nonumber\\
  &\left. +786.763 \ln^2 k_h+126971\ln k_h 
     +27648 \ln^4k_s+2990.4 \ln^3 k_s\right.\nonumber\\
  &\left.-118176 \ln^2k_s-124414 \ln k_s
     +198168\right)-17315.9 \ln^4k_h+55385.1 \ln ^3k_h\nonumber\\
  &-492.747 \ln^2k_h-78953.1 \ln k_h-17315.9 \ln ^4k_s
  +49817.9 \ln^3k_s\nonumber\\
  &  +146950 \ln ^2k_s+48514.7 \ln k_s-77934 \, , 
\\[0.2cm]
\tilde f^{\text{NNLL}(3,2)}_{gg(1)}=&\,
-\frac{43685.6}{\beta }-297750\ln^2\beta+322713\ln\beta-52925.9 \, , \\[0.2cm]
\tilde f^{\text{NNLL}(3,3)}_{gg(1)}=&\,
206398\ln^2\beta+10843.3 \ln\beta-50289.8 \, .
\end{align}

\paragraph{\it Gluon fusion colour-octet channel.} 
The scaling functions for the factorization-scale dependent terms are
\begin{align}
f^{(3,1)}_{gg(8)}=&\,
\frac{1}{\beta ^2}\left(-346.343 \ln\beta-129.429\right)
+\frac{1}{\beta }
\left(30319.4 \ln^3\beta-1092.09 \ln^2\beta-21802.5 \ln\beta\right)
\nonumber\\
&-442368\ln^5\beta+522161\ln ^4\beta+227359\ln^3\beta
+\tilde f^{(3,1)}_{gg(8)}
\,,\\[0.2cm]
f^{(3,2)}_{gg(8)}=&\,\frac{1}{\beta }
\left(-15159.7 \ln ^2\beta +7264.03 \ln\beta\right)
+ 442368 \ln ^4\beta-558901 \ln ^3\beta +\tilde f^{(3,2)}_{gg(8)} \, ,
\\[0.2cm]
f^{(3,3)}_{gg(8)}=&\,
- 147456\ln^3\beta
+\tilde f^{(3,3)}_{gg(8)}\,.
\end{align}
The NNLL approximations for the terms in the scaling
functions not exactly known read 
\begin{align}
  \tilde f^{\text{NNLL}(3,0)}_{gg(8)}=&\,
\frac{1}{\beta }\left(773.485 \ln^2k_C-335.834 \ln k_C
+947.482 \ln^4k_h-2083.05 \ln ^3k_h\right.\nonumber\\
   &\left.-1787.64 \ln^2k_h+3650.77 \ln k_h+947.482 \ln^4k_s
-1275.94 \ln ^3k_s\right.\nonumber\\
   &\left.-6617.62 \ln^2k_s-4868.71 \ln k_s+2424.05\right)
+\ln^2\beta  \left(-27648 \ln ^4k_h\right.\nonumber\\
   &\left.+60784.5 \ln^3k_h+52164.3 \ln^2k_h-106531 \ln k_h
-27648 \ln ^4k_s\right.\nonumber\\
   &\left.-30638.4 \ln^3k_s+81373.3 \ln^2k_s+156108\ln k_s
-347424\right)\nonumber\\
   &+\ln \beta \left(-2152.3 \ln k_C+4715.6 \ln^4k_h-10367.3 \ln^3k_h
-8897.06 \ln^2k_h\right.\nonumber\\
   &\left.+18169.8 \ln k_h+22379.6 \ln^4k_s-78581.5 \ln^3k_s
-289277 \ln^2k_s-205101 \ln k_s\right.\nonumber\\
   &\left.+123426.\right)-4608 \ln ^6k_h+21084.1 \ln^5k_h
-19739.3 \ln ^4k_h-26067.8 \ln^3k_h\nonumber\\
   &+33695.1 \ln ^2k_h+1303.32 \ln k_h+4608 \ln ^6k_s+13547.6 \ln^5k_s
-50831.9 \ln ^4k_s\nonumber\\
   &-118982 \ln^3k_s-16519.8 \ln ^2k_s+62214.4 \ln
   k_s+966.578 \,  ,
\\[0.2cm]
\tilde f^{\text{NNLL}(3,1)}_{gg(8)}=&\,
\frac{5903.02}{\beta }
    -748536\ln^2\beta+\ln\beta
   \left(27648 \ln ^4k_h-60784.5 \ln^3k_h\right.\nonumber\\
   &\left.-52164.3 \ln^2k_h+106531 \ln k_h
     +27648 \ln^4k_s+30638.4 \ln^3k_s\right.\nonumber\\
  &\left. -81373.3 \ln^2k_s-156108\ln k_s+
     399352\right)-17315.9 \ln ^4k_h+38069.2
   \ln ^3k_h\nonumber\\
  &+32670.3 \ln^2k_h-66151.5 \ln k_h-17315.9 \ln^4k_s+
   32502.1 \ln^3k_s\nonumber\\
  &+162669\ln ^2k_s+137140 \ln k_s-61943.1 \,  ,
\\[0.2cm]
\tilde f^{\text{NNLL}(3,2)}_{gg(8)}=&\,\frac{5460.7}{\beta }
-100578\ln^2\beta+284537 \ln\beta-88163.8\, , \\[0.2cm]
\tilde f^{\text{NNLL}(3,3)}_{gg(8)}=&\,\tilde f^{\text{NNLL}(3,3)}_{gg(1)} \, .
\end{align}

\section{Details of Method 2}
\label{app:method2}

\begin{figure}[t]
  \begin{center}
      \includegraphics[width=.6\textwidth]{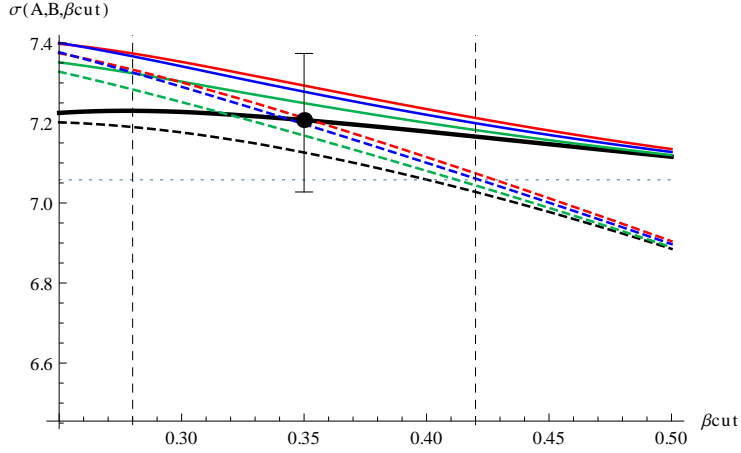}
           \end{center}
           \caption{\sf Determination of $\beta_{\text{cut}}$ and the
             resulting ambiguity for the Tevatron with $k_s=2$. The
             plot shows $ \sigma_{t\bar t}(A_<,B_>,\beta_\text{cut})$ (in pb),
             as defined by convoluting~\eqref{eq:sigma-up-down} with
             the parton luminosity. Solid curves: $A_<=\text{NNLL}_2$,
             dashed curves $A_<=\text{NNLL}_1$, black curves:
             $B_>=\text{NNLL}_2$, red curves:
             $B_>=\text{NNLO}_{\text{app}}$, blue curves:
             $B_>=\text{N$^3$LO}_{\text{B}}$, green curves:
             $B_>=\text{N$^3$LO}_{\text{A}}$. The horizontal dotted
             line is the $\text{NNLO}_{\text{app}}$ cross section,
             the $x$-axis intersects the $y$ axis at the value of the
             NLO cross section.  The point with the error bar is the
             central value of the cross section together with the
             error estimate, obtained by taking the envelope of the
             curves in the $\pm 20\%$ window around the default
             $\beta_{\text{cut}}=0.35$, given by the vertical dashed
             lines. }
\label{fig:method2TeV} 
\end{figure}

\begin{figure}[t]
  \begin{center}
      \includegraphics[width=.6\textwidth]{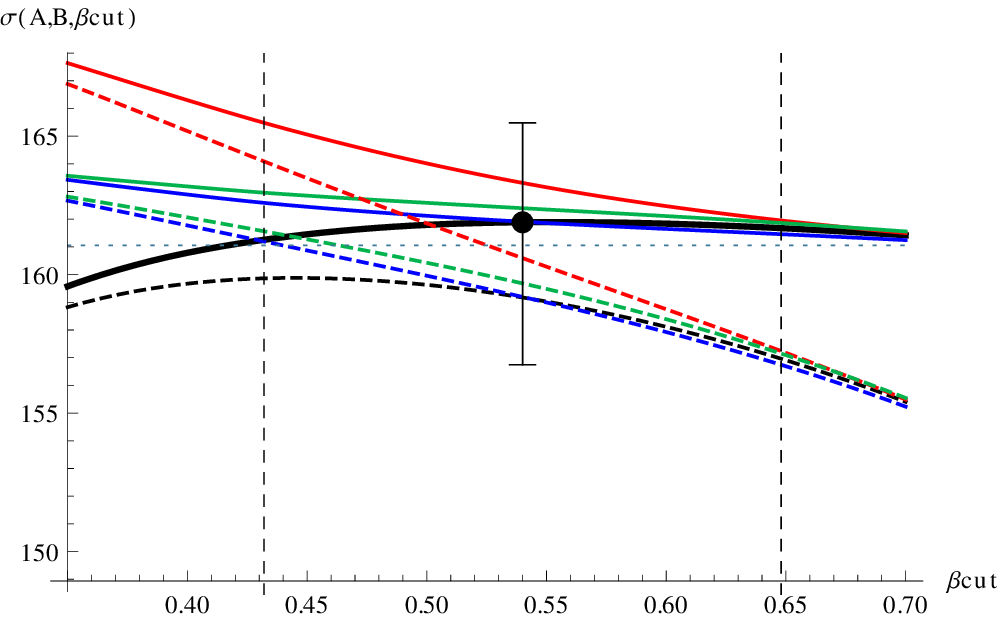} 
      \includegraphics[width=.6\textwidth]{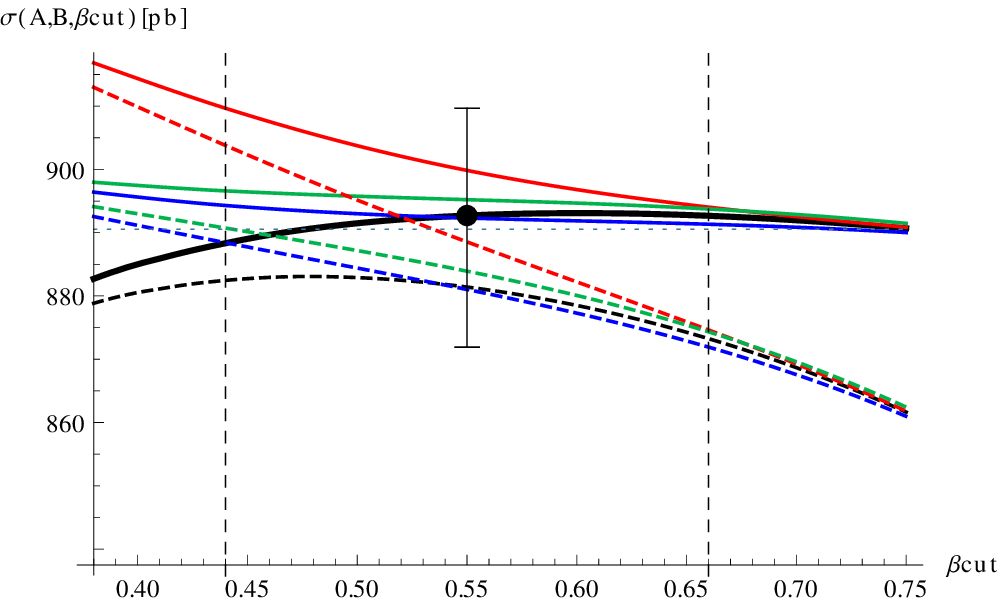}
           \end{center}
           \caption{\sf Determination of $\beta_{\text{cut}}$ and the
             resulting ambiguity for the LHC with $7$ TeV (above)
             and $14$ TeV (below). The curves are the cross section $
             \sigma_{t\bar t}(A_<,B_>,\beta_\text{cut})$, with the
             various approximations defined as in
             Figure~\ref{fig:method2TeV}.}
\label{fig:method27} 
\end{figure}

We provide here the numerical details defining $\beta_{\rm cut}$ 
in Method 2, as discussed in Section~\ref{sec:method2}, and the 
corresponding implementation at the NLL order.  
Figures~\ref{fig:method2TeV} and~\ref{fig:method27} show
the various approximations~\eqref{eq:sigma-up-down} for $k_s=2$ at the
Tevatron and at the LHC, respectively, 
together with the default $\beta_{\text{cut}}$
and the $20\%$-window used in the estimate iii) of the resummation
ambiguity.  The plots show the expected behaviour: for increasing
$\beta_{\text{cut}}$ (and hence $\mu_s^<$) resummation becomes less
effective and the difference between the two implementations used in
the lower interval (the difference between solid and dashed lines of
the same colour) becomes dominant. For smaller $\beta_{\text{cut}}$
the difference of the resummed cross section and perturbative
expansions to different orders used in the upper interval (e.g. the
difference among the solid lines of different colour) becomes sizeable,
while the difference between the two NNLL implementations in the lower
interval becomes negligible.  Note that the default curve (the black
solid line) depends only weakly on the precise value used for
$\beta_{\text{cut}}$. The spread among the different approximations
for varying $\beta_{\text{cut}}$ by $20\%$ leads to an uncertainty of
$\sim 2.5\%$ for both Tevatron and LHC. This is the dominant
contribution to our estimate of the resummation uncertainty. The
criteria i) and ii) discussed in Section~\ref{sec:method2}
both contribute uncertainties $\sim 1.5\%$ or smaller, cf. Table
\ref{Tab:ErrorRes}~.

Finally we also present the values of $\beta_{\text{cut}}$ one obtains
when varying $k_s$ by a factor of 2 around its default value $k_s=2$
for $m_t=173.3$~GeV. We obtain

\newpage
\begin{eqnarray}
&&\hspace{-15mm} \underline{k_s=1}:\nonumber \\
&&\hspace{-15mm} 
\label{eq:betacutks1}
\begin{array}{llll}
\beta_{\text{cut}} (\text{NNLL}):\quad& 
0.52 \,(\text{Tevatron}),& 
0.75\,(\text{LHC}7),&
0.79 \,(\text{LHC}14),\\[0.1cm]
\mu_s^<=m_t\beta^2_{\text{cut}}:&  
47\,\text{GeV}\,(\text{Tevatron}),\quad&
97\,\text{GeV}\, (\text{LHC}7),\quad&
108\,\text{GeV}\, (\text{LHC}14)~.
\end{array}  \\[0.4cm]
&&\hspace{-15mm} \underline{k_s=4}:\nonumber \\
&&\hspace{-15mm} 
\label{eq:betacutks4}
\begin{array}{llll}
\beta_{\text{cut}} (\text{NNLL}):\quad& 
0.23 \,(\text{Tevatron}),& 
0.33\,(\text{LHC}7),&
0.33 \,(\text{LHC}14),\\[0.1cm]
\mu_s^<=4m_t\beta^2_{\text{cut}}:&  
37\,\text{GeV}\,(\text{Tevatron}),\quad&
76\,\text{GeV}\, (\text{LHC}7),\quad&
76\,\text{GeV}\, (\text{LHC}14).
\end{array}
\end{eqnarray}
As one would expect, the corresponding soft scales $\mu_s^<$
lie not very far apart from each other for $k_s=1,2,4$, 
Eq.~(\ref{eq:betacut}) for $k_2=2$.

For resummation at NLL order, we determine $\beta_{\text{cut}}$
anew, using the same prescription to minimize the spread between
different cross sections~\eqref{eq:sigma-up-down} for the
approximations $A_<\in\{\text{NLL}_1,\text{NLL}_2\}$ (defined
in~\eqref{eq:NLL1} and~\eqref{eq:NLL2}) in the lower interval and
$B_>\in\{\text{NLL}_2,\text{NLO},\text{NLL-NNLO}_{\text{A}},
\text{NLL-NNLO}_{\text{B}}\}$
in the upper interval. Here, in analogy to the two N$^3$LO
approximations, NLL-NNLO$_{\text{A}}$ denotes the expansion of the NLL
corrections to order $\alpha_s^2$, keeping all the generated terms,
while only those terms exactly reproduced at NLL accuracy (i.e. the 
terms $\alpha_s^2 \,\{\frac{1}{\beta^2},\frac{\ln^2\beta}{\beta},
\frac{\ln\beta}{\beta},\ln^4\beta ,\ln^3 \beta\})$ are kept in the
NLL-NNLO$_{\text{B}}$ approximation. With this prescription, the 
resulting $\beta_{\text{cut}}$-values at NLL for $k_s=2$
are somewhat smaller than for NNLL resummation:
\begin{equation}
\label{eq:betacut-nll}
\begin{aligned}
&&\beta_{\text{cut}}\text{(NLL)}&:& 0.26 \,(\text{Tevatron}),
\quad 0.32\,(\text{LHC}7),
\quad 0.32 \,(\text{LHC}14)
\end{aligned}
\end{equation}
The remaining uncertainties are estimated as for NNLL.

\bibliography{bib_resum}

\providecommand{\href}[2]{#2}\begingroup\raggedright\begin{thebibliography}{10}

\bibitem{Aaltonen:2010ic}
{\bf CDF} Collaboration, T.~Aaltonen {\em et al.},
  \href{http://dx.doi.org/10.1103/PhysRevLett.105.012001}{{\em Phys. Rev.
  Lett.} {\bf 105} (2010)  012001},
\href{http://arxiv.org/abs/1004.3224}{{\tt arXiv:1004.3224 [hep-ex]}}.

\bibitem{cdf:sigmatt}
{\bf CDF} Collaboration. Conference note 9913, 2009.
\newblock
  \url{http://www-cdf.fnal.gov/physics/new/top/confNotes/cdf9913_ttbarxs4invfb%
.ps}.

\bibitem{Abazov:2011mi}
{\bf D0} Collaboration, V.~M. Abazov {\em et al.},
\href{http://arxiv.org/abs/1101.0124}{{\tt arXiv:1101.0124 [hep-ex]}}.

\bibitem{Abazov:2011cq}
{\bf D0} Collaboration, V.~M. Abazov {\em et al.},
\href{http://arxiv.org/abs/1105.5384}{{\tt arXiv:1105.5384 [hep-ex]}}.

\bibitem{atlas:sigmattlatest}
{\bf ATLAS} Collaboration. Conference note atlas-conf-2011-121, 2011.
\newblock
  \url{https://atlas.web.cern.ch/Atlas/GROUPS/PHYSICS/CONFNOTES/ATLAS-CONF-201%
1-121/ATLAS-CONF-2011-121.pdf}.

\bibitem{CMS:2011yy}
{\bf CMS} Collaboration,
\href{http://arxiv.org/abs/1108.3773}{{\tt arXiv:1108.3773 [hep-ex]}}.

\bibitem{Langenfeld:2009wd}
U.~Langenfeld, S.~Moch, and P.~Uwer,
  \href{http://dx.doi.org/10.1103/PhysRevD.80.054009}{{\em Phys. Rev.} {\bf
  D80} (2009)  054009},
\href{http://arxiv.org/abs/0906.5273}{{\tt arXiv:0906.5273 [hep-ph]}}.

\bibitem{Abazov:2011pt}
{\bf D0} Collaboration, V.~M. Abazov {\em et al.},
\href{http://arxiv.org/abs/1104.2887}{{\tt arXiv:1104.2887 [hep-ex]}}.

\bibitem{atlas:mtt}
{\bf ATLAS} Collaboration. Conference note atlas-conf-2011-54, 2011.
\newblock
  \url{https://atlas.web.cern.ch/Atlas/GROUPS/PHYSICS/CONFNOTES/ATLAS-CONF-201%
1-054/ATLAS-CONF-2011-54.pdf}.

\bibitem{Aaltonen:2011kc}
{\bf CDF} Collaboration, T.~Aaltonen {\em et al.},
\href{http://arxiv.org/abs/1101.0034}{{\tt arXiv:1101.0034 [hep-ex]}}.

\bibitem{Nason:1987xz}
P.~Nason, S.~Dawson, and R.~K. Ellis,
\href{http://dx.doi.org/10.1016/0550-3213(88)90422-1}{{\em Nucl. Phys.} {\bf
  B303} (1988)  607}.

\bibitem{Laenen:1991af}
E.~Laenen, J.~Smith, and W.~L. van Neerven,
\href{http://dx.doi.org/10.1016/0550-3213(92)90279-K}{{\em Nucl. Phys.} {\bf
  B369} (1992)  543--599}.

\bibitem{Catani:1996dj}
S.~Catani, M.~L. Mangano, P.~Nason, and L.~Trentadue,
  \href{http://dx.doi.org/10.1016/0370-2693(96)00387-5}{{\em Phys. Lett.} {\bf
  B378} (1996)  329--336},
\href{http://arxiv.org/abs/hep-ph/9602208}{{\tt arXiv:hep-ph/9602208}}.

\bibitem{Berger:1996ad}
E.~L. Berger and H.~Contopanagos,
  \href{http://dx.doi.org/10.1103/PhysRevD.54.3085}{{\em Phys. Rev.} {\bf D54}
  (1996)  3085--3113},
\href{http://arxiv.org/abs/hep-ph/9603326}{{\tt arXiv:hep-ph/9603326}}.

\bibitem{Kidonakis:1996zd}
N.~Kidonakis, J.~Smith, and R.~Vogt,
  \href{http://dx.doi.org/10.1103/PhysRevD.56.1553}{{\em Phys. Rev.} {\bf D56}
  (1997)  1553--1570},
\href{http://arxiv.org/abs/hep-ph/9608343}{{\tt arXiv:hep-ph/9608343}}.

\bibitem{Bonciani:1998vc}
R.~Bonciani, S.~Catani, M.~L. Mangano, and P.~Nason,
  \href{http://dx.doi.org/10.1016/S0550-3213(98)00335-6}{{\em Nucl. Phys.} {\bf
  B529} (1998)  424--450},
\href{http://arxiv.org/abs/hep-ph/9801375}{{\tt arXiv:hep-ph/9801375}}.

\bibitem{Kidonakis:2001nj}
N.~Kidonakis, E.~Laenen, S.~Moch, and R.~Vogt,
  \href{http://dx.doi.org/10.1103/PhrConvD.64.114001}{{\em Phys. Rev.} {\bf
  D64} (2001)  114001},
\href{http://arxiv.org/abs/hep-ph/0105041}{{\tt arXiv:hep-ph/0105041}}.

\bibitem{Czakon:2008zk}
M.~Czakon, \href{http://dx.doi.org/10.1016/j.physletb.2008.05.028}{{\em Phys.
  Lett.} {\bf B664} (2008)  307--314},
\href{http://arxiv.org/abs/0803.1400}{{\tt arXiv:0803.1400}}.

\bibitem{Bonciani:2008az}
R.~Bonciani, A.~Ferroglia, T.~Gehrmann, D.~Maitre, and C.~Studerus,
  \href{http://dx.doi.org/10.1088/1126-6708/2008/07/129}{{\em JHEP} {\bf 07}
  (2008)  129},
\href{http://arxiv.org/abs/0806.2301}{{\tt arXiv:0806.2301}}.

\bibitem{Bonciani:2009nb}
R.~Bonciani, A.~Ferroglia, T.~Gehrmann, and C.~Studerus,
  \href{http://dx.doi.org/10.1088/1126-6708/2009/08/067}{{\em JHEP} {\bf 08}
  (2009)  067},
\href{http://arxiv.org/abs/0906.3671}{{\tt arXiv:0906.3671}}.

\bibitem{Korner:2008bn}
J.~G. K{\"o}rner, Z.~Merebashvili, and M.~Rogal,
  \href{http://dx.doi.org/10.1103/PhysRevD.77.094011}{{\em Phys. Rev.} {\bf
  D77} (2008)  094011},
\href{http://arxiv.org/abs/0802.0106}{{\tt arXiv:0802.0106}}.

\bibitem{Anastasiou:2008vd}
C.~Anastasiou and S.~M. Aybat,
  \href{http://dx.doi.org/10.1103/PhysRevD.78.114006}{{\em Phys. Rev.} {\bf
  D78} (2008)  114006},
\href{http://arxiv.org/abs/0809.1355}{{\tt arXiv:0809.1355}}.

\bibitem{Kniehl:2008fd}
B.~Kniehl, Z.~Merebashvili, J.~G. K{\"o}rner, and M.~Rogal,
  \href{http://dx.doi.org/10.1103/PhysRevD.78.094013}{{\em Phys. Rev.} {\bf
  D78} (2008)  094013},
\href{http://arxiv.org/abs/0809.3980}{{\tt arXiv:0809.3980}}.

\bibitem{Dittmaier:2007wz}
S.~Dittmaier, P.~Uwer, and S.~Weinzierl,
  \href{http://dx.doi.org/10.1103/PhysRevLett.98.262002}{{\em Phys. Rev. Lett.}
  {\bf 98} (2007)  262002},
\href{http://arxiv.org/abs/hep-ph/0703120}{{\tt arXiv:hep-ph/0703120}}.

\bibitem{Bonciani:2010mn}
R.~Bonciani, A.~Ferroglia, T.~Gehrmann, A.~Manteuffel, and C.~Studerus,
  \href{http://dx.doi.org/10.1007/JHEP01(2011)102}{{\em JHEP} {\bf 01} (2011)
  102},
\href{http://arxiv.org/abs/1011.6661}{{\tt arXiv:1011.6661}}.

\bibitem{Czakon:2011ve}
M.~Czakon, \href{http://dx.doi.org/10.1016/j.nuclphysb.2011.03.020}{{\em
  Nucl.Phys.} {\bf B849} (2011)  250--295},
  \href{http://arxiv.org/abs/1101.0642}{{\tt arXiv:1101.0642 [hep-ph]}}.

\bibitem{Bierenbaum:2011gg}
I.~Bierenbaum, M.~Czakon, and A.~Mitov,
  \href{http://arxiv.org/abs/1107.4384}{{\tt arXiv:1107.4384 [hep-ph]}}.

\bibitem{Kidonakis:2009ev}
N.~Kidonakis, \href{http://dx.doi.org/10.1103/PhysRevLett.102.232003}{{\em
  Phys. Rev. Lett.} {\bf 102} (2009)  232003},
\href{http://arxiv.org/abs/0903.2561}{{\tt arXiv:0903.2561 [hep-ph]}}.

\bibitem{Mitov:2009sv}
A.~Mitov, G.~Sterman, and I.~Sung,
  \href{http://dx.doi.org/10.1103/PhysRevD.79.094015}{{\em Phys. Rev.} {\bf
  D79} (2009)  094015},
\href{http://arxiv.org/abs/0903.3241}{{\tt arXiv:0903.3241 [hep-ph]}}.

\bibitem{Becher:2009kw}
T.~Becher and M.~Neubert,
  \href{http://dx.doi.org/10.1103/PhysRevD.79.125004}{{\em Phys. Rev.} {\bf
  D79} (2009)  125004},
\href{http://arxiv.org/abs/0904.1021}{{\tt arXiv:0904.1021 [hep-ph]}}.

\bibitem{Ferroglia:2009ep}
A.~Ferroglia, M.~Neubert, B.~D. Pecjak, and L.~L. Yang, {\em Phys. Rev. Lett.}
  {\bf 103} (2009)  201601,
\href{http://arxiv.org/abs/0907.4791}{{\tt arXiv:0907.4791 [hep-ph]}}.

\bibitem{Ferroglia:2009ii}
A.~Ferroglia, M.~Neubert, B.~D. Pecjak, and L.~L. Yang,
  \href{http://dx.doi.org/10.1088/1126-6708/2009/11/062}{{\em JHEP} {\bf 11}
  (2009)  062},
\href{http://arxiv.org/abs/0908.3676}{{\tt arXiv:0908.3676 [hep-ph]}}.

\bibitem{Mitov:2010xw}
A.~Mitov, G.~F. Sterman, and I.~Sung,
  \href{http://dx.doi.org/10.1103/PhysRevD.82.034020}{{\em Phys. Rev.} {\bf
  D82} (2010)  034020},
\href{http://arxiv.org/abs/1005.4646}{{\tt arXiv:1005.4646}}.

\bibitem{Beneke:2009rj}
M.~Beneke, P.~Falgari, and C.~Schwinn,
  \href{http://dx.doi.org/10.1016/j.nuclphysb.2009.11.004}{{\em Nucl. Phys.}
  {\bf B828} (2010)  69--101},
\href{http://arxiv.org/abs/0907.1443}{{\tt arXiv:0907.1443 [hep-ph]}}.

\bibitem{Czakon:2009zw}
M.~Czakon, A.~Mitov, and G.~Sterman,
  \href{http://dx.doi.org/10.1103/PhysRevD.80.074017}{{\em Phys. Rev.} {\bf
  D80} (2009)  074017},
\href{http://arxiv.org/abs/0907.1790}{{\tt arXiv:0907.1790 [hep-ph]}}.

\bibitem{Beneke:2009ye}
M.~Beneke, M.~Czakon, P.~Falgari, A.~Mitov, and C.~Schwinn,
  \href{http://dx.doi.org/10.1016/j.physletb.2010.05.038}{{\em Phys. Lett.}
  {\bf B690} (2010)  483--490},
\href{http://arxiv.org/abs/0911.5166}{{\tt arXiv:0911.5166 [hep-ph]}}.

\bibitem{Ahrens:2009uz}
V.~Ahrens, A.~Ferroglia, M.~Neubert, B.~D. Pecjak, and L.~L. Yang,
  \href{http://dx.doi.org/10.1016/j.physletb.2010.03.048}{{\em Phys. Lett.}
  {\bf B687} (2010)  331--337},
\href{http://arxiv.org/abs/0912.3375}{{\tt arXiv:0912.3375 [hep-ph]}}.

\bibitem{Ahrens:2010zv}
V.~Ahrens, A.~Ferroglia, M.~Neubert, B.~D. Pecjak, and L.~L. Yang,
  \href{http://dx.doi.org/10.1007/JHEP09(2010)097}{{\em JHEP} {\bf 09} (2010)
  097},
\href{http://arxiv.org/abs/1003.5827}{{\tt arXiv:1003.5827}}.

\bibitem{Beneke:2010da}
M.~Beneke, P.~Falgari, and C.~Schwinn,
  \href{http://dx.doi.org/10.1016/j.nuclphysb.2010.09.009}{{\em Nucl. Phys.}
  {\bf B842} (2011)  },
\href{http://arxiv.org/abs/1007.5414}{{\tt arXiv:1007.5414 [hep-ph]}}.

\bibitem{Beneke:2010fm}
M.~Beneke, P.~Falgari, S.~Klein, and C.~Schwinn,
  \href{http://dx.doi.org/10.1016/j.nuclphysbps.2010.08.013}{{\em Nucl. Phys.
  Proc. Suppl.} {\bf 205-206} (2010)  20--24},
\href{http://arxiv.org/abs/1009.4011}{{\tt arXiv:1009.4011}}.

\bibitem{Kidonakis:2010dk}
N.~Kidonakis, \href{http://dx.doi.org/10.1103/PhysRevD.82.114030}{{\em Phys.
  Rev.} {\bf D82} (2010)  114030},
\href{http://arxiv.org/abs/1009.4935}{{\tt arXiv:1009.4935}}.

\bibitem{Ahrens:2011mw}
V.~Ahrens, A.~Ferroglia, M.~Neubert, B.~D. Pecjak, and L.~L. Yang,
\href{http://arxiv.org/abs/1103.0550}{{\tt arXiv:1103.0550}}.

\bibitem{Sterman:1986aj}
G.~Sterman,
{\em Nucl. Phys.} {\bf B281} (1987)  310.

\bibitem{Catani:1989ne}
S.~Catani and L.~Trentadue,
{\em Nucl. Phys.} {\bf B327} (1989)  323.

\bibitem{Moch:2008qy}
S.~Moch and P.~Uwer, \href{http://dx.doi.org/10.1103/PhysRevD.78.034003}{{\em
  Phys. Rev.} {\bf D78} (2008)  034003},
\href{http://arxiv.org/abs/0804.1476}{{\tt arXiv:0804.1476 [hep-ph]}}.

\bibitem{Cacciari:2008zb}
M.~Cacciari, S.~Frixione, M.~L. Mangano, P.~Nason, and G.~Ridolfi,
  \href{http://dx.doi.org/10.1088/1126-6708/2008/09/127}{{\em JHEP} {\bf 09}
  (2008)  127},
\href{http://arxiv.org/abs/0804.2800}{{\tt arXiv:0804.2800 [hep-ph]}}.

\bibitem{Kidonakis:2008mu}
N.~Kidonakis and R.~Vogt,
  \href{http://dx.doi.org/10.1103/PhysRevD.78.074005}{{\em Phys. Rev.} {\bf
  D78} (2008)  074005},
\href{http://arxiv.org/abs/0805.3844}{{\tt arXiv:0805.3844 [hep-ph]}}.

\bibitem{Hagiwara:2008df}
K.~Hagiwara, Y.~Sumino, and H.~Yokoya,
  \href{http://dx.doi.org/10.1016/j.physletb.2008.07.006}{{\em Phys. Lett.}
  {\bf B666} (2008)  71--76},
\href{http://arxiv.org/abs/0804.1014}{{\tt arXiv:0804.1014 [hep-ph]}}.

\bibitem{Kiyo:2008bv}
Y.~Kiyo, J.~H. K{\"u}hn, S.~Moch, M.~Steinhauser, and P.~Uwer,
  \href{http://dx.doi.org/10.1140/epjc/s10052-009-0892-7}{{\em Eur. Phys. J.}
  {\bf C60} (2009)  375--386},
\href{http://arxiv.org/abs/0812.0919}{{\tt arXiv:0812.0919 [hep-ph]}}.

\bibitem{Becher:2006nr}
T.~Becher and M.~Neubert, {\em Phys. Rev. Lett.} {\bf 97} (2006)  082001,
\href{http://arxiv.org/abs/hep-ph/0605050}{{\tt hep-ph/0605050}}.

\bibitem{Becher:2006mr}
T.~Becher, M.~Neubert, and B.~D. Pecjak, {\em JHEP} {\bf 01} (2007)  076,
\href{http://arxiv.org/abs/hep-ph/0607228}{{\tt hep-ph/0607228}}.

\bibitem{Becher:2007ty}
T.~Becher, M.~Neubert, and G.~Xu,
  \href{http://dx.doi.org/10.1088/1126-6708/2008/07/030}{{\em JHEP} {\bf 07}
  (2008)  030},
\href{http://arxiv.org/abs/0710.0680}{{\tt arXiv:0710.0680 [hep-ph]}}.

\bibitem{Beneke:1999qg}
M.~Beneke, A.~Signer, and V.~A. Smirnov,
  \href{http://dx.doi.org/10.1016/S0370-2693(99)00343-3}{{\em Phys. Lett.} {\bf
  B454} (1999)  137--146},
\href{http://arxiv.org/abs/hep-ph/9903260}{{\tt arXiv:hep-ph/9903260}}.

\bibitem{Hoang:2000yr}
A.~H. Hoang {\em et al.}, {\em Eur. Phys. J. direct} {\bf C2} (2000)  1,
\href{http://arxiv.org/abs/hep-ph/0001286}{{\tt arXiv:hep-ph/0001286}}.

\bibitem{Hoang:2001mm}
A.~H. Hoang, A.~V. Manohar, I.~W. Stewart, and T.~Teubner,
  \href{http://dx.doi.org/10.1103/PhysRevD.65.014014}{{\em Phys. Rev.} {\bf
  D65} (2002)  014014},
\href{http://arxiv.org/abs/hep-ph/0107144}{{\tt arXiv:hep-ph/0107144}}.

\bibitem{Pineda:2006ri}
A.~Pineda and A.~Signer,
  \href{http://dx.doi.org/10.1016/j.nuclphysb.2006.09.025}{{\em Nucl. Phys.}
  {\bf B762} (2007)  67--94},
\href{http://arxiv.org/abs/hep-ph/0607239}{{\tt arXiv:hep-ph/0607239}}.

\bibitem{Bernreuther:2006vg}
W.~Bernreuther, M.~F{\"u}cker, and Z.-G. Si,
  \href{http://dx.doi.org/10.1103/PhysRevD.74.113005}{{\em Phys. Rev.} {\bf
  D74} (2006)  113005},
\href{http://arxiv.org/abs/hep-ph/0610334}{{\tt arXiv:hep-ph/0610334}}.

\bibitem{Kuhn:2006vh}
J.~H. K{\"u}hn, A.~Scharf, and P.~Uwer,
  \href{http://dx.doi.org/10.1140/epjc/s10052-007-0275-x}{{\em Eur. Phys. J.}
  {\bf C51} (2007)  37--53},
\href{http://arxiv.org/abs/hep-ph/0610335}{{\tt arXiv:hep-ph/0610335}}.

\bibitem{Denner:2010jp}
A.~Denner, S.~Dittmaier, S.~Kallweit, and S.~Pozzorini,
  \href{http://dx.doi.org/10.1103/PhysRevLett.106.052001}{{\em Phys. Rev.
  Lett.} {\bf 106} (2011)  052001},
\href{http://arxiv.org/abs/1012.3975}{{\tt arXiv:1012.3975 [hep-ph]}}.

\bibitem{Bevilacqua:2010qb}
G.~Bevilacqua, M.~Czakon, A.~van Hameren, C.~G. Papadopoulos, and M.~Worek,
  \href{http://dx.doi.org/10.1007/JHEP02(2011)083}{{\em JHEP} {\bf 02} (2011)
  083},
\href{http://arxiv.org/abs/1012.4230}{{\tt arXiv:1012.4230 [hep-ph]}}.

\bibitem{Kidonakis:1997gm}
N.~Kidonakis and G.~Sterman,
  \href{http://dx.doi.org/10.1016/S0550-3213(97)00506-3}{{\em Nucl. Phys.} {\bf
  B505} (1997)  321--348},
\href{http://arxiv.org/abs/hep-ph/9705234}{{\tt arXiv:hep-ph/9705234}}.

\bibitem{Czakon:2008cx}
M.~Czakon and A.~Mitov,
  \href{http://dx.doi.org/10.1016/j.physletb.2009.08.036}{{\em Phys. Lett.}
  {\bf B680} (2009)  154--158},
\href{http://arxiv.org/abs/0812.0353}{{\tt arXiv:0812.0353 [hep-ph]}}.

\bibitem{Bierenbaum:2008yu}
I.~Bierenbaum, J.~Bl{\"u}mlein, S.~Klein, and C.~Schneider,
  \href{http://dx.doi.org/10.1016/j.nuclphysb.2008.05.016}{{\em Nucl.Phys.}
  {\bf B803} (2008)  1--41}, \href{http://arxiv.org/abs/0803.0273}{{\tt
  arXiv:0803.0273 [hep-ph]}}.

\bibitem{Martin:2009iq}
A.~D. Martin, W.~J. Stirling, R.~S. Thorne, and G.~Watt,
  \href{http://dx.doi.org/10.1140/epjc/s10052-009-1072-5}{{\em Eur. Phys. J.}
  {\bf C63} (2009)  189--285},
\href{http://arxiv.org/abs/0901.0002}{{\tt arXiv:0901.0002 [hep-ph]}}.

\bibitem{Czakon:2008ii}
M.~Czakon and A.~Mitov,
  \href{http://dx.doi.org/10.1016/j.nuclphysb.2009.08.020}{{\em Nucl. Phys.}
  {\bf B824} (2010)  111--135},
\href{http://arxiv.org/abs/0811.4119}{{\tt arXiv:0811.4119 [hep-ph]}}.

\bibitem{Aliev:2010zk}
M.~Aliev {\em et al.}, \href{http://dx.doi.org/10.1016/j.cpc.2010.12.040}{{\em
  Comput. Phys. Commun.} {\bf 182} (2011)  1034--1046},
\href{http://arxiv.org/abs/1007.1327}{{\tt arXiv:1007.1327}}.

\bibitem{Gehrmann:2001pz}
T.~Gehrmann and E.~Remiddi,
  \href{http://dx.doi.org/10.1016/S0010-4655(01)00411-8}{{\em
  Comput.Phys.Commun.} {\bf 141} (2001)  296--312},
  \href{http://arxiv.org/abs/hep-ph/0107173}{{\tt arXiv:hep-ph/0107173
  [hep-ph]}}.

\bibitem{Bonvini:2010tp}
M.~Bonvini, S.~Forte, and G.~Ridolfi,
  \href{http://dx.doi.org/10.1016/j.nuclphysb.2011.01.023}{{\em Nucl. Phys.}
  {\bf B847} (2011)  93--159},
\href{http://arxiv.org/abs/1009.5691}{{\tt arXiv:1009.5691 [hep-ph]}}.

\bibitem{Martin:2009bu}
A.~D. Martin, W.~J. Stirling, R.~S. Thorne, and G.~Watt,
  \href{http://dx.doi.org/10.1140/epjc/s10052-009-1164-2}{{\em Eur. Phys. J.}
  {\bf C64} (2009)  653--680},
\href{http://arxiv.org/abs/0905.3531}{{\tt arXiv:0905.3531 [hep-ph]}}.

\bibitem{Ahrens:2011px}
V.~Ahrens, A.~Ferroglia, B.~D. Pecjak, and L.~L. Yang,
\href{http://arxiv.org/abs/1105.5824}{{\tt arXiv:1105.5824 [hep-ph]}}.

\bibitem{JimenezDelgado:2008hf}
P.~Jimenez-Delgado and E.~Reya,
  \href{http://dx.doi.org/10.1103/PhysRevD.79.074023}{{\em Phys. Rev.} {\bf
  D79} (2009)  074023},
\href{http://arxiv.org/abs/0810.4274}{{\tt arXiv:0810.4274 [hep-ph]}}.

\bibitem{Alekhin:2009ni}
S.~Alekhin, J.~Bl{\"u}mlein, S.~Klein, and S.~Moch,
  \href{http://dx.doi.org/10.1103/PhysRevD.81.014032}{{\em Phys. Rev.} {\bf
  D81} (2010)  014032},
\href{http://arxiv.org/abs/0908.2766}{{\tt arXiv:0908.2766 [hep-ph]}}.

\bibitem{Aaron:2009wt}
{\bf H1 and ZEUS} Collaboration, F.~D. Aaron {\em et al.},
  \href{http://dx.doi.org/10.1007/JHEP01(2010)109}{{\em JHEP} {\bf 01} (2010)
  109},
\href{http://arxiv.org/abs/0911.0884}{{\tt arXiv:0911.0884 [hep-ex]}}.

\bibitem{Ball:2011uy}
{\bf The NNPDF} Collaboration, R.~D. Ball {\em et al.},
\href{http://arxiv.org/abs/1107.2652}{{\tt arXiv:1107.2652 [hep-ph]}}.

\bibitem{Watt:2011kp}
G.~Watt,
\href{http://arxiv.org/abs/1106.5788}{{\tt arXiv:1106.5788 [hep-ph]}}.

\bibitem{TopMass2010}
{\bf CDF and D0} Collaboration, {Tevatron Electroweak Working Group},
\href{http://arxiv.org/abs/1007.3178}{{\tt arXiv:1007.3178 [hep-ex]}}.

\bibitem{Beneke:1999zr}
M.~Beneke, \href{http://arxiv.org/abs/hep-ph/9911490}{{\tt
  arXiv:hep-ph/9911490}}.
Proceedings of the 8th International Symposium on Heavy Flavor Physics (Heavy
  Flavors 8), Southampton, England, 25-29 Jul 1999.

\bibitem{Vermaseren:1998uu}
J.~A.~M. Vermaseren, \href{http://dx.doi.org/10.1142/S0217751X99001032}{{\em
  Int. J. Mod. Phys.} {\bf A14} (1999)  2037--2076},
\href{http://arxiv.org/abs/hep-ph/9806280}{{\tt arXiv:hep-ph/9806280}}.

\bibitem{Blumlein:1998if}
J.~Bl{\"u}mlein and S.~Kurth,
  \href{http://dx.doi.org/10.1103/PhysRevD.60.014018}{{\em Phys. Rev.} {\bf
  D60} (1999)  014018},
\href{http://arxiv.org/abs/hep-ph/9810241}{{\tt arXiv:hep-ph/9810241}}.

\bibitem{Blumlein:2009ta}
J.~Bl{\"u}mlein, \href{http://dx.doi.org/10.1016/j.cpc.2009.07.004}{{\em
  Comput.Phys.Commun.} {\bf 180} (2009)  2218--2249},
  \href{http://arxiv.org/abs/0901.3106}{{\tt arXiv:0901.3106 [hep-ph]}}.

\bibitem{Albino:2009ci}
S.~Albino, \href{http://dx.doi.org/10.1016/j.physletb.2009.02.053}{{\em
  Phys.Lett.} {\bf B674} (2009)  41--48},
  \href{http://arxiv.org/abs/0902.2148}{{\tt arXiv:0902.2148 [hep-ph]}}.

\bibitem{Ablinger:2010kw}
J.~Ablinger,
\href{http://arxiv.org/abs/1011.1176}{{\tt arXiv:1011.1176 [math-ph]}}.

\bibitem{Pineda:1997hz}
A.~Pineda and F.~Yndurain,
  \href{http://dx.doi.org/10.1103/PhysRevD.58.094022}{{\em Phys.Rev.} {\bf D58}
  (1998)  094022}, \href{http://arxiv.org/abs/hep-ph/9711287}{{\tt
  arXiv:hep-ph/9711287 [hep-ph]}}.

\bibitem{Beneke:2005hg}
M.~Beneke, Y.~Kiyo, and K.~Schuller,
  \href{http://dx.doi.org/10.1016/j.nuclphysb.2005.02.028}{{\em Nucl.Phys.}
  {\bf B714} (2005)  67--90}, \href{http://arxiv.org/abs/hep-ph/0501289}{{\tt
  arXiv:hep-ph/0501289 [hep-ph]}}.

\end{thebibliography}\endgroup

\end{document}